\documentclass[aps,pra,twocolumn,showpacs,superscriptaddress]{revtex4-1}

\usepackage{amsfonts,mathrsfs,amsmath,amsthm,amssymb}
\usepackage{epsfig,upgreek}
\usepackage{bm}
\usepackage{color}
\usepackage[dvipsnames]{xcolor}

\graphicspath{ {./figs/} }

\usepackage{rotating}
\usepackage{array,multirow,graphicx}
\usepackage{tabularx}

\newcolumntype{M}[1]{>{\centering\arraybackslash}m{#1}}

\makeatletter
\newsavebox{\@brx}
\newcommand{\llangle}[1][]{\savebox{\@brx}{\(\m@th{#1\langle}\)}%
 \mathopen{\copy\@brx\kern-0.5\wd\@brx\usebox{\@brx}}}
\newcommand{\rrangle}[1][]{\savebox{\@brx}{\(\m@th{#1\rangle}\)}%
 \mathclose{\copy\@brx\kern-0.5\wd\@brx\usebox{\@brx}}}
\makeatother

\begin{document}  
\title {\bf 
Universal compilation for quantum state preparation and tomography
}

\author{Vu Tuan Hai}
\affiliation{University of Information Technology, 
Ho Chi Minh City, 700000, Vietnam}
\affiliation{
Vietnam National University, Ho Chi Minh City, 700000, Vietnam}


\author{ Le Bin Ho}
\thanks{Electronic address: binho@riec.tohoku.ac.jp}
\affiliation{Ho Chi Minh City Institute of Physics, 
National Institute of Applied Mechanics and Informatics, 
Vietnam Academy of Science and Technology, Ho Chi Minh City, 700000, Vietnam}
\affiliation{Frontier Research Institute for Interdisciplinary Sciences, 
Tohoku University, Sendai 980-8578, Japan}
\affiliation{Department of Applied Physics, Graduate School of Engineering, Tohoku University, Sendai 980-8579, Japan}

\date{\today}

\begin{abstract}
Universal compilation is 
a 
training process that 
compiles a trainable unitary
into a target unitary 
and it serves vast potential applications
from quantum dynamic simulations to
optimal circuits with deep-compressing, 
device benchmarking, 
quantum error mitigation, and so on.
Here, we propose 
a universal compilation-based 
variational algorithm 
for the preparation and tomography 
of quantum states in low-depth quantum circuits. 
We apply the Fubini-Study distance
to be a trainable cost function 
under various gradient-based optimizers,
including the quantum natural gradient approach.
We evaluate the performance 
of various unitary topologies
and the trainability of 
different 
optimizers 
for getting high efficiency. 
In practice, we address different circuit ansatzes 
in quantum state preparation, 
including the linear and 
graph-based ansatzes 
for preparing different 
entanglement target states 
such as representative GHZ and W states. 
We also discuss the effect of the circuit depth,
barren plateau, 
readout noise in the model, 
and the error mitigation solution. 
We next evaluate the reconstructing efficiency 
in quantum state tomography 
via various popular circuit ansatzes 
and reveal the crucial role 
of the circuit depth 
in the robust fidelity. The results are comparable 
with the shadow tomography method, 
a similar fashion in the field. 
Our work expresses the adequate capacity 
of the universal compilation-based 
variational algorithm to maximize 
the efficiency in the quantum state 
preparation and tomography.
Further, it promises applications 
in quantum metrology and sensing 
and is applicable in the near-term 
quantum computers for verification 
of the circuits fidelity 
and various quantum computing tasks.
\end{abstract}
%
%
\maketitle

\section{Introduction} 

Quantum computers promise 
an excellent computational capacity 
that is intractable for classical computers
to solve challenging problems,
including materials science 
\cite{doi:10.1126/science.abb2823,
PRXQuantum.2.017001,
Ebadi2021},
information science
\cite{Pirandola2015,SPILLER200330},
computer science 
\cite{365700, grover1996fast}, 
mathematical science 
\cite{PhysRevLett.103.150502,
PhysRevApplied.15.034068,
PhysRevA.101.010301},
and others.
However, there are two major challenges
to bringing quantum computers to materialize 
\cite{PRXQuantum.2.017001}:
(i) it is difficult to access full information 
from entangled systems 
because of the state collapse upon measurements,
and (ii) it is difficult to build, control,
and measure quantum states 
with arbitrarily high accuracy. 
In this regard, even though the current state-of-the-art 
quantum computers rely on 
the noisy intermediate-scale (NISQ devices), 
which usually 
prevents high efficiency 
\cite{Preskill2018quantumcomputingin}, 
various hybrid quantum-classical algorithms 
have been proposed and 
actively studied recently \cite{Cerezo2021_r},
and that could be promising for 
quantum speedup in the regime of NISQ devices.
Massive applications include variational quantum eigensolvers 
\cite{Peruzzo2014,PhysRevResearch.1.033062,
Kirby2021contextualsubspace,Gard2020,PRXQuantum.2.020337},
quantum approximate optimization algorithms
\cite{PhysRevX.10.021067},
new frontiers in quantum foundations 
\cite{Arrasmith2019,
PhysRevLett.123.260505,
Koczor_2020,Meyer2021},
and others, have been reported.

Beyond the actively studied VQAs, 
the universal compilation 
has drawn tremendous interest recently.
Its core idea relies on 
a training process to transform 
a trainable unitary into a target unitary
\cite{heya2018variational,
Khatri2019quantumassisted}.
It has demonstrated different applications 
in gate optimization
\cite{heya2018variational},
quantum-assisted itself compiling process
\cite{Khatri2019quantumassisted},
continuous-variable quantum learning
\cite{PRXQuantum.2.040327},
and robust quantum compilation 
\cite{Jones2022robustquantum}.
(We highlight attention that 
does not be mistaken for the term ``quantum compilation" 
which is used for compiling a high-level quantum language
into a hardware description language 
\cite{https://doi.org/10.48550/arxiv.2112.00187}.)
The future of universal quantum compiling
could be circuits depth-compression, 
black-box compiling, 
error mitigation, 
gate-fidelity benchmarking,
and efficient gate synthesis.

In another aspect, recently, 
the preparation and tomography 
of quantum states in quantum circuits 
have attracted significant attention 
owing to the incredible advantages 
of the quantum device 
\cite{aulicino2022state,
PhysRevResearch.4.013091,
Kuzmin2020variationalquantum,
lvovsky2009continuous, 
d2002quantum, 
takeda2021quantum}. 
Quantum computers allow
to efficiently prepare quantum states 
with high confidence,
fully control the Hamiltonian 
for the state evolution,
and directly access 
the measurement results. 
Early works on the 
quantum state preparation direction 
can be grouped into 
``with" and ``without" ancillary qubits.
Without ancillary qubits, 
however, a major obstacle is 
the invertible exponential growth 
of the circuits depth, 
ie.,  $\mathcal{O}[2^N]$ 
\cite{10.5555/2011670.2011675,
1629135,PhysRevA.93.032318,
PhysRevA.83.032302}, and
$\mathcal{O}[2^{N}/N]$
\cite{sun2021asymptotically}.
With ancillary qubits, 
the circuit depth significantly reduces
to sub-exponential scaling 
\cite{sun2021asymptotically,
PhysRevResearch.3.043200,
rosenthal2022query,Araujo2021},
e.g. $\mathcal{O}[2^{N/2}]$,
but those still require an exponential
number of ancillary qubits in the worse cases.

Similarly, for quantum state tomography (QST,)
the standard approach requires 
an exponentially growing 
number of measurements, 
which is intractable for large systems. 
Numerous methods have been proposed
for improving the QST
in terms of efficiency \cite{Palmieri2020,
Cramer2010,PhysRevA.92.042312,Moroder_2012}, 
methodology \cite{PhysRevResearch.3.033278,
PhysRevLett.105.250403,Torlai2018-sm,
Blume_Kohout_2010,PRXQuantum.2.020303,
PhysRevLett.105.150401,Flammia_2012}, 
quantum dynamic \cite{Czerwinski2020,
PhysRevX.10.011006,PhysRevA.100.042109},
and so on.
Recently, in terms of quantum circuits-based 
QST, 
a variational approach
\cite{PhysRevA.101.052316}
and single-shot measurements
\cite{PhysRevA.98.052302,
PhysRevLett.126.170504},
have been investigated. 

Despite recent achievements 
on 
quantum state preparation
and quantum state tomography, 
it is still challenging to 
implement them in the NISQ devices.
In this work, we introduce a promising application 
of the universal compilation 
on quantum state preparation 
and quantum state tomography. 
Our main idea is to use a trainable unitary 
that acts upon a known fiducial state 
to prepare a target state 
or reconstruct an unknown state.
The target state and unknown state 
are created using a suitable target unitary 
that acts upon the fiducial state. 
Then, the solution returns
to the compilation from the 
trainable unitary to the target unitary.
The advantage of this method is that 
it requires low-depth trainable unitaries 
and few measurements to realize 
the target unitaries that significantly 
reduce the complexity. Furthermore, 
the flexibility of the trainable unitaries 
is more elevated than 
that of the target unitaries, 
resulting in a better fault-tolerant capacity 
and thus allowing high efficiency 
for the trainable quantum circuits.

Concretely, we first introduce the general 
framework of the universal compilation-based
quantum variational algorithm (UC-VQA)
and then apply it to particular 
quantum state preparation (QSP) 
and quantum state tomography (QST).
We also introduce several gradient-based optimizers,
including the standard gradient descent (SGD), 
the Adam, and the quantum natural gradient descent (QNG).

We discuss the numerical experiment results
for these two processes. For the QSP, we
propose several structures for the trainable unitary
such as linear, graph-based polygon, and 
graph-based star ansatzes 
for preparing different entanglement 
target states such as representative 
GHZ and W states. 
Here, we reduce the circuit depth 
from exponential to polynomial, 
i.e., $\mathcal{O}(N)$, 
$N$ the qubit numbers.
We also discuss the accuracy
under the effect of 
the circuit depth, barren plateau, 
readout noise, 
and 
the error mitigation solution.
The result is applicable 
to any arbitrary target state. 
For the QST, we first examine the simple 
case of single-qubit tomography, 
and then evaluate 
the reconstructing efficiency 
of unknown Haar random states
via various popular circuit ansatzes.
We find that the circuit depth plays 
a crucial role in the robust fidelity,
i.e., by choosing a proper circuit depth 
via the number of layers in the quantum circuit,
we get high fidelity at any qubit numbers.
We finally compare the results 
with the shadow tomography method
\cite{https://doi.org/10.48550/arxiv.1711.01053,
Huang2020}, 
a similar fashion in the field. 

The study reveals that 
the accuracy mainly relies on
(i) the ansatz topologies 
with the optimal circuit depth 
and (ii) the significant impact of
different optimizers such as 
the standard gradient descent, the Adam, 
and the quantum natural gradient descent. 
Our study can further promise applications 
in quantum metrology and sensing 
and is applicable in the near-term 
quantum computers for verification 
of the circuits fidelity 
and various quantum computing tasks.


\section{Framework}{\label{secii}}
We introduce a variational quantum algorithm that
bases on universal compilation \cite{Jones2022robustquantum,PRXQuantum.2.040327,
Khatri2019quantumassisted,heya2018variational} 
to translate a given state into another one, 
and apply it to quantum state preparation
and quantum state tomography.

\subsection{Universal compilation-based variational quantum algorithm}
\begin{figure*}[t]
\centering
\includegraphics[width=\textwidth]{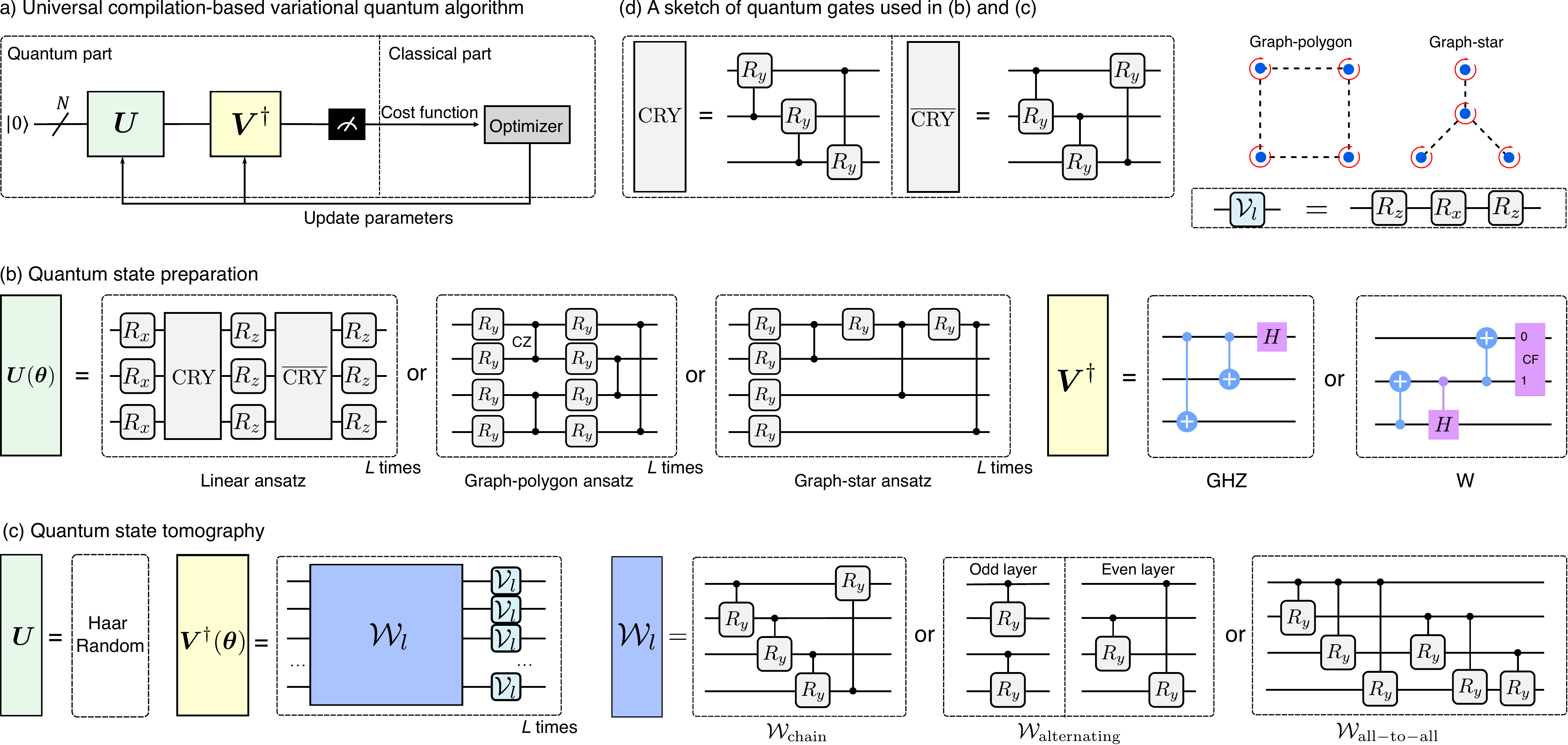}
\caption{
(a) Universal compilation-based 
variational quantum algorithm (UC-VQA).
It consists of a quantum part and a classical part.
In the quantum part, the final state 
$|\psi_f\rangle$ is created
by applying a set of quantum gates $\bm U$ 
followed by $\bm V^\dagger$
onto the initial circuit
and then be measured.
In the classical part, 
we compute the appropriate cost function, 
use an optimizer to compute new parameters, 
and update the scheme until it converges.
(b) Quantum state preparation. 
The unitary $\bm U$
is parameterized into $\bm U(\bm\theta)$ 
with several ansatzes,
including the linear, graph-polygon, 
and graph-star ansatzes.
The unitary $\bm V$ is fixed for generating 
entangled GHZ and W classes.
(c) Quantum state tomography. 
The unitary $\bm U$
is a Haar random generator while 
$\bm V^\dagger$ is parameterized 
into $\bm V^\dagger(\bm\theta)$ 
and broken out into
entangled gates $\mathcal{W}$ 
and local rotation gates $\mathcal{V}$
with several structures as shown in the figure.
(d) A sketch of some quantum gates used in (b) and (c).
Other notations: $N$ the number of qubits, $L$ the number of layers,
$R_j, j = \{x, y, z\}$ the rotation gate around the $j$-axis,
CZ controlled-rotation-$Z$ gate, $H$ Hadamard gate,
CF controlled-F gate \cite{diker2016deterministic}.
}
\label{fig:1}
\end{figure*}

A universal compilation-based 
variational quantum algorithm (UC-VQA) 
consists of a quantum 
and a classical part, as shown in Fig.~\ref{fig:1} (a).
The quantum part is a circuit with parameterizable ansatzes. 
Let $\bm U(\bm\theta)$ and 
$\bm V^\dagger(\bm\theta')$ are two unitary ansatzes 
(sets of quantum gates with some parameters 
$\bm\theta, \bm\theta'$) 
that act onto the circuit and transform
an initial state $|\psi_0\rangle$ 
into a final state $|\psi_f\rangle$ as
\begin{align}\label{eq:fin}
	|\psi_{f}\rangle
	= \bm V^\dagger(\bm\theta')
	\bm U(\bm\theta)|\psi_0\rangle.
\end{align}
The transition probability yields
\begin{align}\label{eq:trans}
	p(\psi_0\to\psi_f) 
	= \big|\langle\psi_0|\psi_f\rangle\big|^2
	= \big|\langle\psi_0|\bm V^\dagger(\bm\theta')
	\bm U(\bm\theta)|\psi_0\rangle\big|^2.
\end{align} 
Our task is to maximize the
transition probability $p_{\rm max}(\psi_0\to\psi_f)$,
such that a state $|\psi(\bm\theta)\rangle
\equiv \bm U(\bm\theta)|\psi_0\rangle$
is compiled to
$|\phi(\bm\theta')\rangle
\equiv \bm V(\bm\theta')|\psi_0\rangle$.
We thus name this method is 
``universal compilation-based 
variational quantum algorithm."
Alternatively, we introduce a ``quantum kernel" 
for the transition probability as
\begin{align}\label{eq:kernel}
	\mathcal{K}(\bm\theta,\bm\theta') = 
	\big|\langle\phi(\bm\theta')|\psi(\bm\theta)\rangle\big|^2
	= p(\psi_0\to\psi_f),
\end{align}
such that
\begin{align}\label{eq:pmax}
	p_{\rm max}(\psi_0\to\psi_f) = 
	\operatorname*{argmax}_{\bm\theta, \bm\theta'}
   	 \mathcal{K}(\bm{\theta}, \bm\theta').
\end{align}
The maximization of the transition probability,
i.e., $p(\psi_0\to\psi_f) = 1$,  implies
$|\psi(\bm\theta)\rangle = |\phi(\bm\theta')\rangle$
in the quantum kernel,
which can be applied to the quantum state preparation
and quantum state tomography as we will describe below.

In the classical part, 
we define a cost function that relates to 
the kernel $\mathcal{K}(\bm{\theta}, \bm\theta')$
and use it
to compute new parameters 
$\bm\theta,\bm\theta'$ 
via an appropriate optimizer (see Sec.~\ref{seciiD}).
These new parameters are 
iteratively fed back 
to the quantum circuit until the convergence 
criteria are satisfied.


\subsection{Quantum state preparation}
Quantum states preparation (QSP) 
performs
a set of controllable evolutions  
on a quantum system to 
transform its initial state into a target state.
This is an essential step 
in quantum computation and quantum information processing. 
Here, we present a UC-VQA scheme 
that transforms an initial quantum register 
into an arbitrary target state with
high accuracy and
well against the noise under mitigation aid.

Starting from an initial register 
$|\bm 0\rangle \equiv |0\rangle^{\otimes N}$
in the quantum circuit as shown in Fig.~\ref{fig:1} (a), 
where $N$ is the number of qubits,
the task is to transform this state into 
a target quantum state, i.e., 
$|\tau\rangle$. 
First, we transform the initial register 
into a variational state 
\begin{align}\label{eq:psi_var}
|\upsilon(\bm\theta)\rangle = 
\bm U(\bm\theta)|\bm 0\rangle,
\end{align}
under a trainable unitary $\bm U(\bm\theta)$,
where $\bm\theta = \{\theta_1, 
\theta_2,\cdots, \theta_M\}$ can be 
adaptively updated during a training process, 
$M$ is the number of trainable parameters.
The target state can be expressed 
in terms of a quantum circuit as 
\begin{align}\label{eq:psi_tar}
	|\tau\rangle = \bm V|\bm0\rangle,
\end{align}	
with a target (known) unitary $\bm V$.
To qualify how closed the two states are,
we consider the Fubini-Study distance as \cite{Kuzmak2021}
\begin{align}\label{eq:FB_dis}
	d(\upsilon(\bm\theta),\tau)
	= \sqrt{1-\big|\langle \tau |\upsilon(\bm\theta)\rangle\big|^2}
    = \sqrt{1-p_0(\bm \theta)},
\end{align}
where $p_0(\bm \theta) = |\langle \tau|\upsilon(\bm\theta)\rangle|^2
= |\langle \bm 0|
\bm V^\dagger\bm U(\bm\theta) |\bm 0\rangle|^2
$ is the probability 
for getting the outcome $|\bm 0\rangle$.
In the quantum circuit, 
we apply a sequence of $\bm U(\bm\theta)$
followed by $\bm V^\dagger$ 
onto the initial state $|\bm 0\rangle$
to get the final state $\bm V^\dagger\bm 
U(\bm\theta) |\bm 0\rangle$ 
and then measure a projective operator
$\bm P_0 = |\bm 0\rangle\langle\bm 0|$, 
which yields the probability $p_0(\bm\theta)$. 

The variational state becomes the target state
if their distance reaches zero. 
In the variational circuit, we thus use the 
Fubini-Study distance as a cost function
that needs to minimize, i.e.,
$\mathcal{C(\bm\theta}) = 
d(\upsilon(\bm\theta),\tau)$, such that
\begin{align}\label{eq:optimize-theta}
    \bm\theta^*=\operatorname*{argmin}_{\bm\theta}
    \mathcal{C}(\bm{\theta}).
\end{align}
By training the variational circuit until it converges,
we obtain $\bm\theta^*$, 
and the variational state
$|\upsilon(\bm\theta^*)\rangle$
reaches the target state $|\tau\rangle$.

In Sec.~\ref{seciii_i}, we show the results for 
preparing some representative target states, 
such as entangled GHZ and W classes.
We use several ansatzes 
$\bm U(\bm\theta)$, including the linear ansatz,
the graph-polygon ansatz, and the graph-star ansatz,
while the structure of $\bm V$ is fixed so that 
it can generate the desired target states. 
The corresponding $\bm U(\bm\theta)$ 
and $\bm V$ structures are shown in Fig.~\ref{fig:1} (b).
As mentioned beforehand in general, 
it requires 
a circuit depth at exponential 
or sub-exponential
to prepare an arbitrary quantum state.
However, using those proposed ansatzes,
we efficiently reduce the circuit depth to 
$\mathcal{O}[(2N+3)L], \mathcal{O}[5L] $, 
and $\mathcal{O}[(2N-2)L]$
for the linear, graph-polygon, 
and graph-star ansatzes,
respectively,
with $N$ is the number of quits
and $L$ is the number of layers.

\subsection{Quantum state tomography}
Quantum state tomography (QST) 
is a measurement process 
performed on many identical copies of a system
to extract its state's information
\cite{paris2004quantum}.
In general, for a given unknown quantum state
$|\mu\rangle$ in a complex Hilbert space 
of $d$-dimension, it 
requires a set of $2^d-1$ measurements
on different bases
to reproduce the state completely. 
Here, we introduce an alternative QST framework 
based on the UC-VQA, 
that would show high accuracy
and comparable with the shadow tomography,
a similar fashion for QST using 
few-measurement quantum circuits.

Starting from a random (unknown) quantum state
\begin{align}\label{eq:unknown}
    |\mu\rangle = \bm U|\bm 0\rangle,
\end{align}
where a Haar random unitary $\bm U$ 
\cite{white2020mana}
acting on the initial state 
$|\bm 0\rangle$. 
To reconstruct this state,
we apply a trainable unitary evolution 
$\bm V^\dagger(\bm\theta)$ 
that can learn the role of $\bm U$,
such that a reconstructed state 
\begin{align}\label{eq:re_state}
|\gamma(\bm\theta)\rangle 
= \bm V(\bm\theta)|\bm 0\rangle,
\end{align}
resembles to the unknown state $|\mu\rangle$.
The efficiency of the tomography process
is given by the Fubini-Study distance as
\begin{align}\label{eq:FB_tomo}
   d(\mu,\gamma(\bm\theta)) = 
    \sqrt{1-|\langle\gamma(\bm\theta)|\mu\rangle|^2}
    = \sqrt{1-p_0(\bm \theta)},
\end{align}
where $p_0(\bm \theta) = 
|\langle\gamma(\bm\theta)|\mu\rangle|^2
= |\langle \bm 0|\bm V^\dagger(\bm\theta)\bm U|\bm 0\rangle|^2$.

Similar to the above, we also use 
the Fubini-Study distance
as the cost function and train 
the model to get the reconstructed state 
$|\gamma(\bm\theta^*)\rangle 
= \bm V(\bm\theta^*)|\bm 0\rangle$
with an optimal $\bm\theta^*$.
For the training process, 
we generate the unknown state
following the Haar random, 
while the ansatz $\bm V^\dagger(\bm\theta)$ 
is broken out into
\begin{align}\label{eq:V_theta}
	\bm V^\dagger(\bm\theta)
	= \prod_{l = 1}^L \mathcal{V}_l(\bm\theta_l)
	\mathcal{W}_l(\bm\theta_l); \text{ with }
	\mathcal{V}_l = R_zR_xR_z,
\end{align}
and $\mathcal{W}_l$ includes the chain, alternating, and all-to-all
structures \cite{PRXQuantum.2.040309} 
as shown in Fig.~\ref{fig:1}(c). 
The training results are shown in Sec.~\ref{seciii_ii}.

We emphasize that both $\bm U(\bm\theta)$ 
in the QSP and the entangled gates $\mathcal{W}$ here 
consist of two-qubit controlled $y$-rotation gates, 
which differs from previous works
\cite{PRXQuantum.2.040309}.
We refer to these gates as 
``parameter-dependent entanglement gate."
They are useful for preparing 
variational states metrology \cite{Koczor_2020}
and rapid entangled circuits \cite{schuld2020circuit},
for testing of the expressibility and entangling capability 
\cite{sim2019expressibility},
and so on.

\subsection{Training}
\label{seciiD}
The training process is a hybrid 
protocol as illustrated in figure~\ref{fig:1} (a):
a set of unitary gates $\bm U$
followed by $\bm V^\dagger$  apply onto the circuit
and measure the final state. 
The results are sent to the 
classical counterpart to compute 
the corresponding cost function
and then update new parameters $\bm\theta$ 
using a suitable optimizer protocol
until it reaches convergence. 

We use gradient-based optimizations 
to iteratively update the parameters $\bm\theta$ 
and minimize the cost function. 
To do that, we need to calculate the derivative
$\partial \mathcal{C}(\bm\theta)/
\partial\theta_{j}$  w.r.t $\theta_j$ 
in the $j^{\rm th}$ gate for every $\theta_j\in\bm\theta$. 
We compute two cases as follows.
First, if the $j^{\rm th}$ gate is a single-qubit rotation gate,
i.e., $\exp(-i\theta_j\sigma_k/2), k \in \{x,y,z\}$, then
using the standard (two-term) parameter-shift rule 
\cite{mitarai2018quantum, schuld2019evaluating}, 
we have
\begin{align}\label{eq:psr}
\notag \dfrac{\partial \mathcal{C}(\bm\theta)}
{\partial\theta_j} &=
-\dfrac{1}{2\mathcal{C}(\bm\theta)}
\dfrac{\partial p_{0}(\bm\theta)}{\partial \theta_j}\\
&=-\frac{1}{2\mathcal{C}(\bm\theta)}
\frac{1}{2\sin(s)}
\Big[p_0(\bm\theta + s\bm e_j) - 
p_0(\bm\theta - s\bm e_j)\Big],
\end{align}
where $s$ denotes an arbitrary shift,
and $\bm e_j$ is the $j^{\rm th}$ unit vector, 
or in other words, we only add $s$ to $\theta_j$.
Second, if the $j^{\rm th}$ gate is a controlled rotation gate,
i.e., $CR_y(\theta_j)$, then
using the four-term parameter-shift rule
\cite{Anselmetti_2021}, we partially compute
\begin{align}\label{eq:4psr}
\notag \dfrac{\partial p_{0}(\bm\theta)}{\partial \theta_j}
& = d_+ \Big[p_0(\bm\theta + \alpha\bm e_j) - 
p_0(\bm\theta - \alpha\bm e_j)\Big] - \\
& \hspace{1.2cm}
d_- \Big[p_0(\bm\theta + \beta\bm e_j) - 
p_0(\bm\theta - \beta\bm e_j)\Big],
\end{align}
where $d_\pm = (\sqrt{2}\pm1)/4\sqrt{2};\
\alpha = \pi/2;\ \beta = 3\pi/2$.
Then, we get
$\frac{\partial \mathcal{C}(\bm\theta)}{
\partial\theta_j} = -\frac{1}{2\mathcal{C}(\bm\theta)}
 \frac{\partial p_{0}(\bm\theta)}{\partial \theta_j}$.

To compute new parameters,
we use several optimizers 
in all experiments: Standard gradient descent (SGD),
 Adam gradient descent \cite{kingma2014adam}, and
 Quantum natural gradient (QNG) \cite{stokes2020quantum}.
 %
The formula for SGD reads
\begin{align}\label{eq:sgd}
	\bm{\theta}^{t+1}
	=\bm{\theta}^{t}-\alpha\nabla_{\bm\theta}\mathcal{C}(\bm\theta),
\end{align}
where 
$\nabla_{\bm\theta}
\mathcal{C}(\bm\theta)
= \big(
\partial_{\theta_1}\mathcal{C}(\bm\theta),
\partial_{\theta_2}\mathcal{C}(\bm\theta),
\cdots, 
\partial_{\theta_M}\mathcal{C}(\bm\theta)
\big)^{\rm T}$ for $M$ training parameters,
and $\alpha$ is the learning rate
that we fix at 0.2.
In comparison, Adam is a non-local 
averaging optimizer that allows
adapting the learning rate 
but requires more steps than the SGD
\begin{align}\label{eq:adam}
&\bm{\theta}^{t+1}=\bm{\theta}^{t}
-\alpha\frac{\hat{m}_{t}}{\sqrt{\hat{v}_{t}} + \epsilon},
\end{align}
where $m_{t}=\beta_{1} m_{t-1}
+\left(1-\beta_{1}\right) 
\nabla_{\bm\theta}\mathcal{C}(\bm\theta), 
v_{t}=\beta_{2} v_{t-1}+(1-\beta_{2}) 
\nabla_{\bm\theta}^2\mathcal{C}(\bm\theta),
\hat{m}_{t}=m_{t} /\left(1-\beta_{1}^{t}\right),
\hat{v}_{t}=v_{t} /\left(1-\beta_{2}^{t}\right),
$
with the hyper-parameters 
are chosen as 
$\alpha = 0.2, \beta_1 = 0.8, 
\beta_2 = 0.999$ 
and $\epsilon = 10^{-8}$. 
Finally, the QNG is defined by
\begin{align}\label{eq:QuanNat}
    \bm{\theta}^{t+1}=\bm{\theta}^{t}-\alpha 
    g^+\nabla_{\bm\theta}\mathcal{C}(\bm\theta),
\end{align}
where 
$g^+$ is the pseudo-inverse of a 
Fubini-Study metric tensor $g$ \cite{harrow2021low}.
Assume that we can group $\bm\theta$ into
$\mathcal{L}$ layers, i.e., 
$\bm\theta = \bm\theta^{(1)}\oplus \bm\theta^{(2)}\oplus
\cdots\oplus \bm\theta^{(\mathcal{L})}$,
so that in each layer 
$\bm\theta^{(\ell)} = \{\theta^{(\ell)}_1, 
\theta^{(\ell)}_2,\cdots,\theta^{(\ell)}_{M^{(\ell)}}
\big|\ \sum_\ell M^{(\ell)} = M\}$, 
any two of unitaries satisfy 
$[G_i^{(\ell)},
G_j^{(\ell)}]=\delta_{ij}$.
Then, the metric tensor $g$ gives 
\cite{Stokes2020quantumnatural}
\begin{align}\label{eq:g}
g = 
\left(\begin{array}{cccc}
\left[\begin{array}{c}
g^{(1)}
\end{array}\right] &  &  &  \bm 0\\
 & \left[\begin{array}{c}
g^{(2)}
\end{array}\right]
 & \\
 &  & \ddots\\
\bm 0 &  &  &  
\left[\begin{array}{c}
g^{(\mathcal{L})}
\end{array}\right]
\end{array}\right)
\end{align}
where an element 
$g_{ij}^{(\ell)}$ of $g^{(\ell)}$ reads
\begin{align}\label{eq:gij}
g_{i j}^{(\ell)} = {\rm Re}\big[
\langle\partial_i\psi_{\ell}
| \partial_j \psi_{\ell}\rangle-
\langle\partial_i\psi_{\ell}| \psi_{\ell}
\rangle\langle\psi_{\ell}|\partial_j \psi_{\ell}\rangle\big],
\end{align}
where $|\psi_{\ell}\rangle$ is the quantum state
at the $\ell^{\rm th}$ layer. For unitary $G_i^{(\ell)} 
= e^{-i\theta_i^{(\ell)} K_i^{(\ell)}}$,
e.g., a rotation gate,
such that $[G_i^{(\ell)}, K_i^{(\ell)}] = 0$,
then $g_{i j}^{(\ell)}$ is recast as
 \cite{Stokes2020quantumnatural}
\begin{align}\label{eq:gij_re}
\notag g_{i j}^{(\ell)} &= {\rm Re}\big[\langle\psi_{\ell-1}
|K_{i} K_{j}| \psi_{\ell-1}\rangle \\ 
&\hspace{1cm}-
\langle\psi_{\ell-1}|K_{i}| \psi_{\ell-1}
\rangle\langle\psi_{\ell-1}|K_{j}| \psi_{\ell-1}\rangle\big].
\end{align}
See a detailed example of computing a tensor metric $g$
in Appendix~\ref{appA}.

Each optimizer has its own pros and cons:
(i) the SQD is simple but low coverage, 
one must choose a proper learning rate 
to achieve the best result, 
(ii) the Adam allows to 
automatically adapt the learning rate and fast coverage 
but it is noisy near the optimal point,
and (iii) the QNG is better than other optimizers 
but also requires more computational cost 
regards to quantum circuits.

In terms of complexity, to execute the
parameter-shift rule in Eq.~\eqref{eq:psr},
the quantum circuit executes $2M + 1$ times,
one $M$ times to compute $p_0(\bm\theta + s\bm e_i)$,
one $M$ times to compute $p_0(\bm\theta - s\bm e_i)$,
and one time to compute $p_0(\bm\theta)$.
Furthermore, a single evaluation requires executing the circuit 
for a constant number of shots to reach a certain precision, 
and each execution involves around $G$ gate operations. 
So, the complexity of each iteration is 
$\mathcal{O}[(2M+1)G]$. 
Similarly, the complexity for an 
iteration with four-term parameter-shift rule
is $\mathcal{O}[(4M+1)G]$. 
Ideally, after each step, the cost function
will decrease with a linear or logarithmic speed 
regarding the number of iterations. 
However, the variational circuit always offers 
a lower bound of the loss value 
during the training process. In particular, 
this bound increases by 
the number of qubits $N$, 
which means the problem will be harder 
according to the size of the system
\begin{align}\label{eq:Ccomp}
\mathcal{C}(\bm{\theta}) \geq \textit{poly}(N).
\end{align}

This work implements the numerical experiments 
using various configurations described above 
to train the variational models and compare them together.
The numerical results are executed by 
Qiskit open-source package, version 0.24.0, 
which is available to run on all platforms. 
For each experiment, to get the probability
$p_0$ we execute $10^4$ shots using the
{\it qasm} simulator backend.
The number of iterations for every training 
process is fixed at 400, 
except for others shown in the text.
The experiments are then scaled up to 
10 qubits for quantum state preparation 
and 6 qubits for quantum state tomography
to demonstrate the scalability.
The results are shown in Sec.~\ref{seciii}
for both quantum state preparation 
and quantum state tomography. 


\section{Results}\label{seciii}
\subsection{Quantum state preparation}\label{seciii_i}
We apply the UC-VQA approach to prepare
various quantum states. 
As mentioned aforehand, we aim to prepare 
a family of entangled GHZ and W states 
\cite{PhysRevA.62.062314}
using several
variational anstazs as shown in Fig.~\ref{fig:1} (b):
the linear ansatz and graph-based ansatzes with 
polygon and star structures. 

The linear ansatz contains a 
linear number of parameters 
w.r.t. the number of qubits 
that is sufficient for entanglement state preparation.
A similar structure has been proposed 
for the variational state metrology \cite{Koczor_2020}.
Each layer contains a set of single-qubit rotation gates $R_x$
applied on each qubit,
followed by an entangled gate CRY,
a set of $R_z$ gates, 
an entangled gate $\overline{\rm CRY}$,
and finally end up by a set of $R_z$ gates. 
The single-qubit rotation is
$R_j^{(n)} = \exp(-i\frac{\theta}{2}\sigma_j^{(n)}), 
\ j\in\{x, y, z\},$ and $\sigma_j^{(n)}$ is a Pauli matrix
applied on the $n^{\rm th}$ qubit.
The structures of CRY and $\overline{\rm CRY}$ 
are detailed in Fig.~\ref{fig:1} (d).
They are entanglement gates that consist of
a set of controlled $y$-rotation gates,
which are parameter-dependent. 
In total, the number of parameters is $M= 5NL$.

\begin{figure*}[t]
\centering
\includegraphics[width=\textwidth]{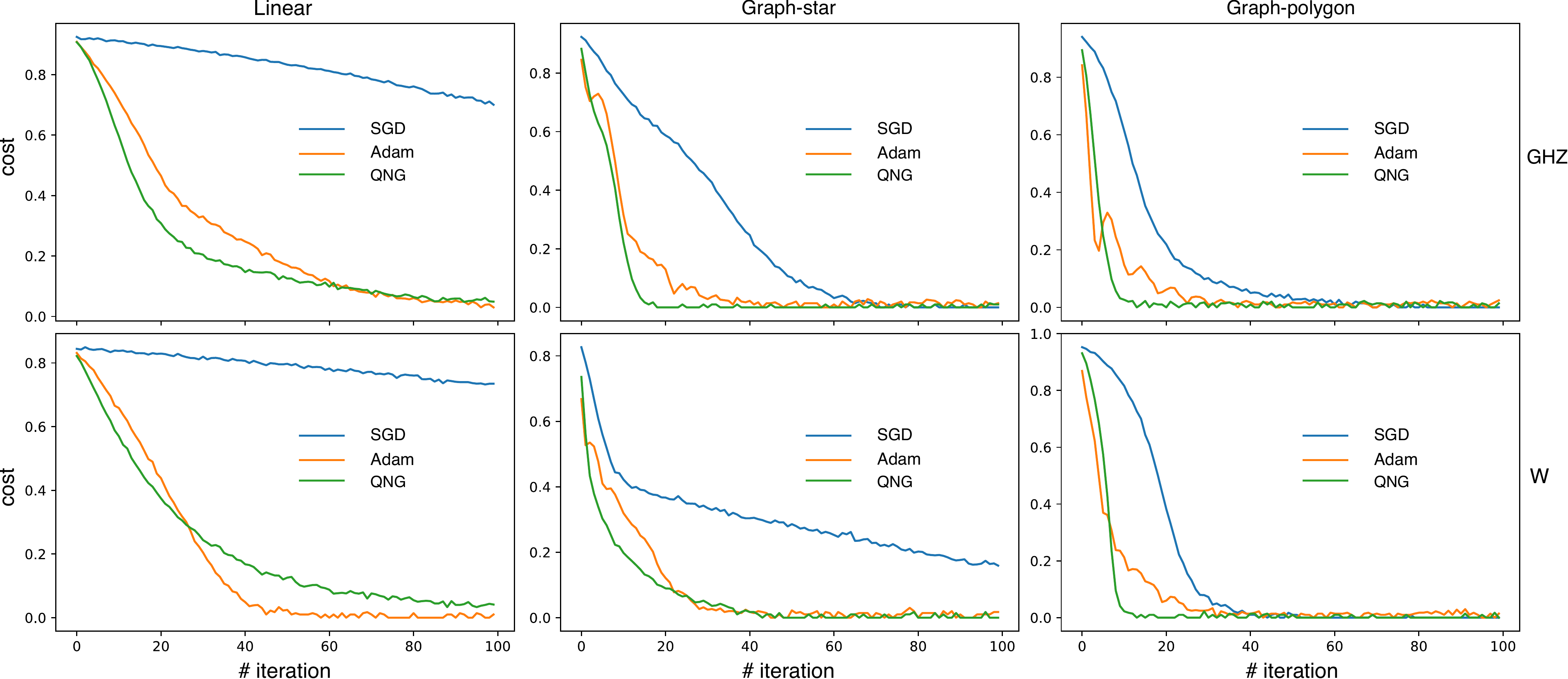}
\caption{
Comparison of different optimizers: standard gradient descent (SGD),
Adam gradient descent, and quantum natural gradient descent (QNG). 
From left to right: linear ansatz, graph-star ansatz, and graph-polygon ansatz.
The upper row is the plot for the target state's case is GHZ, 
and the lower row is the plot for that of the W state.
The results are plotted at the number of qubits $N = 3$,
and the number of layers $L = 2$.
}
\label{fig:2}
\end{figure*}

\begin{figure*}[t]
\centering
\includegraphics[width=\textwidth]{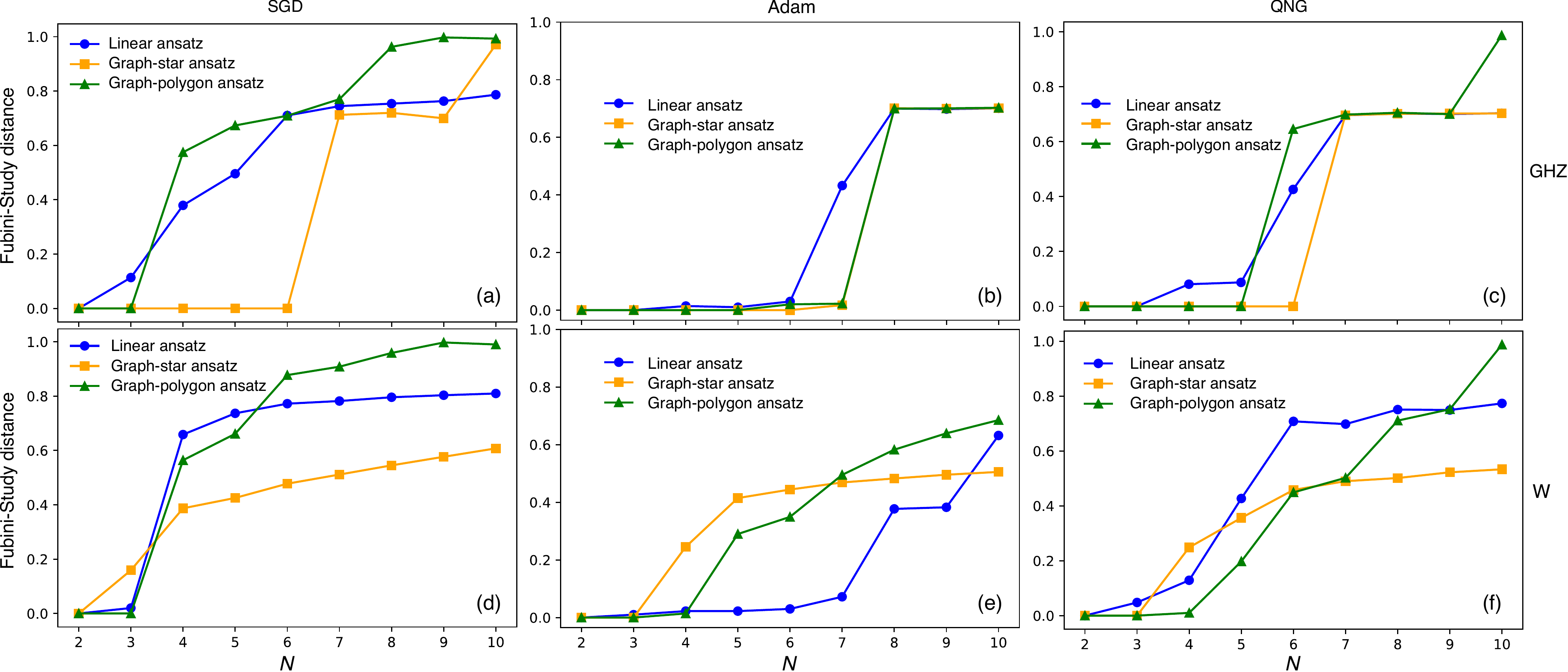}
\caption{
Funibi-Study distance between the target states (GHZ and W)
and the variational prepared ansatzes (linear, graph-star, graph-polygon).
From left to right, we plot for different optimizers: SGD, Adam, and QNG,
respectively. The upper row (a,b,c) is the plot of the GHZ state, 
and the lower row (d,e,f) is the plot of the W state. 
Here, we fixed $L = 2$.
}
\label{fig:3}
\end{figure*}

Besides, we also propose a graph-based ansatz, 
wherein we parameterize a graph state using
single-qubit rotation gates.
A graph state is a multi-qubit state 
that consists of vertexes and edges that connect
in a graph structure.
Each vertex represents a qubit,
and an edge links an interacting pair 
of qubits using a CZ gate.
See Fig.~\ref{fig:1} (d) for examples of graph states. 
To prepare a graph-based ansatz,
we apply $R_y$ gates following the rule: 
each CZ gate is always 
surrounded by four $R_y$ gates.  
See Fig.~\ref{fig:1} (b) for details.
This structure is similar to the one in
Ref.~\cite{Cerezo2021}.
Here, we focus on two specific structures 
of the graph-based ansatz: 
polygon and star ansatzes, 
where the number of parameters are
$M = 2NL$ and $2NL-2L$, respectively. 

\begin{figure*}
\begin{minipage}[c]{0.48 \linewidth}
\includegraphics[width=8.6cm]{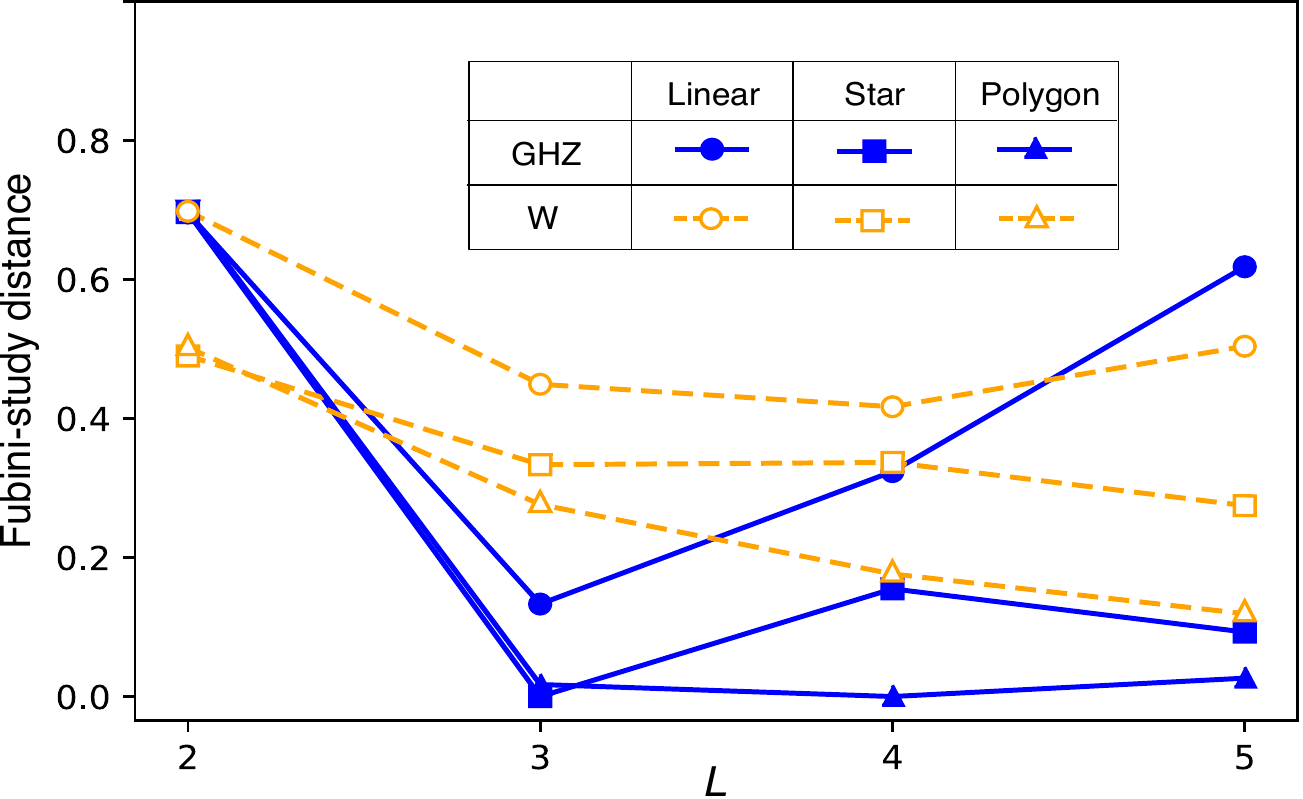}
\caption{Fubini-Study distance versus 
the number of layers $L$.
For the GHZ class, the distance reaches the minimum
at $L = 3$, while for the W class, 
the distance continues to reduce
up to $L = 5$ for the graph-based ansatzes.
Here, we fixed $N = 7$ and the QNG optimizer.}
\label{fig:4}
\end{minipage}
\hfill
\begin{minipage}[c]{0.48 \linewidth}
\includegraphics[width=8.5cm]{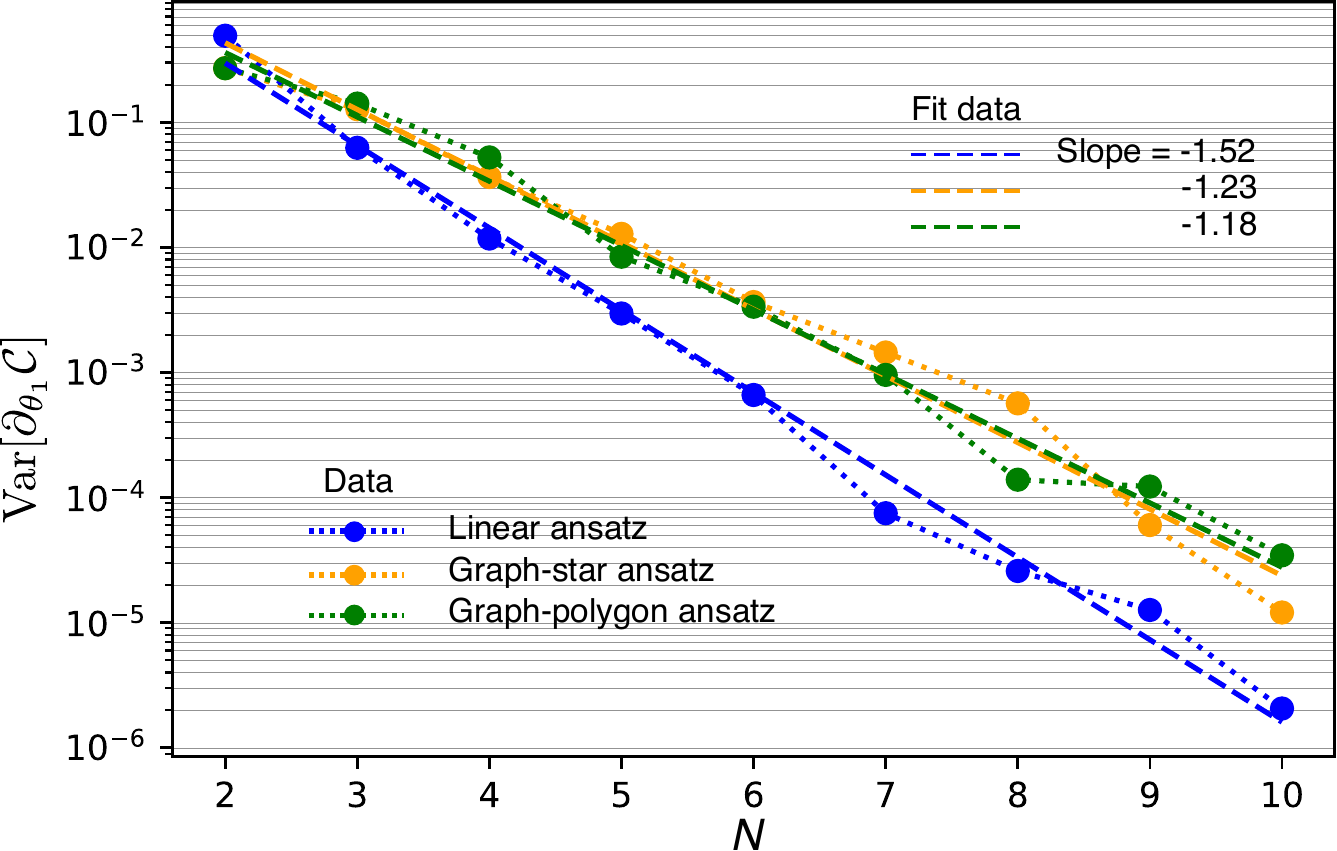}
\caption{A semi-log plot 
of the variance
${\rm Var}[(\partial_{\theta_1}\mathcal{C})]$ 
as a function of the number of qubits. 
As predicted, exponential decay is observed 
as proof of the existence of the barren plateaus.
The slope of a fit line indicates 
the decay rate.
Here we fixed $L = 2$ and GHZ target state.}
\label{fig:5}
\end{minipage}%
\end{figure*}

In Fig.~\ref{fig:2}, we examine the cost function
versus the number of iterations for different optimizers. 
It can be seen that for most of the ansatzes 
exhibited here (except the linear ansatz for the W state,)
the QNG optimizer (green line) requires 
the smallest number of iterations
among the three optimizers (SGD, Adam, QNG)
to achieve the convergence. 
This is the natural feature of the QNG optimizer
because its gradient points toward to the optimal direction
\cite{Stokes2020quantumnatural,haug2021natural}.  
Nevertheless, we emphasize that the Adam 
is not stable at the optimal point 
(as mentioned in the pros and cons section,)
while the QNG is more stable. 
Both the Adam and QNG can achieve the minimum
at a certain number of iterations; 
however, the SGD (blue line) 
is hard to get the optimal point in some cases
(see Fig.~\ref{fig:2}.)

Figure \ref{fig:3} displays the results for 
the Funibi-Study distance
between the target state (GHZ and W)
and the variational prepared state.
We discuss the accuracy and the stability
of different ansatzes under various optimizers.
In general, the Funibi-Study distance increases 
as the number of qubits increases, 
which reveals that the accuracy is low 
for larger system preparation.

Particularly, 
we first observe a strong dependence 
of the resulting linear ansatz 
(Fig.~\ref{fig:3}, \textcolor{blue}{$\bullet$})
on different optimizers for both GHZ and W classes. 
For the SGD and QNG optimizers, 
the Funibi-Study distance 
increases as $N$ increases,
while for the Adam optimizer, the distance 
remains small and stable up to $N = 6$
(for both GHZ and W.) 
This observation 
exhibits the instability of the linear ansatz
under different optimizers, 
and its best optimizer is Adam 
(in agreement with the 
previous results in Fig.~\ref{fig:2}.)

For the graph-star ansatz,
the Funibi-Study distance  
remains small up to $N = 6$ at all optimizers 
for the target is the GHZ state 
(Fig.~\ref{fig:3} (a,b,c), 
\textcolor{orange}{\tiny$\blacksquare$}),
which implies a high accuracy and well stability.
Moreover, the resulting graph-polygon ansatz
also achieves high accuracy under 
the Adam and QNG optimizers 
up to $N = 6$ and 5, respectively
(Fig.~\ref{fig:3} (b,c), \textcolor{PineGreen}
{\footnotesize$\blacktriangle$}), 
while its accuracy gradually 
decreases (larger distance) when increasing $N$ 
under the SGD optimizer 
(Fig.~\ref{fig:3} (a), \textcolor{PineGreen}
{\footnotesize$\blacktriangle$}).

So far, for the target state is the W state
[Fig.~\ref{fig:3} (d,e,f)],
all the ansatzes strongly depend on the optimizers 
as we can see the Fubini-Study distance gradually 
increases when increasing $N$.
This observation can be explained 
when noting that
W states belong to 
the multipartite entanglement class
\footnote{For W states, if one qubit is lost, 
the remaining system still entangles,
from which contrasts with GHZ states, 
that fully separable after disentangle one qubit.
See also Ref.~\cite{PhysRevA.62.062314}.},
and thus it is more challenging to prepare the W states
in comparison to the GHZ states of entanglement. 

It is reported that the number of layers $L$ 
can significantly affect
the accuracy \cite{Steinbrecher2019,Giacomo2020,
PhysRevA.104.042601}.
In Fig.~\ref{fig:4}, we show the dependence of 
the Fubini-Study distance on the number of layers.
Here, we fix $N = 7$ and apply the QNG optimizer.
For the GHZ class target, the Fubini-Study distance
reaches the minimum at $L = 3$ and then gradually
increases when $L > 3$. 
Among the three ansatzes,
the linear one 
(\textcolor{blue}{$\bullet$})
increases fastest 
while the graph-star 
(\textcolor{blue}{\tiny$\blacksquare$})
slowly increases,
and the graph-polygon 
(\textcolor{blue}{\footnotesize$\blacktriangle$})
nearly saturates (slightly increases). 
This observation is a hind for the barren plateaus,
i.e., the accuracy of the training process reduces 
when increasing the parameters space, 
and it agrees with the barren plateaus 
analyses in Fig.~\ref{fig:5} below. 
Furthermore, in the W class, 
the distance for the linear ansatz 
(\textcolor{orange}{$\circ$})
reaches the minimum
at $L = 4$, while the distances for graph-star 
(\textcolor{orange}{\tiny$\square$})
and graph-polygon
(\textcolor{orange}{\scriptsize$\triangle$})
ansatzes continue to reduce up to $L = 5$. 
This result agrees with the previous observation 
about the entangled features of the GHZ and W classes,
as the W class requires more computational cost
to achieve the same accuracy as the GHZ class. 

%

Those results in Figs.~(\ref{fig:3}, \ref{fig:4}) 
suggest a sign of the existence of 
the barren plateau effect (BP) 
in the training landscapes \cite{McClean2018},
which frequently appears in the 
Variational Quantum Algorithms
and Quantum Neural Networks.
When the BP appears, 
the cost function $\mathcal{C}(\bm\theta)$ derivative
exponentially vanishes with 
the circuit size \cite{McClean2018}.
Primary sources for the BP effect include 
the random parameter initialization 
\cite{McClean2018},
shallow depth with global cost functions \cite{Cerezo2021},
highly expressive ansatzes \cite{PRXQuantum.3.010313},
the entanglement-induced \cite{PRXQuantum.2.040316},
and the noise-induced \cite{Wang2021},
among others. 
So far, it has been reported 
several promising strategies for 
avoiding the BP using 
local cost functions \cite{Cerezo2021,Uvarov_2021},
correlated parameters \cite{Volkoff_2021}, 
pre-training by the classical neural network \cite{verdon2019learning}, 
and layer-by-layer training \cite{grant2019initialization}. 

To examine the vanish of 
the derivative cost function 
$\mathcal{C}(\bm\theta)$, i.e.,
the existence of the BP, 
we derive the variance
\begin{align}\label{eq:varC}
{\rm Var}[\partial_k\mathcal{C}] = 
\langle(\partial_k\mathcal{C})^2\rangle
-\langle\partial_k\mathcal{C}\rangle^2,
\end{align}
where we have shorthanded $\mathcal{C}(\bm\theta)$
to $\mathcal{C}$, $\partial_k\mathcal{C} := 
\partial\mathcal{C}/\partial\theta_k$,
and the expectation value takes over the final state.
Numerical results ${\rm Var}[\partial_{\theta_1}\mathcal{C}]$
for a representative first parameter $\theta_1$
are given in Fig.~\ref{fig:5},
where we observe exponential decays 
as functions of the number of qubits $N$
with slopes of -1.52, -1.23, and -1.18, 
for the linear ansatz
(\textcolor{blue}{$\bullet$}), 
graph-star ansatz
(\textcolor{orange}{$\bullet$}),
and graph-polygon ansatz
(\textcolor{PineGreen}{$\bullet$})
, respectively. 
Those are the evidence for the existence of the BP in the
parameters spaces in our model and agrees
with our prediction in Figs.~(\ref{fig:3}, \ref{fig:4}).
The results are shown 
for the GHZ class and $L = 2$. 

Finally, we account for the effect of noise.
In real hardware that belongs to the current era of 
noisy intermediate-scale quantum computers (NISQ), 
noisy qubits bias the results and thus limit
the applications \cite{Nachman2020}. 
One of the important classes of the noisy qubits 
is the readout error, which typically arises from
(i) the qubits decoherence, i.e., qubits decay,
phase change,..., 
during the measurement time,
and (ii) the incomplete of the measuring devices,
i.e., overlap between the measured bases.
Here, we model the noise channel 
by the readout error probability 
for each qubit in the circuit as
\begin{align}\label{eq:prob}
	\begin{pmatrix}
		p^{(\epsilon)}_0\\ p^{(\epsilon)}_1
	\end{pmatrix}
	=
	\begin{pmatrix}
	1-\epsilon & \epsilon\\
	\epsilon & 1-\epsilon
	\end{pmatrix}	
	\begin{pmatrix}
		p_0\\ p_1
	\end{pmatrix},
\end{align}
where $p_0, p_1$ are the true 
probabilities when measuring
the bases $|0\rangle$ and $|1\rangle$
of the qubit, $p^{(\epsilon)}_0, 
p^{(\epsilon)}_1$ are the 
readout error probabilities, respectively,
$\epsilon$ is the error rate.

\begin{figure}  
\includegraphics[width=8.2cm]{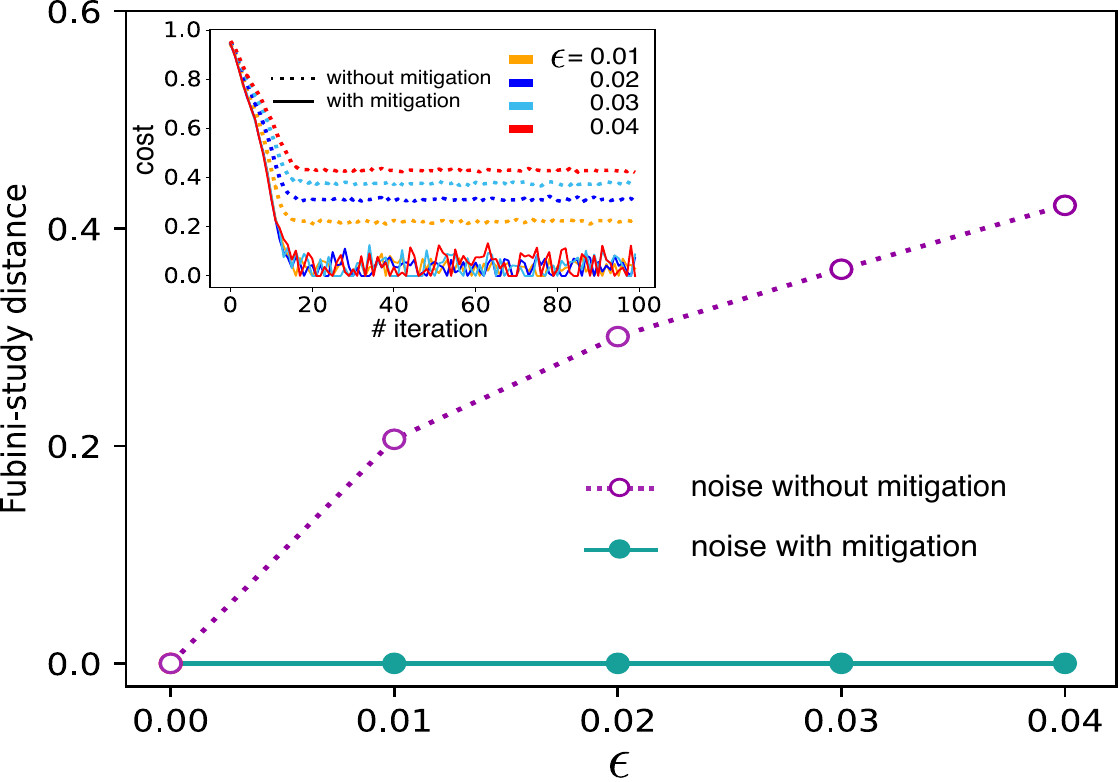}
\caption{Fubini-Study distance 
under the presence of noise 
($\epsilon$)
with and without mitigation. 
Inset: plot of cost versus iteration
for two cases of with and without mitigation 
with various error rates $\epsilon$.
We fixed $N = 5, L = 2$, QNG optimizer, 
and plot for graph-star ansatz with 
the GHZ state's target.
}
\label{fig:6}
\end{figure}

As shown in Fig.~\ref{fig:6}
(open circle, \textcolor{Plum}{$\circ$}),
the Fubini-Study distance rapidly increases 
when gradually raising the error rate. 
In the inset Fig.~\ref{fig:6}, the dotted lines
are the cost functions regarding different $\epsilon$
(the colors match with the inset figure's legend.)
The cost function rapidly reduces and reaches
convergence around 20 iterations. 
However, the optimal value limits 
in a range $[0.2, 0.4]$ for $\epsilon \in [0.01, 0.04]$,
which results in the increase of 
the Fubini-Study distance
in the main figure,
and thus loses the accuracy under noise. 

\begin{figure}  
\includegraphics[width=\linewidth]{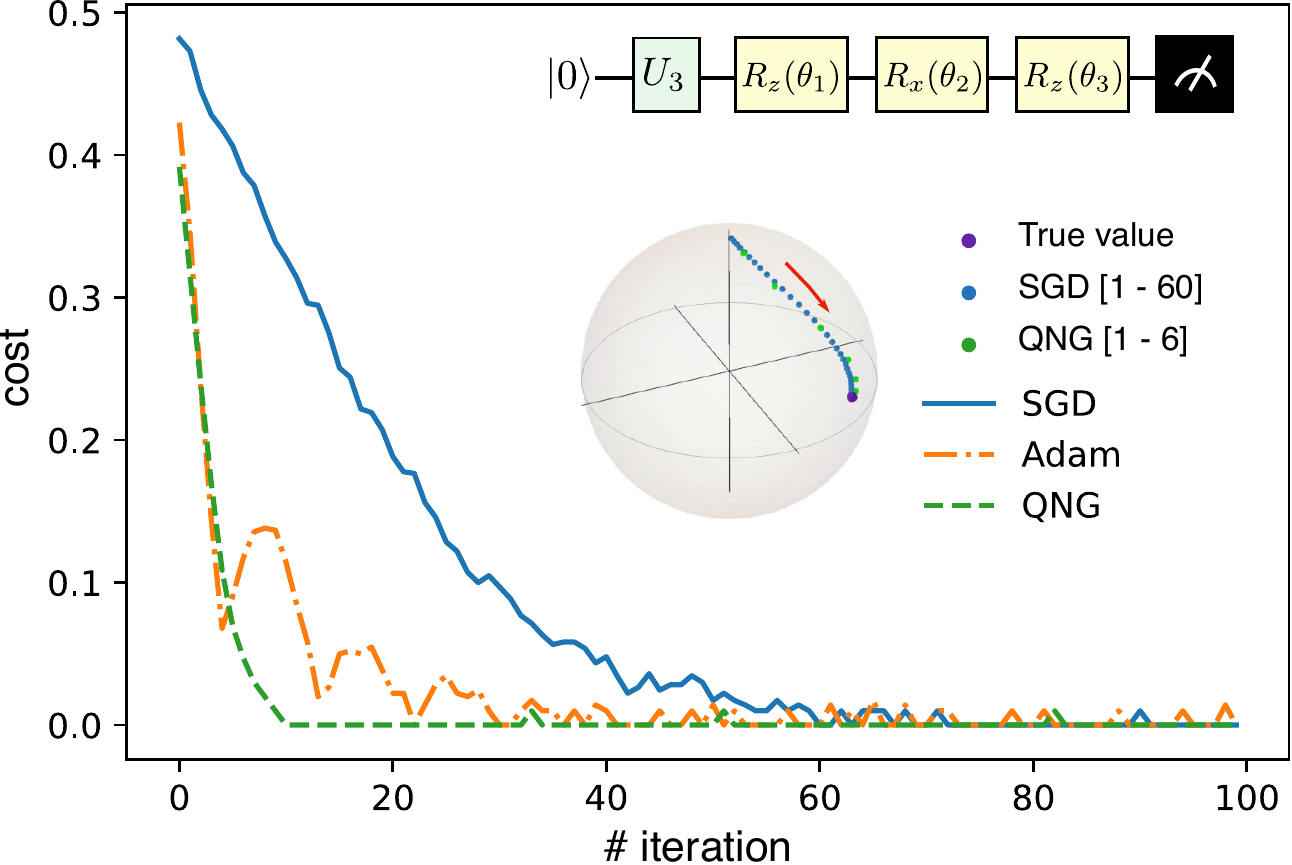}
\caption{Main: plot of cost function versus 
the number of iterations for difference optimizers
as shown in the figure. 
Inset 1: quantum circuit for a 
single-qubit tomography 
where an unknown qubit state is 
generated by a random unitary 
$\bm U_3\left(\theta,\phi,\lambda\right)$, 
and $V^\dagger(\bm{\theta})$ is made up 
of $R_z(\theta_{1})$, $R_x(\theta_{2})$, 
and $R _z(\theta_{3})$ gates. 
Inset 2: Bloch sphere represents the qubit states:
\textcolor{Purple}{$\bullet$} the true state,
\textcolor{blue}{$\bullet$} the trajectory of 
the reconstructed state
under the SGD optimizer for the  
iterations run from 1 to 60,
\textcolor{JungleGreen}{$\bullet$}
the trajectory under the QNG optimizer for the 
iteration runs from 1 to 6.
}
\label{fig:7}
\end{figure}

To suppress the noise, we apply 
the measurement error mitigation
\cite{doi:10.7566/JPSJ.90.032001,
Czarnik2021errormitigation,
PRXQuantum.2.040326,
Maciejewski2020mitigationofreadout}.
The basic technique for the measurement-mitigation is
using a $(2^N\times2^N)$ 
calibration matrix
$ \mathcal{M}$, such that
\begin{align}\label{eq:M}
	\bm p^{(\epsilon)} =  \mathcal{M}\bm p,
\end{align}
where $\bm p = (p_0, p_1, 
\cdots, p_{2^N-1})^{\rm T}$
a vector represents the true probabilities, 
and similar for $\bm p^{(\epsilon)}$ 
a vector represents 
the readout probabilities.
The calibration matrix contains 
all the probabilities transitions 
$\mathcal{M}_{ij} = {\rm Prob}
(p^{(\epsilon)}_i\to p_j)$.
One direct approach for obtaining the
mitigated probabilities is to invert the 
calibration matrix and get
$\bm p^{\rm mitigated} = \mathcal{M}^{-1}
\bm p^{(\epsilon)}$.
Other developed methods, such as 
the least square \cite{Geller_2020}, 
truncated Neumann series 
\cite{wang2021measurement},
and unfolding methods,
have been demonstrated \cite{Nachman2020}.
Finally, we construct $\mathcal{M}$ 
by running $2^N$ circuits corresponding to  
$2^N$ elements in the computation basis 
$\{|00\cdots0\rangle, |00\cdots1\rangle, 
\cdots, |11\cdots1\rangle\}$.

The mitigation results
are shown in Fig.~\ref{fig:6}
(filled circle, \textcolor{JungleGreen}{$\bullet$}).
In this example, we can totally eliminate 
the effect of noise
after the mitigation and thus the accuracy 
reaches the original value (without noise).

\begin{figure*}
\includegraphics[width=\textwidth]{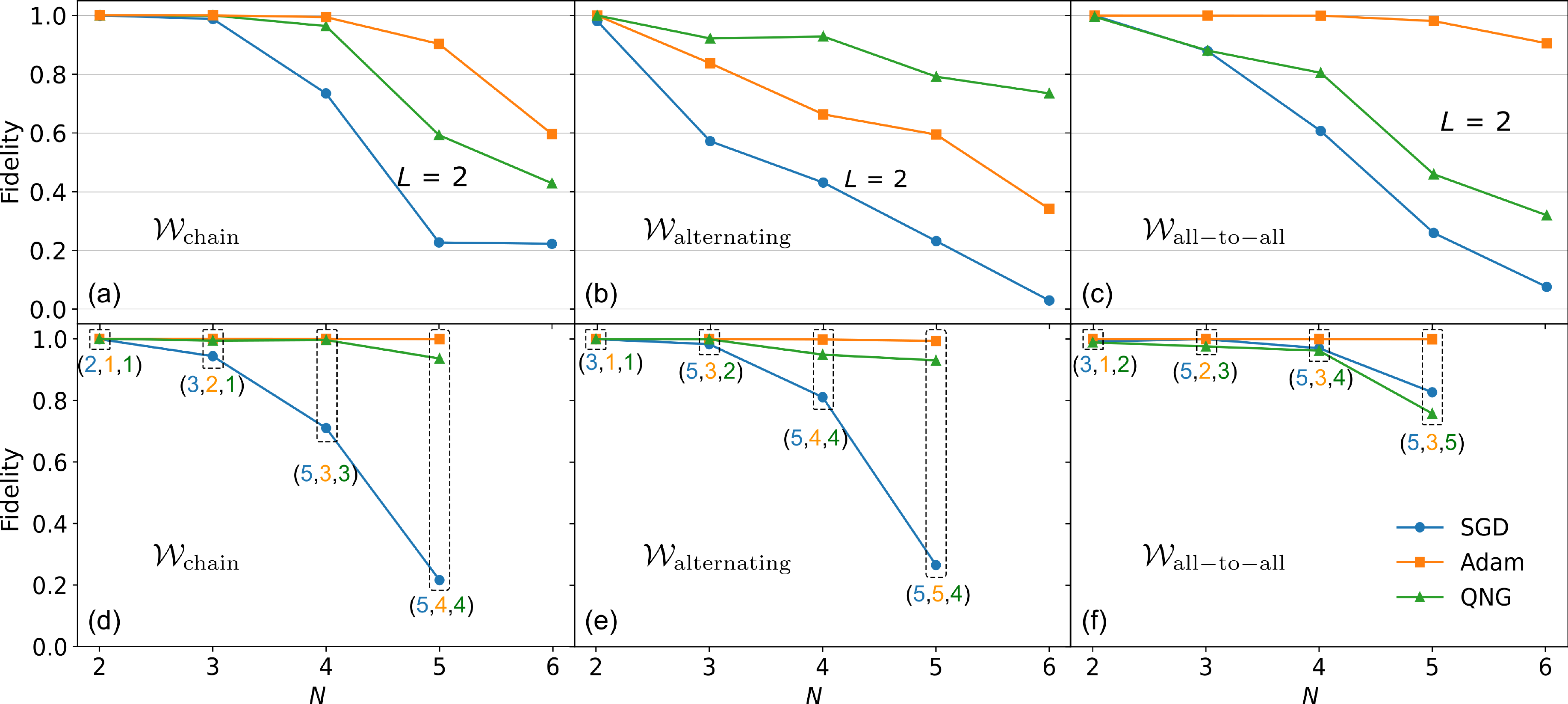}
\caption{(a,b,c) Plot of the fidelity between an unknown 
Haar random state $|\mu\rangle$ and its reconstructed
state $|\gamma(\bm\theta)\rangle$ for different 
$\mathcal{W}$ structures: $\mathcal{W}_{\rm chain}$ (a),
$\mathcal{W}_{\rm alternating}$ (b), and 
$\mathcal{W}_{\rm all-to-all}$ (c). 
For each case, we show 
the results for different optimizers: 
SGD, Adam, and QNG.
Here we fixed $L = 2$. 
(d,e,f) Plot of the fidelity similar as above for different $L$
as shown in the colored parentheses 
$(\textcolor{RoyalBlue}{\star},
\textcolor{orange}\dagger,
\textcolor{Green}\ddagger)$. 
Here, $\textcolor{RoyalBlue}{\star}$ is
the optimal number of layers for the SGD,
$\textcolor{orange}\dagger$ is
the optimal number of layers for the Adam,
and $\textcolor{Green}\ddagger$ is
the optimal number of layers for the QNG.
We choose the appropriate  
$(\textcolor{RoyalBlue}{\star},
\textcolor{orange}\dagger,
\textcolor{Green}\ddagger)$
for each $N$ and each $\mathcal{W}$
structure so that the fidelity gets 
its (possible) highest accuracy. 
}
\label{fig:8}
\end{figure*}

\subsection{Quantum state tomography}\label{seciii_ii}
This section applies the UC-VQA framework
to quantum state tomography. 
We first consider reconstructing 
an abstract 1-qubit state
encodes in a quantum circuit 
as shown in the inset Fig.~\ref{fig:7}. 
We randomly generate 
an unknown quantum state 
$|\mu\rangle = \bm U_3|0\rangle$,
where
\begin{align}\label{eq:u3}
\bm U_{3}(\theta, \phi, \lambda)=
   \begin{pmatrix}
	\cos \frac{\theta}{2} & 
	-e^{i \lambda} \sin \frac{\theta}{2} \\ 
	e^{i \phi} \sin \frac{\theta}{2} & 
	e^{i(\phi+\theta)} \cos \frac{\theta}{2}
   \end{pmatrix},
\end{align}
where we have set random with Haar measure 
$\sin(\theta)/2$, $\phi$,
and $\lambda$.
To reconstruct $|\mu\rangle$, we set 
the unitary $\bm V^\dagger(\bm{\theta})
= R_z(\theta_3)R_x(\theta_2)R_z(\theta_1)$. 
We train the scheme with 100 iterations 
using various optimizers 
and show the cost function versus 
iteration in the main Fig.~\ref{fig:7}.
Here, the QNG optimizer gives the best optimization. 
In the inset figure, we also show 
the trajectory in the Bloch sphere
of the reconstructed state 
$|\gamma(\bm\theta)\rangle$ under the 
update of $\bm\theta$ for 
two cases of SGD and QNG optimizers.
For the former one, it needs around 60 iterations
for the reconstructed state to reach the true state,
while it only requires around 6 iterations for the latter. 

Now, we focus on a general random Haar state, 
i.e., $|\mu\rangle = \bm U_{\rm Haar}|\bm0\rangle$,
as shown in Fig.~\ref{fig:1} (c).
To reconstruct the state, we use several ansatzes for
the entangled gate $\mathcal{W}$ in $\bm V^\dagger(\bm\theta)$, 
including the $\mathcal{W}_{\rm chain}, 
\mathcal{W}_{\rm alternating}$,
and $\mathcal{W}_{\rm all-to-all}$ structures. 
Refer Fig.~\ref{fig:1} (c) for details of these structures, 
where we have used the parameter-dependent 
controlled $y$-rotation gates to construct them.
The trainable parameters for these ansatzes
are $M = 4NL, \lfloor NL/2 \rfloor + 3NL$,
and $N(N+5)L/2$, respectively. 

The results are shown in Fig.~\ref{fig:8}.
Let us consider the fidelity between the true Haar state
and the reconstructed state as
\begin{align}\label{eq:F}
F(\mu, \gamma(\bm\theta)) = 
\big|\langle\gamma(\bm\theta)|\mu\rangle\big|^2,
\end{align}
which is the overlap between these two states.
In Fig.~\ref{fig:8} (a,b,c), we show the fidelities 
for different structures of $\mathcal{W}$.
For each case, we fix $L = 2$ and examine the 
three optimizers 
SGD (\textcolor{RoyalBlue}{$\bullet$}),
Adam (\textcolor{orange}{\tiny$\blacksquare$}), 
and QNG (\textcolor{Green}{\footnotesize$\blacktriangle$}).
We first observe that the SGD optimizer 
is not good for all $\mathcal{W}$ structures 
and needs to choose an appropriate learning rate.
The fidelities reduce with the increasing $N$
and nearly vanish at $N = 6$. 
In contrast, the Adam optimizer exhibit high
fidelities up to $N = 4$ for 
$\mathcal{W}_{\rm chain}$ (a), 
$N = 5$ for $\mathcal{W}_{\rm all-to-all}$ (c),
and gradually reduces from $N = 2$ for 
$\mathcal{W}_{\rm alternating}$ (b).
Even though it
is not stable near the optimal point,
the Adam is remarkable for achieving high accuracy 
in the QST. Furthermore, the QNG optimizer
also allows for getting such high accuracy 
up to $N = 4$ for $\mathcal{W}_{\rm chain}$ (a)
and even better than the Adam for 
$\mathcal{W}_{\rm alternating}$ (b),
while it gradually reduces for 
$\mathcal{W}_{\rm all-to-all}$ (c).
This observation can be explained 
by these own structures: the 
$\mathcal{W}_{\rm all-to-all}$ 
contains the most 
number of parameters via 
the controlled $y$-rotation gates 
compared
to the others, which results 
in the low accuracy. 
It is apparent that the 
QNG optimizer is sensitive to the 
controlled $y$-rotation gates. 

Next, to achieve high accuracy for
any qubit number $N$,
we increase the number of layers $L$
while paying attention to the barren plateau. 
This is the pros and also the cons
of the UC-VQA method.
Figure~\ref{fig:8} (d,e,f)
plot the fidelities versus $N$,
where for each $N$, 
the corresponding $L$ is shown
in the colored parenthesis 
$(\textcolor{RoyalBlue}{\star},
\textcolor{orange}\dagger,
\textcolor{Green}\ddagger)$, 
for the SGD, Adam, and QNG, respectively. 
The number of layers shown in the parenthesis
is the smallest (optimal) $L$ required
for achieving such high accuracy
before it goes down due to the barren plateau.
As can be seen from the figure,
the Adam method 
allows for reaching the 
maximum fidelity (results are shown up to $N = 5$
for all $\mathcal{W}$ structures) 
with a suitable $L$ as shown
in the middle position of the parenthesis.
Similarly, we can reach high accuracy 
with the QNG optimizer up to $N = 4$ 
when choosing an appropriate $L$ 
as shown in the last position of the parenthesis. 
For the SGD, it is intractable for achieving high accuracy.
Even though the relation between $N$ 
and the required $L$ is not clear, interestingly,
we can see from the results up to $N = 5$,
the required $L$ is also around 5
(more $L$ is redundancy or may reduce 
the accuracy due to the barren plateau,
see detailed in Appendix~\ref{appB}.)

Finally, we address the merit 
of our UC-VQA approach 
and the shadow tomography protocal
\cite{https://doi.org/10.48550/arxiv.1711.01053,
Huang2020}, 
a recent promising method in this regime. 
A shadow tomography protocol 
is given as follows \cite{Huang2020}:
(i) initially prepare a 
random unknown quantum state $\rho$,
and the task ahead is to 
predict a target function underlying 
the state from its shadow,
(ii) randomly pick up a unitary $\bm U_k$
in a $T$-tuple $\mathcal{U}$, i.e., $\mathcal{U} 
= \{\bm U_1, \bm U_2, \cdots, \bm U_T\}$
then apply it to the initial state 
to tranform $\rho\mapsto \bm U \rho \bm U^\dagger$,
(iii) measure the evolved state in the computational basis
$|b\rangle = \{|0\rangle, |1\rangle\}^N$.
The procedure is repeated for 
a certain number of measurements. 
For each measurement, 
we get a random classical snapshot 
\begin{align}\label{eq:classsnapshot}
	\sigma_{k,b} = \bm U^\dagger_k |b\rangle
	\langle b|\bm U_k.
\end{align}	
We then define an invertible channel matrix
\begin{align}\label{eq:Mmax}
	\mathcal{M}(\rho) =\mathbb{E}_k \sum_b
	{\rm Tr}(\sigma_{k,b}\ \rho) \cdot \sigma_{k,b},
\end{align}
where $\mathbb{E}_k$ is the average over $\bm U_k$, 
with a corresponding pick-up probability.
Let $\mathcal{M}^{-1}$ exists, and let $p_k$ is the probability 
of picking up a unitary $\bm U_k$, then
we can reconstruct a (non-normalized) state as
\begin{align}\label{eq:rho_re}
	\check\rho = \sum_{k}p_k
	\sum_b {\rm Tr}(\sigma_{k,b}\ \rho)\cdot
	\mathcal{M}^{-1} 
	(\sigma_{k,b}),
\end{align}
which is the classical shadow of 
the original unknown state $\rho$.
For the transformation $\bm U$ 
belongs to a family of the global Clifford gates,
i.e., $\bm U\in\mathcal{U}_{C} = 
\{ {\rm CNOT, Hadamard, S\_gate, T\_gate}\}$,
refer to {\it Random Clifford measurements}, 
the  reconstructed state explicitly yields \cite{Huang2020}
\begin{align}\label{eq:rho_re_Cliffird}
	\check\rho = (2^N+1)
	\bm U^\dagger |b\rangle\langle b|\bm U
	-\bm I.
\end{align}
For the transformation $\bm U$ 
belongs to the random Pauli gates, 
such as $\bm U \in \mathcal{U}_P 
= \{\bm X, \bm Y, \bm Z, \cdots\}$, 
refer to {\it Random Pauli measurements}, 
it straightforwardly yields \cite{Huang2020}
\begin{align}\label{eq:rho_re_Pauli}
	\check\rho = \bigotimes_{j = 1}^N 
	\Bigl(
	3\bm U_j^\dagger |b_j\rangle\langle b_j|\bm U_j
	-\bm I
	\Bigr),
\end{align}
for $b = (b_1, \cdots, b_N) \in \{0,1\}^N$.

\begin{figure}
\includegraphics[width=8.6cm]{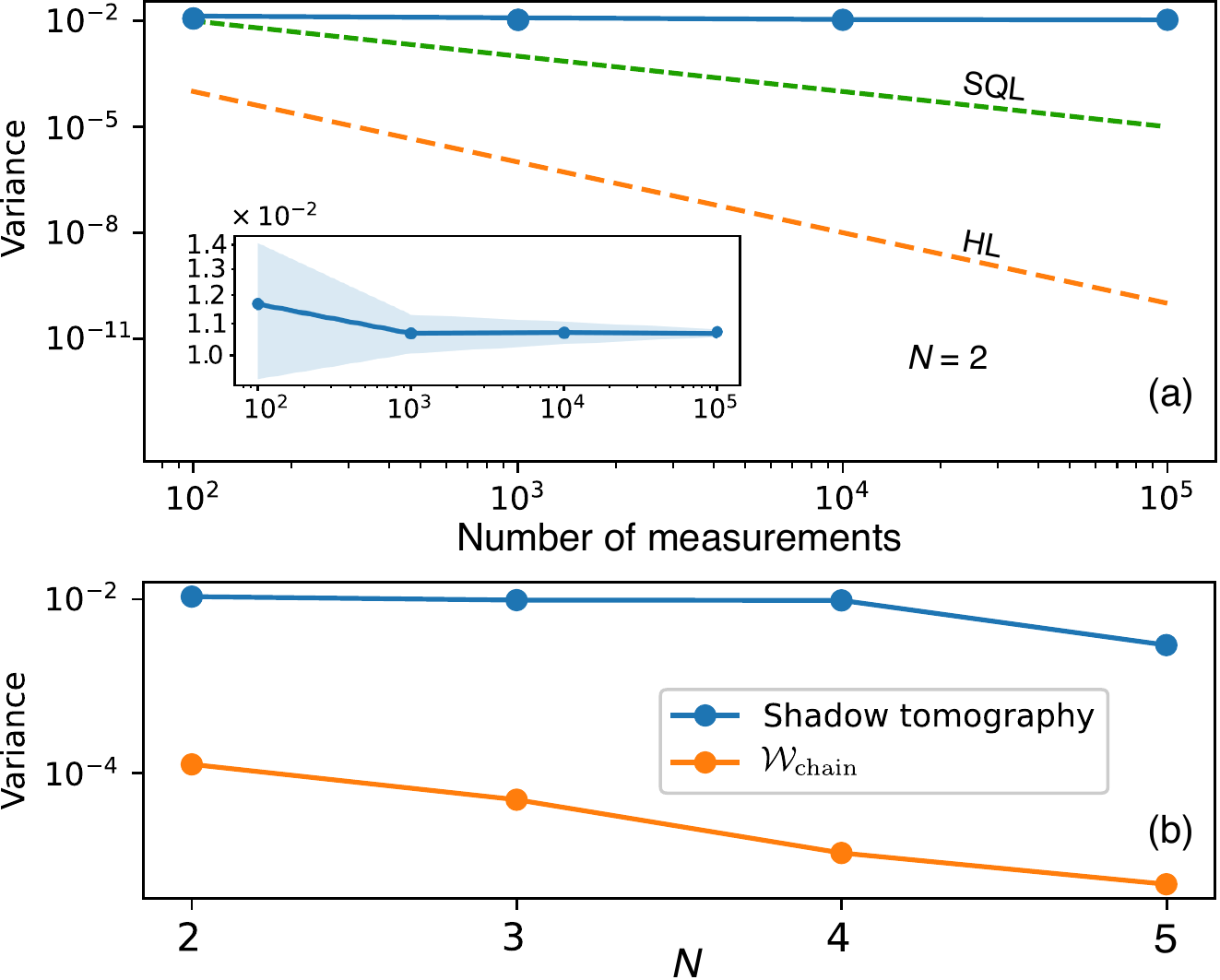}
\caption{(a) Log-log plot of the variance Var$[\check z]$ versus
the number of measurements using the shadow tomography.
The standard quantum limit (SQL) and Heisenberg limit (HL)
are shown for comparison purposes. The number of qubits
is fixed at $N = 2$. Inset: Zoom-in the variance, 
where the blue area is the standard deviation after talking
ten runs. 
(b) Comparison between the shadow tomography and 
UC-VQA. For the shadow tomography,
we fix the number of measurements at $10^5$,
and for the UC-VQA, we use the structure of 
$\mathcal{W}_{\rm chain}$ and the Adam optimizer,
the number of shots is $10^4$.
}
\label{fig:9}
\end{figure}

For comparing the shadow tomography with 
the UC-VQA scheme, we consider the 
prediction of a linear function 
as a figure of merit for the 
accuracy.
We apply the Random Pauli measurement.
As a concrete representative, we 
measure a global observable, i.e., 
$\mathcal{\bm Z} \equiv \bm Z^{\otimes N}$, 
$\bm Z$ is a Pauli matrix, 
which gives the predicted (linear) expectation value
\begin{align}\label{eq:expZ} 
	 {\check z} = {\rm Tr}
	(\mathcal{\bm Z}\check \rho),
	\text{ that obeys } 
	\mathbb{E}[\check z]
	= {\rm Tr}(\mathcal{\bm Z}\rho).
\end{align}
%
The fluctuation (distribution around 
the true expectation value) 
of the predicted expectation value
is given by the variance Var$[\check z]$ as
\begin{align}\label{eq:Varz}
{\rm Var}[\check z]
	 = \mathbb{E}
	\bigl[\bigl(\check z - 
	\mathbb{E}[\check z]
	\bigl)^2\bigr] 
	 = 
	\bigl[\bigl(
	{\rm Tr}(\mathcal{\bm Z}\check\rho)
	- {\rm Tr}(\mathcal{\bm Z}\rho)
	\bigl)^2\bigr] .
\end{align}

In Fig.~\ref{fig:9} (a), we show
the variance Var$[\check z]$ as a function of 
the number of measurements for the 
shadow tomography. 
The variance slightly decreases when increasing the 
number of measurements from $10^2$ to $10^5$.
See the inset figure for detailed zoom-in. 
The result is compared with the standard quantum limit
(SQL), i.e., SQL = 1/ Number of measurements, 
and the Heisenberg limit (HL), i.e., HL = 
1/ (Number of measurements)$^2$. 
Here, the variance does not beat the SQL nor HL.

In Fig.~\ref{fig:9} (b), we  compare the variances obtained
from the shadow tomography and the UC-VQA for the different 
number of qubits $N$. 
For the shadow tomography, we fix the
number of measurements at $10^5$.
For the UC-VQA, we consider the 
$\mathcal{W}_{\rm chain}$ structure with 
the Adam optimizer as an example. 
The number of shots (measurements) is 
fixed at $10^4$.
It can be seen that the UC-VQA 
offers a better result over 100 times
than the resulting shadow tomography.

We further discuss some 
features of these two approaches.
The shadow tomography only
allows for predicting target functions, 
such as expectation values, entanglement entropies,
correlation functions, and so on \cite{Huang2020},
whereas the UC-VQA allows for reconstructing 
the entire quantum state up to a phase shift.
Both schemes allow for predicting properties 
of quantum states or quantum states 
with fewer measurements 
compared to standard quantum tomography. 
Another remarkable feature is that 
the efficiency of the shadow tomography protocol 
depends on the random choice 
of the unitaries in an ensemble
$\mathcal{U}$, while the efficiency of the UC-VQA scheme
relies on the choice of different ansatzes and optimizes. 
Finally, we emphasize that the comparison in this section
only provides a very first glance about the two approaches. 
We need to further characterize these features 
in future works for more concrete evaluation.

\section{Conclusion}
We have proposed a universal compilation-based 
variational quantum algorithm (UC-VQA) 
for compiling a given quantum state 
to another one and its applications 
to quantum state preparation (QSP) 
and quantum state tomography (QST). 
We have considered the compiling probability 
as a quantum kernel that needs to maximize, 
and thus, it serves as a cost function. 
We have conducted various numerical experiments 
using different ansatz structures and optimizers
to analyze the algorithm's performance 
for the aforementioned subjects. 
The key feature of the UC-VQA is that 
it provides low-depth circuits and 
only requires fewer measurements 
than conventional methods 
to achieve high accuracy. 
We highlight the advantages of the proposed algorithm
that are comparable to other similar approaches.

In the QSP, we showed that 
the entanglement class of 
the GHZ and W target states
could be prepared with excellent efficiency  
and well against the noise, such as 
using an error mitigation protocol. 
This result is applicable 
to any arbitrary target state.
We also addressed the effect of 
the barren plateau that is unavoidable 
due to the increase of the system's space. 
In the QST, we gain high fidelity for reconstructing 
an entire unknown random state
by choosing a proper circuit depth 
via the number of layers 
in the quantum circuit. Besides, 
our method can compare with 
a similar scheme based on
classical shadow tomography 
and exhibit a better result.

The proposed UC-VQA can further 
promise applications to quantum metrology, 
quantum sensing, and new frontier foundation aspects. 
Moreover, it is possible to implement 
the algorithm on near-term quantum computers, 
and thus it could be a valuable technique 
for verifying the fidelity of quantum circuits 
and studying quantum computing.

\begin{acknowledgments}
This work is supported by 
the Vietnam Academy of 
Science and Technology (VAST) 
under the grant number 
CSCL14.01/22-23, and 
the VNUHCM-University of Information Technology’s 
Scientific Research Support Fund.
\end{acknowledgments}
\vspace{0.25cm}
\noindent {\bf Code availability}:
The codes used for this study are available 
in https://github.com/vutuanhai237/UC-VQA.


\appendix
\setcounter{equation}{0}
\renewcommand{\theequation}{A.\arabic{equation}}
\section{Fubini-Study tensor metric}\label{appA}
\begin{figure}[!h]
\includegraphics[width=8cm]{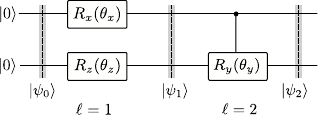}
\caption{An example circuit.}
\label{fig:1app}
\end{figure}

We provide a practical example of
how to compute a Fubini-Study tensor metric. 
Let us consider a concrete 
circuit as shown in Fig.~\ref{fig:1app}.
It consists of $R_x = \exp(-i\frac{\theta_x}{2}\sigma_x)$, 
$R_z =  \exp(-i\frac{\theta_z}{2}\sigma_z)$, and
 $CR_y = |0\rangle\langle0|\otimes I_2 + 
 |1\rangle\langle1|\otimes\exp(-i\frac{\theta_y}{2}\sigma_y)$.
Since $[R_x, R_z] = 0$ (because they act on different qubits),
we can group them into one layer (layer 1),
with $\bm\theta^{(1)} = \{\theta^{(1)}_0, \theta^{(1)}_1\}
=  \{\theta_x, \theta_z\}$,
and put $CR_y$ into another layer (layer 2), with 
$\bm\theta^{(2)} = \{\theta^{(2)}_0\} = \{\theta_y\}$.
The tensor metric $g$ explicitly yields
\begin{align}\label{eq:tensor_app}
g = 
	\begin{pmatrix}
		g_{xx}^{(1)} & g_{xz}^{(1)} & 0\\
		g_{zx}^{(1)} & g_{zz}^{(1)} & 0\\
		0 & 0 & g_{yy}^{(2)}
	\end{pmatrix}.
\end{align}

The quantum states are explicitly expressed as
\begin{align}\label{eq:psi_app}
|\psi_0\rangle &= |00\rangle,\;\\
|\psi_1\rangle &= 
e^{-i\frac{\theta_x}{2}\sigma_x\otimes I_2}
e^{-i\frac{\theta_z}{2}I_2\otimes\sigma_z} |\psi_0\rangle,\\
|\psi_2\rangle &= \big[|0\rangle\langle0|\otimes I_2 + 
|1\rangle\langle1|\otimes 
e^{-i\frac{\theta_y}{2}\sigma_y}\big]
|\psi_1\rangle.
\end{align}

The elements $g^{(1)}_{ij}$ 
is given through Eq.~\eqref{eq:gij_re} 
as
\begin{align}
\notag g^{(1)}_{xx} &= \langle\psi_0|K_x^2|\psi_0\rangle
-\langle\psi_0|K_x|\psi_0\rangle^2
= \dfrac{1}{4}\;,\\
\notag g^{(1)}_{xz} &= \langle\psi_0|K_xK_z|\psi_0\rangle
-\langle\psi_0|K_x|\psi_0\rangle
\langle\psi_0|K_z|\psi_0\rangle
= 0\;,\\
\notag g^{(1)}_{zx} &= \langle\psi_0|K_zK_x|\psi_0\rangle
-\langle\psi_0|K_z|\psi_0\rangle
\langle\psi_0|K_x|\psi_0\rangle = 0\;,\\
\notag g^{(1)}_{zz} &= \langle\psi_0|K_z^2|\psi_0\rangle
-\langle\psi_0|K_z|\psi_0\rangle^2
= 0\;,
\end{align}
where $K_x = \frac{\sigma_x\otimes I_2}{2}$
and $K_z = \frac{I_2\otimes \sigma_z}{2}$.

Next, we calculate $g^{(2)}_{yy}$.
Starting from Eq.~\eqref{eq:gij} in the main text,
we derive
\begin{align}\label{eq:par_app}
	|\partial_{\theta_y}\psi_2\rangle
	= -i|1\rangle\langle 1|
	\otimes\dfrac{\sigma_y}{2}e^{-i\frac{\theta_y}{2}\sigma_y}
	|\psi_1\rangle.
\end{align}
Then, we get
\begin{align}\label{eq:gyy_app}
	\notag g^{(2)}_{yy} &= \langle\psi_1|K_y^2|\psi_1\rangle
	-\langle\psi_1|K_y|\psi_1\rangle^2\\
	& = \dfrac{1}{4}\sin^2\big(\textstyle\frac{\theta_x}{2}\big),
\end{align}
where $K_y = |1\rangle\langle1|\otimes\frac{\sigma_y}{2}$.
To derive expectation values in Eq.~\eqref{eq:gyy_app},
we prepare $|\psi_1\rangle$ as in Fig.~\ref{fig:1app}, 
then measure $\langle\psi_1|K_y^2|\psi_1\rangle 
= \frac{1}{4}\langle\psi_1|\big(|1\rangle\langle1
|\otimes I_2\big)|\psi_1\rangle$
and $\langle\psi_1|K_y|\psi_1\rangle 
= \frac{1}{2}\langle\psi_1|\big(|1\rangle\langle1
|\otimes \sigma_y\big)|\psi_1\rangle$.
Finally, we obtain the tensor metric $g$
\begin{align}\label{eq:tensor_app}
g = 
	\begin{pmatrix}
		\frac{1}{4} & 0& 0\\
		0 & 0 & 0\\
		0 & 0 & \frac{1}{4}\sin^2(\frac{\theta_x}{2})
	\end{pmatrix}.
\end{align}

\setcounter{equation}{0}
\renewcommand{\theequation}{B.\arabic{equation}}
\section{Supported data for QST}\label{appB}
We discuss more data supporting the 
results in Fig.~\ref{fig:8} (d,e,f) in the main text.
As we discussed above, the accuracy can be improved when 
increasing the number of layers $L$.
However, we cannot increase $L$ arbitrarily large
and need to stop at an optimal point. 
We define the optimal $L$ 
as the smallest number of layers that, 
at the next layer, the 
accuracy saturates or starts to reduce. 
In Fig.~\ref{fig:2app} below, we discuss the optimal $L$
for various cases, where we mark the optimal $L$ with
colored arrows.
See also  Table~\ref{tab:1} below.
\begin{figure}[!h]
\includegraphics[width=8.6cm]{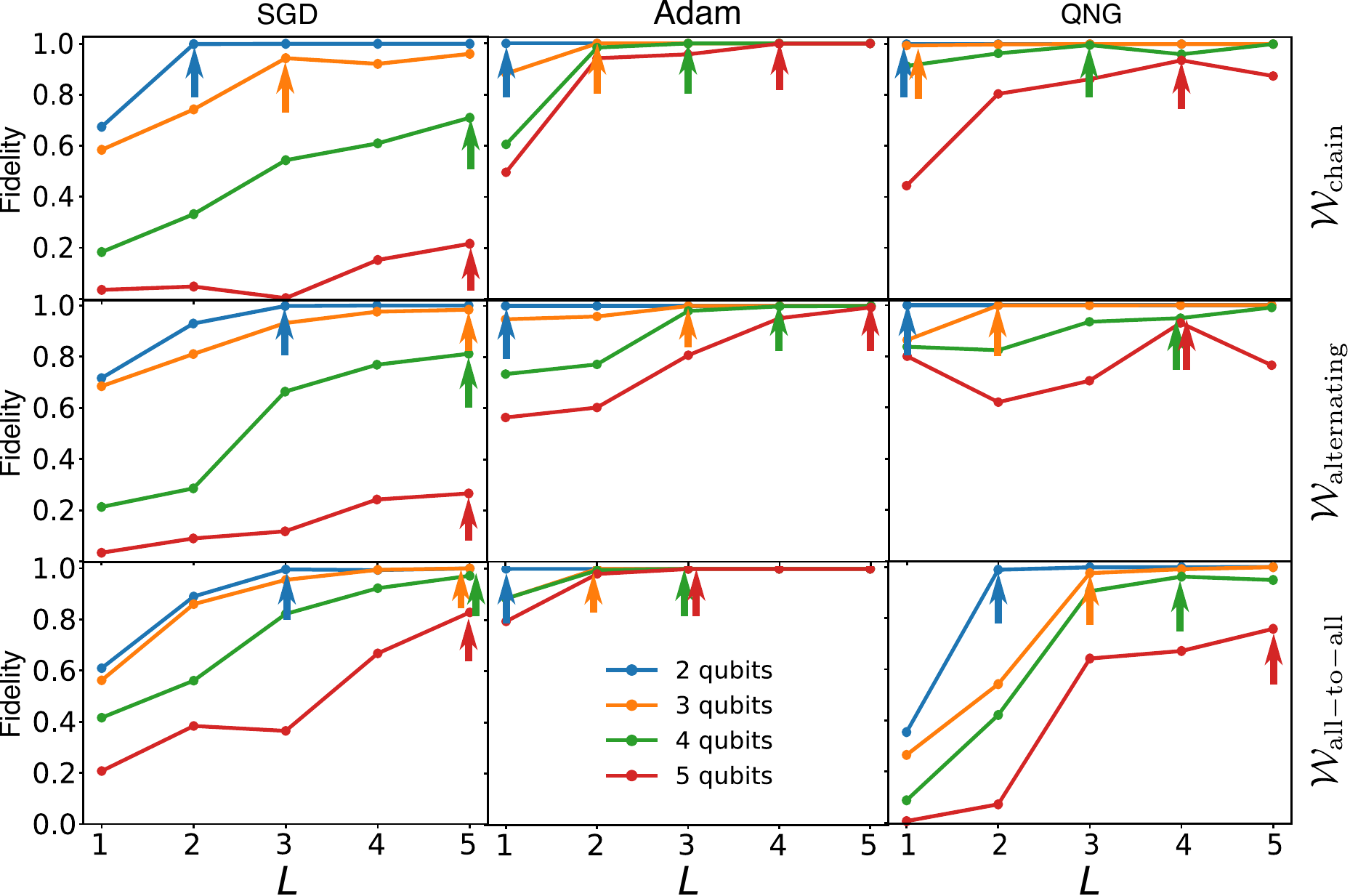}
\caption{Plot of fidelity as a function of $L$
for different $\mathcal{W}$ structures (different rows)
and different optimizers (different columns.)}
\label{fig:2app}
\end{figure}

\begin{table}[!h]
  \centering
    \caption{Number of optimal layers taking from Fig.~\ref{fig:2app}}
  \begin{tabular}{|c|c|c|c|c|c|}
    \hline
    \multicolumn{1}{|c|}{Structure}&\multicolumn{1}{c|}{Optimizer}
    &\multicolumn{4}{c|}{$N$} \\
     \cline{3-6}
     \multicolumn{1}{|c|}{} & \multicolumn{1}{c|}{} & 2 & 3 & 4 & 5\\\hline
     \multirow{5}{*}{$\mathcal{W}_{\rm chain}$}  
     & SGD & $L = 2$ & $L = 3$ &  $L = 5$ &  $L = 5$\\\cline{2-6}    
     &Adam& $L = 1$ & $L = 2$ &  $L = 3$ &  $L = 4$ \\\cline{2-6}
     &QNG&  $L = 1$ & $L = 1$ &  $L = 3$ &  $L = 4$\\\cline{2-6}
     &Fig~\ref{fig:8}(d)&  (\textcolor{RoyalBlue}{2},
					\textcolor{orange}1,
					\textcolor{Green}1)
				     & (\textcolor{RoyalBlue}{3},
					\textcolor{orange}2,
					\textcolor{Green}1)
				   &  (\textcolor{RoyalBlue}{5},
					\textcolor{orange}3,
					\textcolor{Green}3)
				   &  (\textcolor{RoyalBlue}{5},
					\textcolor{orange}4,
					\textcolor{Green}4)\\
     \hline\hline
     \multirow{5}{*}{$\mathcal{W}_{\rm alternating}$}  
     &SGD &  $L = 3$ & $L = 5$ &  $L = 5$ &  $L = 5$\\\cline{2-6}    
     &Adam& $L = 1$ & $L = 3$ &  $L = 4$ &  $L = 5$ \\\cline{2-6}
     &QNG&  $L = 1$ & $L = 2$ &  $L = 4$ &  $L = 4$\\\cline{2-6}
     &Fig~\ref{fig:8}(e)&  (\textcolor{RoyalBlue}{3},
					\textcolor{orange}1,
					\textcolor{Green}1)
				     & (\textcolor{RoyalBlue}{5},
					\textcolor{orange}3,
					\textcolor{Green}2)
				   &  (\textcolor{RoyalBlue}{5},
					\textcolor{orange}4,
					\textcolor{Green}4)
				   &  (\textcolor{RoyalBlue}{5},
					\textcolor{orange}5,
					\textcolor{Green}4)\\
     \hline\hline  
     \multirow{5}{*}{$\mathcal{W}_{\rm all-to-all}$}  
     &SGD &  $L = 3$ & $L = 5$ &  $L = 5$ &  $L = 5$\\\cline{2-6}    
     &Adam& $L = 1$ & $L = 2$ &  $L = 3$ &  $L = 3$ \\\cline{2-6}
     &QNG&  $L = 2$ & $L = 3$ &  $L = 4$ &  $L = 5$\\\cline{2-6}
     &Fig~\ref{fig:8}(f)&  (\textcolor{RoyalBlue}{3},
					\textcolor{orange}1,
					\textcolor{Green}2)
				     & (\textcolor{RoyalBlue}{5},
					\textcolor{orange}2,
					\textcolor{Green}3)
				   &  (\textcolor{RoyalBlue}{5},
					\textcolor{orange}3,
					\textcolor{Green}4)
				   &  (\textcolor{RoyalBlue}{5},
					\textcolor{orange}3,
					\textcolor{Green}5)\\
     \hline
   \end{tabular}
  \label{tab:1}
\end{table}
From the results here, 
we trace out the optimal $L$ as shown in Fig.~\ref{fig:8}
in the main text.

\bibliography{refs}

\begin{thebibliography}{96}%
\makeatletter
\providecommand \@ifxundefined [1]{%
 \@ifx{#1\undefined}
}%
\providecommand \@ifnum [1]{%
 \ifnum #1\expandafter \@firstoftwo
 \else \expandafter \@secondoftwo
 \fi
}%
\providecommand \@ifx [1]{%
 \ifx #1\expandafter \@firstoftwo
 \else \expandafter \@secondoftwo
 \fi
}%
\providecommand \natexlab [1]{#1}%
\providecommand \enquote  [1]{``#1''}%
\providecommand \bibnamefont  [1]{#1}%
\providecommand \bibfnamefont [1]{#1}%
\providecommand \citenamefont [1]{#1}%
\providecommand \href@noop [0]{\@secondoftwo}%
\providecommand \href [0]{\begingroup \@sanitize@url \@href}%
\providecommand \@href[1]{\@@startlink{#1}\@@href}%
\providecommand \@@href[1]{\endgroup#1\@@endlink}%
\providecommand \@sanitize@url [0]{\catcode `\\12\catcode `\$12\catcode
  `\&12\catcode `\#12\catcode `\^12\catcode `\_12\catcode `\%12\relax}%
\providecommand \@@startlink[1]{}%
\providecommand \@@endlink[0]{}%
\providecommand \url  [0]{\begingroup\@sanitize@url \@url }%
\providecommand \@url [1]{\endgroup\@href {#1}{\urlprefix }}%
\providecommand \urlprefix  [0]{URL }%
\providecommand \Eprint [0]{\href }%
\providecommand \doibase [0]{http://dx.doi.org/}%
\providecommand \selectlanguage [0]{\@gobble}%
\providecommand \bibinfo  [0]{\@secondoftwo}%
\providecommand \bibfield  [0]{\@secondoftwo}%
\providecommand \translation [1]{[#1]}%
\providecommand \BibitemOpen [0]{}%
\providecommand \bibitemStop [0]{}%
\providecommand \bibitemNoStop [0]{.\EOS\space}%
\providecommand \EOS [0]{\spacefactor3000\relax}%
\providecommand \BibitemShut  [1]{\csname bibitem#1\endcsname}%
\let\auto@bib@innerbib\@empty
\bibitem [{\citenamefont {de~Leon}\ \emph {et~al.}(2021)\citenamefont
  {de~Leon}, \citenamefont {Itoh}, \citenamefont {Kim}, \citenamefont {Mehta},
  \citenamefont {Northup}, \citenamefont {Paik}, \citenamefont {Palmer},
  \citenamefont {Samarth}, \citenamefont {Sangtawesin},\ and\ \citenamefont
  {Steuerman}}]{doi:10.1126/science.abb2823}%
  \BibitemOpen
  \bibfield  {author} {\bibinfo {author} {\bibfnamefont {N.~P.}\ \bibnamefont
  {de~Leon}}, \bibinfo {author} {\bibfnamefont {K.~M.}\ \bibnamefont {Itoh}},
  \bibinfo {author} {\bibfnamefont {D.}~\bibnamefont {Kim}}, \bibinfo {author}
  {\bibfnamefont {K.~K.}\ \bibnamefont {Mehta}}, \bibinfo {author}
  {\bibfnamefont {T.~E.}\ \bibnamefont {Northup}}, \bibinfo {author}
  {\bibfnamefont {H.}~\bibnamefont {Paik}}, \bibinfo {author} {\bibfnamefont
  {B.~S.}\ \bibnamefont {Palmer}}, \bibinfo {author} {\bibfnamefont
  {N.}~\bibnamefont {Samarth}}, \bibinfo {author} {\bibfnamefont
  {S.}~\bibnamefont {Sangtawesin}}, \ and\ \bibinfo {author} {\bibfnamefont
  {D.~W.}\ \bibnamefont {Steuerman}},\ }\href {\doibase
  10.1126/science.abb2823} {\bibfield  {journal} {\bibinfo  {journal}
  {Science}\ }\textbf {\bibinfo {volume} {372}},\ \bibinfo {pages} {eabb2823}
  (\bibinfo {year} {2021})}\BibitemShut {NoStop}%
\bibitem [{\citenamefont {Alexeev}\ \emph {et~al.}(2021)\citenamefont
  {Alexeev}, \citenamefont {Bacon}, \citenamefont {Brown}, \citenamefont
  {Calderbank}, \citenamefont {Carr}, \citenamefont {Chong}, \citenamefont
  {DeMarco}, \citenamefont {Englund}, \citenamefont {Farhi}, \citenamefont
  {Fefferman}, \citenamefont {Gorshkov}, \citenamefont {Houck}, \citenamefont
  {Kim}, \citenamefont {Kimmel}, \citenamefont {Lange}, \citenamefont {Lloyd},
  \citenamefont {Lukin}, \citenamefont {Maslov}, \citenamefont {Maunz},
  \citenamefont {Monroe}, \citenamefont {Preskill}, \citenamefont {Roetteler},
  \citenamefont {Savage},\ and\ \citenamefont
  {Thompson}}]{PRXQuantum.2.017001}%
  \BibitemOpen
  \bibfield  {author} {\bibinfo {author} {\bibfnamefont {Y.}~\bibnamefont
  {Alexeev}}, \bibinfo {author} {\bibfnamefont {D.}~\bibnamefont {Bacon}},
  \bibinfo {author} {\bibfnamefont {K.~R.}\ \bibnamefont {Brown}}, \bibinfo
  {author} {\bibfnamefont {R.}~\bibnamefont {Calderbank}}, \bibinfo {author}
  {\bibfnamefont {L.~D.}\ \bibnamefont {Carr}}, \bibinfo {author}
  {\bibfnamefont {F.~T.}\ \bibnamefont {Chong}}, \bibinfo {author}
  {\bibfnamefont {B.}~\bibnamefont {DeMarco}}, \bibinfo {author} {\bibfnamefont
  {D.}~\bibnamefont {Englund}}, \bibinfo {author} {\bibfnamefont
  {E.}~\bibnamefont {Farhi}}, \bibinfo {author} {\bibfnamefont
  {B.}~\bibnamefont {Fefferman}}, \bibinfo {author} {\bibfnamefont {A.~V.}\
  \bibnamefont {Gorshkov}}, \bibinfo {author} {\bibfnamefont {A.}~\bibnamefont
  {Houck}}, \bibinfo {author} {\bibfnamefont {J.}~\bibnamefont {Kim}}, \bibinfo
  {author} {\bibfnamefont {S.}~\bibnamefont {Kimmel}}, \bibinfo {author}
  {\bibfnamefont {M.}~\bibnamefont {Lange}}, \bibinfo {author} {\bibfnamefont
  {S.}~\bibnamefont {Lloyd}}, \bibinfo {author} {\bibfnamefont {M.~D.}\
  \bibnamefont {Lukin}}, \bibinfo {author} {\bibfnamefont {D.}~\bibnamefont
  {Maslov}}, \bibinfo {author} {\bibfnamefont {P.}~\bibnamefont {Maunz}},
  \bibinfo {author} {\bibfnamefont {C.}~\bibnamefont {Monroe}}, \bibinfo
  {author} {\bibfnamefont {J.}~\bibnamefont {Preskill}}, \bibinfo {author}
  {\bibfnamefont {M.}~\bibnamefont {Roetteler}}, \bibinfo {author}
  {\bibfnamefont {M.~J.}\ \bibnamefont {Savage}}, \ and\ \bibinfo {author}
  {\bibfnamefont {J.}~\bibnamefont {Thompson}},\ }\href {\doibase
  10.1103/PRXQuantum.2.017001} {\bibfield  {journal} {\bibinfo  {journal} {PRX
  Quantum}\ }\textbf {\bibinfo {volume} {2}},\ \bibinfo {pages} {017001}
  (\bibinfo {year} {2021})}\BibitemShut {NoStop}%
\bibitem [{\citenamefont {Ebadi}\ \emph {et~al.}(2021)\citenamefont {Ebadi},
  \citenamefont {Wang}, \citenamefont {Levine}, \citenamefont {Keesling},
  \citenamefont {Semeghini}, \citenamefont {Omran}, \citenamefont {Bluvstein},
  \citenamefont {Samajdar}, \citenamefont {Pichler}, \citenamefont {Ho},
  \citenamefont {Choi}, \citenamefont {Sachdev}, \citenamefont {Greiner},
  \citenamefont {Vuleti{\'{c}}},\ and\ \citenamefont {Lukin}}]{Ebadi2021}%
  \BibitemOpen
  \bibfield  {author} {\bibinfo {author} {\bibfnamefont {S.}~\bibnamefont
  {Ebadi}}, \bibinfo {author} {\bibfnamefont {T.~T.}\ \bibnamefont {Wang}},
  \bibinfo {author} {\bibfnamefont {H.}~\bibnamefont {Levine}}, \bibinfo
  {author} {\bibfnamefont {A.}~\bibnamefont {Keesling}}, \bibinfo {author}
  {\bibfnamefont {G.}~\bibnamefont {Semeghini}}, \bibinfo {author}
  {\bibfnamefont {A.}~\bibnamefont {Omran}}, \bibinfo {author} {\bibfnamefont
  {D.}~\bibnamefont {Bluvstein}}, \bibinfo {author} {\bibfnamefont
  {R.}~\bibnamefont {Samajdar}}, \bibinfo {author} {\bibfnamefont
  {H.}~\bibnamefont {Pichler}}, \bibinfo {author} {\bibfnamefont {W.~W.}\
  \bibnamefont {Ho}}, \bibinfo {author} {\bibfnamefont {S.}~\bibnamefont
  {Choi}}, \bibinfo {author} {\bibfnamefont {S.}~\bibnamefont {Sachdev}},
  \bibinfo {author} {\bibfnamefont {M.}~\bibnamefont {Greiner}}, \bibinfo
  {author} {\bibfnamefont {V.}~\bibnamefont {Vuleti{\'{c}}}}, \ and\ \bibinfo
  {author} {\bibfnamefont {M.~D.}\ \bibnamefont {Lukin}},\ }\href {\doibase
  10.1038/s41586-021-03582-4} {\bibfield  {journal} {\bibinfo  {journal}
  {Nature}\ }\textbf {\bibinfo {volume} {595}},\ \bibinfo {pages} {227}
  (\bibinfo {year} {2021})}\BibitemShut {NoStop}%
\bibitem [{\citenamefont {Pirandola}\ \emph {et~al.}(2015)\citenamefont
  {Pirandola}, \citenamefont {Eisert}, \citenamefont {Weedbrook}, \citenamefont
  {Furusawa},\ and\ \citenamefont {Braunstein}}]{Pirandola2015}%
  \BibitemOpen
  \bibfield  {author} {\bibinfo {author} {\bibfnamefont {S.}~\bibnamefont
  {Pirandola}}, \bibinfo {author} {\bibfnamefont {J.}~\bibnamefont {Eisert}},
  \bibinfo {author} {\bibfnamefont {C.}~\bibnamefont {Weedbrook}}, \bibinfo
  {author} {\bibfnamefont {A.}~\bibnamefont {Furusawa}}, \ and\ \bibinfo
  {author} {\bibfnamefont {S.~L.}\ \bibnamefont {Braunstein}},\ }\href
  {\doibase 10.1038/nphoton.2015.154} {\bibfield  {journal} {\bibinfo
  {journal} {Nature Photonics}\ }\textbf {\bibinfo {volume} {9}},\ \bibinfo
  {pages} {641} (\bibinfo {year} {2015})}\BibitemShut {NoStop}%
\bibitem [{\citenamefont {Spiller}(2003)}]{SPILLER200330}%
  \BibitemOpen
  \bibfield  {author} {\bibinfo {author} {\bibfnamefont {T.~P.}\ \bibnamefont
  {Spiller}},\ }\href {\doibase https://doi.org/10.1016/S1369-7021(03)00130-5}
  {\bibfield  {journal} {\bibinfo  {journal} {Materials Today}\ }\textbf
  {\bibinfo {volume} {6}},\ \bibinfo {pages} {30} (\bibinfo {year}
  {2003})}\BibitemShut {NoStop}%
\bibitem [{\citenamefont {Shor}(1994)}]{365700}%
  \BibitemOpen
  \bibfield  {author} {\bibinfo {author} {\bibfnamefont {P.}~\bibnamefont
  {Shor}},\ }in\ \href {\doibase 10.1109/SFCS.1994.365700} {\emph {\bibinfo
  {booktitle} {Proceedings 35th Annual Symposium on Foundations of Computer
  Science}}}\ (\bibinfo {year} {1994})\ pp.\ \bibinfo {pages}
  {124--134}\BibitemShut {NoStop}%
\bibitem [{\citenamefont {Grover}(1996)}]{grover1996fast}%
  \BibitemOpen
  \bibfield  {author} {\bibinfo {author} {\bibfnamefont {L.~K.}\ \bibnamefont
  {Grover}},\ }in\ \href@noop {} {\emph {\bibinfo {booktitle} {Proceedings of
  the twenty-eighth annual ACM symposium on Theory of computing}}}\ (\bibinfo
  {year} {1996})\ pp.\ \bibinfo {pages} {212--219}\BibitemShut {NoStop}%
\bibitem [{\citenamefont {Harrow}\ \emph {et~al.}(2009)\citenamefont {Harrow},
  \citenamefont {Hassidim},\ and\ \citenamefont
  {Lloyd}}]{PhysRevLett.103.150502}%
  \BibitemOpen
  \bibfield  {author} {\bibinfo {author} {\bibfnamefont {A.~W.}\ \bibnamefont
  {Harrow}}, \bibinfo {author} {\bibfnamefont {A.}~\bibnamefont {Hassidim}}, \
  and\ \bibinfo {author} {\bibfnamefont {S.}~\bibnamefont {Lloyd}},\ }\href
  {\doibase 10.1103/PhysRevLett.103.150502} {\bibfield  {journal} {\bibinfo
  {journal} {Phys. Rev. Lett.}\ }\textbf {\bibinfo {volume} {103}},\ \bibinfo
  {pages} {150502} (\bibinfo {year} {2009})}\BibitemShut {NoStop}%
\bibitem [{\citenamefont {Xu}\ \emph {et~al.}(2021)\citenamefont {Xu},
  \citenamefont {Benjamin},\ and\ \citenamefont
  {Yuan}}]{PhysRevApplied.15.034068}%
  \BibitemOpen
  \bibfield  {author} {\bibinfo {author} {\bibfnamefont {X.}~\bibnamefont
  {Xu}}, \bibinfo {author} {\bibfnamefont {S.~C.}\ \bibnamefont {Benjamin}}, \
  and\ \bibinfo {author} {\bibfnamefont {X.}~\bibnamefont {Yuan}},\ }\href
  {\doibase 10.1103/PhysRevApplied.15.034068} {\bibfield  {journal} {\bibinfo
  {journal} {Phys. Rev. Applied}\ }\textbf {\bibinfo {volume} {15}},\ \bibinfo
  {pages} {034068} (\bibinfo {year} {2021})}\BibitemShut {NoStop}%
\bibitem [{\citenamefont {Lubasch}\ \emph {et~al.}(2020)\citenamefont
  {Lubasch}, \citenamefont {Joo}, \citenamefont {Moinier}, \citenamefont
  {Kiffner},\ and\ \citenamefont {Jaksch}}]{PhysRevA.101.010301}%
  \BibitemOpen
  \bibfield  {author} {\bibinfo {author} {\bibfnamefont {M.}~\bibnamefont
  {Lubasch}}, \bibinfo {author} {\bibfnamefont {J.}~\bibnamefont {Joo}},
  \bibinfo {author} {\bibfnamefont {P.}~\bibnamefont {Moinier}}, \bibinfo
  {author} {\bibfnamefont {M.}~\bibnamefont {Kiffner}}, \ and\ \bibinfo
  {author} {\bibfnamefont {D.}~\bibnamefont {Jaksch}},\ }\href {\doibase
  10.1103/PhysRevA.101.010301} {\bibfield  {journal} {\bibinfo  {journal}
  {Phys. Rev. A}\ }\textbf {\bibinfo {volume} {101}},\ \bibinfo {pages}
  {010301} (\bibinfo {year} {2020})}\BibitemShut {NoStop}%
\bibitem [{\citenamefont {Preskill}(2018)}]{Preskill2018quantumcomputingin}%
  \BibitemOpen
  \bibfield  {author} {\bibinfo {author} {\bibfnamefont {J.}~\bibnamefont
  {Preskill}},\ }\href {\doibase 10.22331/q-2018-08-06-79} {\bibfield
  {journal} {\bibinfo  {journal} {{Quantum}}\ }\textbf {\bibinfo {volume}
  {2}},\ \bibinfo {pages} {79} (\bibinfo {year} {2018})}\BibitemShut {NoStop}%
\bibitem [{\citenamefont {Cerezo}\ \emph
  {et~al.}(2021{\natexlab{a}})\citenamefont {Cerezo}, \citenamefont
  {Arrasmith}, \citenamefont {Babbush}, \citenamefont {Benjamin}, \citenamefont
  {Endo}, \citenamefont {Fujii}, \citenamefont {McClean}, \citenamefont
  {Mitarai}, \citenamefont {Yuan}, \citenamefont {Cincio},\ and\ \citenamefont
  {Coles}}]{Cerezo2021_r}%
  \BibitemOpen
  \bibfield  {author} {\bibinfo {author} {\bibfnamefont {M.}~\bibnamefont
  {Cerezo}}, \bibinfo {author} {\bibfnamefont {A.}~\bibnamefont {Arrasmith}},
  \bibinfo {author} {\bibfnamefont {R.}~\bibnamefont {Babbush}}, \bibinfo
  {author} {\bibfnamefont {S.~C.}\ \bibnamefont {Benjamin}}, \bibinfo {author}
  {\bibfnamefont {S.}~\bibnamefont {Endo}}, \bibinfo {author} {\bibfnamefont
  {K.}~\bibnamefont {Fujii}}, \bibinfo {author} {\bibfnamefont {J.~R.}\
  \bibnamefont {McClean}}, \bibinfo {author} {\bibfnamefont {K.}~\bibnamefont
  {Mitarai}}, \bibinfo {author} {\bibfnamefont {X.}~\bibnamefont {Yuan}},
  \bibinfo {author} {\bibfnamefont {L.}~\bibnamefont {Cincio}}, \ and\ \bibinfo
  {author} {\bibfnamefont {P.~J.}\ \bibnamefont {Coles}},\ }\href {\doibase
  10.1038/s42254-021-00348-9} {\bibfield  {journal} {\bibinfo  {journal}
  {Nature Reviews Physics}\ }\textbf {\bibinfo {volume} {3}},\ \bibinfo {pages}
  {625} (\bibinfo {year} {2021}{\natexlab{a}})}\BibitemShut {NoStop}%
\bibitem [{\citenamefont {Peruzzo}\ \emph {et~al.}(2014)\citenamefont
  {Peruzzo}, \citenamefont {McClean}, \citenamefont {Shadbolt}, \citenamefont
  {Yung}, \citenamefont {Zhou}, \citenamefont {Love}, \citenamefont
  {Aspuru-Guzik},\ and\ \citenamefont {O'Brien}}]{Peruzzo2014}%
  \BibitemOpen
  \bibfield  {author} {\bibinfo {author} {\bibfnamefont {A.}~\bibnamefont
  {Peruzzo}}, \bibinfo {author} {\bibfnamefont {J.}~\bibnamefont {McClean}},
  \bibinfo {author} {\bibfnamefont {P.}~\bibnamefont {Shadbolt}}, \bibinfo
  {author} {\bibfnamefont {M.-H.}\ \bibnamefont {Yung}}, \bibinfo {author}
  {\bibfnamefont {X.-Q.}\ \bibnamefont {Zhou}}, \bibinfo {author}
  {\bibfnamefont {P.~J.}\ \bibnamefont {Love}}, \bibinfo {author}
  {\bibfnamefont {A.}~\bibnamefont {Aspuru-Guzik}}, \ and\ \bibinfo {author}
  {\bibfnamefont {J.~L.}\ \bibnamefont {O'Brien}},\ }\href {\doibase
  10.1038/ncomms5213} {\bibfield  {journal} {\bibinfo  {journal} {Nature
  Communications}\ }\textbf {\bibinfo {volume} {5}},\ \bibinfo {pages} {4213}
  (\bibinfo {year} {2014})}\BibitemShut {NoStop}%
\bibitem [{\citenamefont {Nakanishi}\ \emph {et~al.}(2019)\citenamefont
  {Nakanishi}, \citenamefont {Mitarai},\ and\ \citenamefont
  {Fujii}}]{PhysRevResearch.1.033062}%
  \BibitemOpen
  \bibfield  {author} {\bibinfo {author} {\bibfnamefont {K.~M.}\ \bibnamefont
  {Nakanishi}}, \bibinfo {author} {\bibfnamefont {K.}~\bibnamefont {Mitarai}},
  \ and\ \bibinfo {author} {\bibfnamefont {K.}~\bibnamefont {Fujii}},\ }\href
  {\doibase 10.1103/PhysRevResearch.1.033062} {\bibfield  {journal} {\bibinfo
  {journal} {Phys. Rev. Research}\ }\textbf {\bibinfo {volume} {1}},\ \bibinfo
  {pages} {033062} (\bibinfo {year} {2019})}\BibitemShut {NoStop}%
\bibitem [{\citenamefont {Kirby}\ \emph {et~al.}(2021)\citenamefont {Kirby},
  \citenamefont {Tranter},\ and\ \citenamefont
  {Love}}]{Kirby2021contextualsubspace}%
  \BibitemOpen
  \bibfield  {author} {\bibinfo {author} {\bibfnamefont {W.~M.}\ \bibnamefont
  {Kirby}}, \bibinfo {author} {\bibfnamefont {A.}~\bibnamefont {Tranter}}, \
  and\ \bibinfo {author} {\bibfnamefont {P.~J.}\ \bibnamefont {Love}},\ }\href
  {\doibase 10.22331/q-2021-05-14-456} {\bibfield  {journal} {\bibinfo
  {journal} {{Quantum}}\ }\textbf {\bibinfo {volume} {5}},\ \bibinfo {pages}
  {456} (\bibinfo {year} {2021})}\BibitemShut {NoStop}%
\bibitem [{\citenamefont {Gard}\ \emph {et~al.}(2020)\citenamefont {Gard},
  \citenamefont {Zhu}, \citenamefont {Barron}, \citenamefont {Mayhall},
  \citenamefont {Economou},\ and\ \citenamefont {Barnes}}]{Gard2020}%
  \BibitemOpen
  \bibfield  {author} {\bibinfo {author} {\bibfnamefont {B.~T.}\ \bibnamefont
  {Gard}}, \bibinfo {author} {\bibfnamefont {L.}~\bibnamefont {Zhu}}, \bibinfo
  {author} {\bibfnamefont {G.~S.}\ \bibnamefont {Barron}}, \bibinfo {author}
  {\bibfnamefont {N.~J.}\ \bibnamefont {Mayhall}}, \bibinfo {author}
  {\bibfnamefont {S.~E.}\ \bibnamefont {Economou}}, \ and\ \bibinfo {author}
  {\bibfnamefont {E.}~\bibnamefont {Barnes}},\ }\href {\doibase
  10.1038/s41534-019-0240-1} {\bibfield  {journal} {\bibinfo  {journal} {npj
  Quantum Information}\ }\textbf {\bibinfo {volume} {6}},\ \bibinfo {pages}
  {10} (\bibinfo {year} {2020})}\BibitemShut {NoStop}%
\bibitem [{\citenamefont {Tkachenko}\ \emph {et~al.}(2021)\citenamefont
  {Tkachenko}, \citenamefont {Sud}, \citenamefont {Zhang}, \citenamefont
  {Tretiak}, \citenamefont {Anisimov}, \citenamefont {Arrasmith}, \citenamefont
  {Coles}, \citenamefont {Cincio},\ and\ \citenamefont
  {Dub}}]{PRXQuantum.2.020337}%
  \BibitemOpen
  \bibfield  {author} {\bibinfo {author} {\bibfnamefont {N.~V.}\ \bibnamefont
  {Tkachenko}}, \bibinfo {author} {\bibfnamefont {J.}~\bibnamefont {Sud}},
  \bibinfo {author} {\bibfnamefont {Y.}~\bibnamefont {Zhang}}, \bibinfo
  {author} {\bibfnamefont {S.}~\bibnamefont {Tretiak}}, \bibinfo {author}
  {\bibfnamefont {P.~M.}\ \bibnamefont {Anisimov}}, \bibinfo {author}
  {\bibfnamefont {A.~T.}\ \bibnamefont {Arrasmith}}, \bibinfo {author}
  {\bibfnamefont {P.~J.}\ \bibnamefont {Coles}}, \bibinfo {author}
  {\bibfnamefont {L.}~\bibnamefont {Cincio}}, \ and\ \bibinfo {author}
  {\bibfnamefont {P.~A.}\ \bibnamefont {Dub}},\ }\href {\doibase
  10.1103/PRXQuantum.2.020337} {\bibfield  {journal} {\bibinfo  {journal} {PRX
  Quantum}\ }\textbf {\bibinfo {volume} {2}},\ \bibinfo {pages} {020337}
  (\bibinfo {year} {2021})}\BibitemShut {NoStop}%
\bibitem [{\citenamefont {Zhou}\ \emph {et~al.}(2020)\citenamefont {Zhou},
  \citenamefont {Wang}, \citenamefont {Choi}, \citenamefont {Pichler},\ and\
  \citenamefont {Lukin}}]{PhysRevX.10.021067}%
  \BibitemOpen
  \bibfield  {author} {\bibinfo {author} {\bibfnamefont {L.}~\bibnamefont
  {Zhou}}, \bibinfo {author} {\bibfnamefont {S.-T.}\ \bibnamefont {Wang}},
  \bibinfo {author} {\bibfnamefont {S.}~\bibnamefont {Choi}}, \bibinfo {author}
  {\bibfnamefont {H.}~\bibnamefont {Pichler}}, \ and\ \bibinfo {author}
  {\bibfnamefont {M.~D.}\ \bibnamefont {Lukin}},\ }\href {\doibase
  10.1103/PhysRevX.10.021067} {\bibfield  {journal} {\bibinfo  {journal} {Phys.
  Rev. X}\ }\textbf {\bibinfo {volume} {10}},\ \bibinfo {pages} {021067}
  (\bibinfo {year} {2020})}\BibitemShut {NoStop}%
\bibitem [{\citenamefont {Arrasmith}\ \emph {et~al.}(2019)\citenamefont
  {Arrasmith}, \citenamefont {Cincio}, \citenamefont {Sornborger},
  \citenamefont {Zurek},\ and\ \citenamefont {Coles}}]{Arrasmith2019}%
  \BibitemOpen
  \bibfield  {author} {\bibinfo {author} {\bibfnamefont {A.}~\bibnamefont
  {Arrasmith}}, \bibinfo {author} {\bibfnamefont {L.}~\bibnamefont {Cincio}},
  \bibinfo {author} {\bibfnamefont {A.~T.}\ \bibnamefont {Sornborger}},
  \bibinfo {author} {\bibfnamefont {W.~H.}\ \bibnamefont {Zurek}}, \ and\
  \bibinfo {author} {\bibfnamefont {P.~J.}\ \bibnamefont {Coles}},\ }\href
  {\doibase 10.1038/s41467-019-11417-0} {\bibfield  {journal} {\bibinfo
  {journal} {Nature Communications}\ }\textbf {\bibinfo {volume} {10}},\
  \bibinfo {pages} {3438} (\bibinfo {year} {2019})}\BibitemShut {NoStop}%
\bibitem [{\citenamefont {Kaubruegger}\ \emph {et~al.}(2019)\citenamefont
  {Kaubruegger}, \citenamefont {Silvi}, \citenamefont {Kokail}, \citenamefont
  {van Bijnen}, \citenamefont {Rey}, \citenamefont {Ye}, \citenamefont
  {Kaufman},\ and\ \citenamefont {Zoller}}]{PhysRevLett.123.260505}%
  \BibitemOpen
  \bibfield  {author} {\bibinfo {author} {\bibfnamefont {R.}~\bibnamefont
  {Kaubruegger}}, \bibinfo {author} {\bibfnamefont {P.}~\bibnamefont {Silvi}},
  \bibinfo {author} {\bibfnamefont {C.}~\bibnamefont {Kokail}}, \bibinfo
  {author} {\bibfnamefont {R.}~\bibnamefont {van Bijnen}}, \bibinfo {author}
  {\bibfnamefont {A.~M.}\ \bibnamefont {Rey}}, \bibinfo {author} {\bibfnamefont
  {J.}~\bibnamefont {Ye}}, \bibinfo {author} {\bibfnamefont {A.~M.}\
  \bibnamefont {Kaufman}}, \ and\ \bibinfo {author} {\bibfnamefont
  {P.}~\bibnamefont {Zoller}},\ }\href {\doibase
  10.1103/PhysRevLett.123.260505} {\bibfield  {journal} {\bibinfo  {journal}
  {Phys. Rev. Lett.}\ }\textbf {\bibinfo {volume} {123}},\ \bibinfo {pages}
  {260505} (\bibinfo {year} {2019})}\BibitemShut {NoStop}%
\bibitem [{\citenamefont {Koczor}\ \emph {et~al.}(2020)\citenamefont {Koczor},
  \citenamefont {Endo}, \citenamefont {Jones}, \citenamefont {Matsuzaki},\ and\
  \citenamefont {Benjamin}}]{Koczor_2020}%
  \BibitemOpen
  \bibfield  {author} {\bibinfo {author} {\bibfnamefont {B.}~\bibnamefont
  {Koczor}}, \bibinfo {author} {\bibfnamefont {S.}~\bibnamefont {Endo}},
  \bibinfo {author} {\bibfnamefont {T.}~\bibnamefont {Jones}}, \bibinfo
  {author} {\bibfnamefont {Y.}~\bibnamefont {Matsuzaki}}, \ and\ \bibinfo
  {author} {\bibfnamefont {S.~C.}\ \bibnamefont {Benjamin}},\ }\href {\doibase
  10.1088/1367-2630/ab965e} {\bibfield  {journal} {\bibinfo  {journal} {New
  Journal of Physics}\ }\textbf {\bibinfo {volume} {22}},\ \bibinfo {pages}
  {083038} (\bibinfo {year} {2020})}\BibitemShut {NoStop}%
\bibitem [{\citenamefont {Meyer}\ \emph {et~al.}(2021)\citenamefont {Meyer},
  \citenamefont {Borregaard},\ and\ \citenamefont {Eisert}}]{Meyer2021}%
  \BibitemOpen
  \bibfield  {author} {\bibinfo {author} {\bibfnamefont {J.~J.}\ \bibnamefont
  {Meyer}}, \bibinfo {author} {\bibfnamefont {J.}~\bibnamefont {Borregaard}}, \
  and\ \bibinfo {author} {\bibfnamefont {J.}~\bibnamefont {Eisert}},\ }\href
  {\doibase 10.1038/s41534-021-00425-y} {\bibfield  {journal} {\bibinfo
  {journal} {npj Quantum Information}\ }\textbf {\bibinfo {volume} {7}},\
  \bibinfo {pages} {89} (\bibinfo {year} {2021})}\BibitemShut {NoStop}%
\bibitem [{\citenamefont {Heya}\ \emph {et~al.}(2018)\citenamefont {Heya},
  \citenamefont {Suzuki}, \citenamefont {Nakamura},\ and\ \citenamefont
  {Fujii}}]{heya2018variational}%
  \BibitemOpen
  \bibfield  {author} {\bibinfo {author} {\bibfnamefont {K.}~\bibnamefont
  {Heya}}, \bibinfo {author} {\bibfnamefont {Y.}~\bibnamefont {Suzuki}},
  \bibinfo {author} {\bibfnamefont {Y.}~\bibnamefont {Nakamura}}, \ and\
  \bibinfo {author} {\bibfnamefont {K.}~\bibnamefont {Fujii}},\ }\href@noop {}
  {\bibfield  {journal} {\bibinfo  {journal} {arXiv preprint arXiv:1810.12745}\
  } (\bibinfo {year} {2018})}\BibitemShut {NoStop}%
\bibitem [{\citenamefont {Khatri}\ \emph {et~al.}(2019)\citenamefont {Khatri},
  \citenamefont {LaRose}, \citenamefont {Poremba}, \citenamefont {Cincio},
  \citenamefont {Sornborger},\ and\ \citenamefont
  {Coles}}]{Khatri2019quantumassisted}%
  \BibitemOpen
  \bibfield  {author} {\bibinfo {author} {\bibfnamefont {S.}~\bibnamefont
  {Khatri}}, \bibinfo {author} {\bibfnamefont {R.}~\bibnamefont {LaRose}},
  \bibinfo {author} {\bibfnamefont {A.}~\bibnamefont {Poremba}}, \bibinfo
  {author} {\bibfnamefont {L.}~\bibnamefont {Cincio}}, \bibinfo {author}
  {\bibfnamefont {A.~T.}\ \bibnamefont {Sornborger}}, \ and\ \bibinfo {author}
  {\bibfnamefont {P.~J.}\ \bibnamefont {Coles}},\ }\href {\doibase
  10.22331/q-2019-05-13-140} {\bibfield  {journal} {\bibinfo  {journal}
  {{Quantum}}\ }\textbf {\bibinfo {volume} {3}},\ \bibinfo {pages} {140}
  (\bibinfo {year} {2019})}\BibitemShut {NoStop}%
\bibitem [{\citenamefont {Volkoff}\ \emph {et~al.}(2021)\citenamefont
  {Volkoff}, \citenamefont {Holmes},\ and\ \citenamefont
  {Sornborger}}]{PRXQuantum.2.040327}%
  \BibitemOpen
  \bibfield  {author} {\bibinfo {author} {\bibfnamefont {T.}~\bibnamefont
  {Volkoff}}, \bibinfo {author} {\bibfnamefont {Z.}~\bibnamefont {Holmes}}, \
  and\ \bibinfo {author} {\bibfnamefont {A.}~\bibnamefont {Sornborger}},\
  }\href {\doibase 10.1103/PRXQuantum.2.040327} {\bibfield  {journal} {\bibinfo
   {journal} {PRX Quantum}\ }\textbf {\bibinfo {volume} {2}},\ \bibinfo {pages}
  {040327} (\bibinfo {year} {2021})}\BibitemShut {NoStop}%
\bibitem [{\citenamefont {Jones}\ and\ \citenamefont
  {Benjamin}(2022)}]{Jones2022robustquantum}%
  \BibitemOpen
  \bibfield  {author} {\bibinfo {author} {\bibfnamefont {T.}~\bibnamefont
  {Jones}}\ and\ \bibinfo {author} {\bibfnamefont {S.~C.}\ \bibnamefont
  {Benjamin}},\ }\href {\doibase 10.22331/q-2022-01-24-628} {\bibfield
  {journal} {\bibinfo  {journal} {{Quantum}}\ }\textbf {\bibinfo {volume}
  {6}},\ \bibinfo {pages} {628} (\bibinfo {year} {2022})}\BibitemShut {NoStop}%
\bibitem [{\citenamefont {Maronese}\ \emph {et~al.}(2021)\citenamefont
  {Maronese}, \citenamefont {Moro}, \citenamefont {Rocutto},\ and\
  \citenamefont {Prati}}]{https://doi.org/10.48550/arxiv.2112.00187}%
  \BibitemOpen
  \bibfield  {author} {\bibinfo {author} {\bibfnamefont {M.}~\bibnamefont
  {Maronese}}, \bibinfo {author} {\bibfnamefont {L.}~\bibnamefont {Moro}},
  \bibinfo {author} {\bibfnamefont {L.}~\bibnamefont {Rocutto}}, \ and\
  \bibinfo {author} {\bibfnamefont {E.}~\bibnamefont {Prati}},\ }\href
  {\doibase 10.48550/ARXIV.2112.00187} {\enquote {\bibinfo {title} {Quantum
  compiling},}\ } (\bibinfo {year} {2021})\BibitemShut {NoStop}%
\bibitem [{\citenamefont {Aulicino}\ \emph {et~al.}(2022)\citenamefont
  {Aulicino}, \citenamefont {Keen},\ and\ \citenamefont
  {Peng}}]{aulicino2022state}%
  \BibitemOpen
  \bibfield  {author} {\bibinfo {author} {\bibfnamefont {J.~C.}\ \bibnamefont
  {Aulicino}}, \bibinfo {author} {\bibfnamefont {T.}~\bibnamefont {Keen}}, \
  and\ \bibinfo {author} {\bibfnamefont {B.}~\bibnamefont {Peng}},\ }\href@noop
  {} {\bibfield  {journal} {\bibinfo  {journal} {International Journal of
  Quantum Chemistry}\ }\textbf {\bibinfo {volume} {122}},\ \bibinfo {pages}
  {e26853} (\bibinfo {year} {2022})}\BibitemShut {NoStop}%
\bibitem [{\citenamefont {Ashhab}(2022)}]{PhysRevResearch.4.013091}%
  \BibitemOpen
  \bibfield  {author} {\bibinfo {author} {\bibfnamefont {S.}~\bibnamefont
  {Ashhab}},\ }\href {\doibase 10.1103/PhysRevResearch.4.013091} {\bibfield
  {journal} {\bibinfo  {journal} {Phys. Rev. Research}\ }\textbf {\bibinfo
  {volume} {4}},\ \bibinfo {pages} {013091} (\bibinfo {year}
  {2022})}\BibitemShut {NoStop}%
\bibitem [{\citenamefont {Kuzmin}\ and\ \citenamefont
  {Silvi}(2020)}]{Kuzmin2020variationalquantum}%
  \BibitemOpen
  \bibfield  {author} {\bibinfo {author} {\bibfnamefont {V.~V.}\ \bibnamefont
  {Kuzmin}}\ and\ \bibinfo {author} {\bibfnamefont {P.}~\bibnamefont {Silvi}},\
  }\href {\doibase 10.22331/q-2020-07-06-290} {\bibfield  {journal} {\bibinfo
  {journal} {{Quantum}}\ }\textbf {\bibinfo {volume} {4}},\ \bibinfo {pages}
  {290} (\bibinfo {year} {2020})}\BibitemShut {NoStop}%
\bibitem [{\citenamefont {Lvovsky}\ and\ \citenamefont
  {Raymer}(2009)}]{lvovsky2009continuous}%
  \BibitemOpen
  \bibfield  {author} {\bibinfo {author} {\bibfnamefont {A.~I.}\ \bibnamefont
  {Lvovsky}}\ and\ \bibinfo {author} {\bibfnamefont {M.~G.}\ \bibnamefont
  {Raymer}},\ }\href@noop {} {\bibfield  {journal} {\bibinfo  {journal}
  {Reviews of modern physics}\ }\textbf {\bibinfo {volume} {81}},\ \bibinfo
  {pages} {299} (\bibinfo {year} {2009})}\BibitemShut {NoStop}%
\bibitem [{\citenamefont {D'Ariano}\ \emph {et~al.}(2002)\citenamefont
  {D'Ariano}, \citenamefont {De~Laurentis}, \citenamefont {Paris},
  \citenamefont {Porzio},\ and\ \citenamefont {Solimeno}}]{d2002quantum}%
  \BibitemOpen
  \bibfield  {author} {\bibinfo {author} {\bibfnamefont {G.~M.}\ \bibnamefont
  {D'Ariano}}, \bibinfo {author} {\bibfnamefont {M.}~\bibnamefont
  {De~Laurentis}}, \bibinfo {author} {\bibfnamefont {M.~G.}\ \bibnamefont
  {Paris}}, \bibinfo {author} {\bibfnamefont {A.}~\bibnamefont {Porzio}}, \
  and\ \bibinfo {author} {\bibfnamefont {S.}~\bibnamefont {Solimeno}},\
  }\href@noop {} {\bibfield  {journal} {\bibinfo  {journal} {Journal of Optics
  B: Quantum and Semiclassical Optics}\ }\textbf {\bibinfo {volume} {4}},\
  \bibinfo {pages} {S127} (\bibinfo {year} {2002})}\BibitemShut {NoStop}%
\bibitem [{\citenamefont {Takeda}\ \emph {et~al.}(2021)\citenamefont {Takeda},
  \citenamefont {Noiri}, \citenamefont {Nakajima}, \citenamefont {Yoneda},
  \citenamefont {Kobayashi},\ and\ \citenamefont
  {Tarucha}}]{takeda2021quantum}%
  \BibitemOpen
  \bibfield  {author} {\bibinfo {author} {\bibfnamefont {K.}~\bibnamefont
  {Takeda}}, \bibinfo {author} {\bibfnamefont {A.}~\bibnamefont {Noiri}},
  \bibinfo {author} {\bibfnamefont {T.}~\bibnamefont {Nakajima}}, \bibinfo
  {author} {\bibfnamefont {J.}~\bibnamefont {Yoneda}}, \bibinfo {author}
  {\bibfnamefont {T.}~\bibnamefont {Kobayashi}}, \ and\ \bibinfo {author}
  {\bibfnamefont {S.}~\bibnamefont {Tarucha}},\ }\href@noop {} {\bibfield
  {journal} {\bibinfo  {journal} {Nature Nanotechnology}\ ,\ \bibinfo {pages}
  {1}} (\bibinfo {year} {2021})}\BibitemShut {NoStop}%
\bibitem [{\citenamefont {M\"{o}tt\"{o}nen}\ \emph {et~al.}(2005)\citenamefont
  {M\"{o}tt\"{o}nen}, \citenamefont {Vartiainen}, \citenamefont {Bergholm},\
  and\ \citenamefont {Salomaa}}]{10.5555/2011670.2011675}%
  \BibitemOpen
  \bibfield  {author} {\bibinfo {author} {\bibfnamefont {M.}~\bibnamefont
  {M\"{o}tt\"{o}nen}}, \bibinfo {author} {\bibfnamefont {J.~J.}\ \bibnamefont
  {Vartiainen}}, \bibinfo {author} {\bibfnamefont {V.}~\bibnamefont
  {Bergholm}}, \ and\ \bibinfo {author} {\bibfnamefont {M.~M.}\ \bibnamefont
  {Salomaa}},\ }\href@noop {} {\bibfield  {journal} {\bibinfo  {journal}
  {Quantum Info. Comput.}\ }\textbf {\bibinfo {volume} {5}},\ \bibinfo {pages}
  {467–473} (\bibinfo {year} {2005})}\BibitemShut {NoStop}%
\bibitem [{\citenamefont {Shende}\ \emph {et~al.}(2006)\citenamefont {Shende},
  \citenamefont {Bullock},\ and\ \citenamefont {Markov}}]{1629135}%
  \BibitemOpen
  \bibfield  {author} {\bibinfo {author} {\bibfnamefont {V.}~\bibnamefont
  {Shende}}, \bibinfo {author} {\bibfnamefont {S.}~\bibnamefont {Bullock}}, \
  and\ \bibinfo {author} {\bibfnamefont {I.}~\bibnamefont {Markov}},\ }\href
  {\doibase 10.1109/TCAD.2005.855930} {\bibfield  {journal} {\bibinfo
  {journal} {IEEE Transactions on Computer-Aided Design of Integrated Circuits
  and Systems}\ }\textbf {\bibinfo {volume} {25}},\ \bibinfo {pages} {1000}
  (\bibinfo {year} {2006})}\BibitemShut {NoStop}%
\bibitem [{\citenamefont {Iten}\ \emph {et~al.}(2016)\citenamefont {Iten},
  \citenamefont {Colbeck}, \citenamefont {Kukuljan}, \citenamefont {Home},\
  and\ \citenamefont {Christandl}}]{PhysRevA.93.032318}%
  \BibitemOpen
  \bibfield  {author} {\bibinfo {author} {\bibfnamefont {R.}~\bibnamefont
  {Iten}}, \bibinfo {author} {\bibfnamefont {R.}~\bibnamefont {Colbeck}},
  \bibinfo {author} {\bibfnamefont {I.}~\bibnamefont {Kukuljan}}, \bibinfo
  {author} {\bibfnamefont {J.}~\bibnamefont {Home}}, \ and\ \bibinfo {author}
  {\bibfnamefont {M.}~\bibnamefont {Christandl}},\ }\href {\doibase
  10.1103/PhysRevA.93.032318} {\bibfield  {journal} {\bibinfo  {journal} {Phys.
  Rev. A}\ }\textbf {\bibinfo {volume} {93}},\ \bibinfo {pages} {032318}
  (\bibinfo {year} {2016})}\BibitemShut {NoStop}%
\bibitem [{\citenamefont {Plesch}\ and\ \citenamefont
  {Brukner}(2011)}]{PhysRevA.83.032302}%
  \BibitemOpen
  \bibfield  {author} {\bibinfo {author} {\bibfnamefont {M.}~\bibnamefont
  {Plesch}}\ and\ \bibinfo {author} {\bibfnamefont {i.~c.~v.}\ \bibnamefont
  {Brukner}},\ }\href {\doibase 10.1103/PhysRevA.83.032302} {\bibfield
  {journal} {\bibinfo  {journal} {Phys. Rev. A}\ }\textbf {\bibinfo {volume}
  {83}},\ \bibinfo {pages} {032302} (\bibinfo {year} {2011})}\BibitemShut
  {NoStop}%
\bibitem [{\citenamefont {Sun}\ \emph {et~al.}(2021)\citenamefont {Sun},
  \citenamefont {Tian}, \citenamefont {Yang}, \citenamefont {Yuan},\ and\
  \citenamefont {Zhang}}]{sun2021asymptotically}%
  \BibitemOpen
  \bibfield  {author} {\bibinfo {author} {\bibfnamefont {X.}~\bibnamefont
  {Sun}}, \bibinfo {author} {\bibfnamefont {G.}~\bibnamefont {Tian}}, \bibinfo
  {author} {\bibfnamefont {S.}~\bibnamefont {Yang}}, \bibinfo {author}
  {\bibfnamefont {P.}~\bibnamefont {Yuan}}, \ and\ \bibinfo {author}
  {\bibfnamefont {S.}~\bibnamefont {Zhang}},\ }\href@noop {} {\enquote
  {\bibinfo {title} {Asymptotically optimal circuit depth for quantum state
  preparation and general unitary synthesis},}\ } (\bibinfo {year} {2021}),\
  \Eprint {http://arxiv.org/abs/2108.06150} {arXiv:2108.06150 [quant-ph]}
  \BibitemShut {NoStop}%
\bibitem [{\citenamefont {Zhang}\ \emph {et~al.}(2021)\citenamefont {Zhang},
  \citenamefont {Yung},\ and\ \citenamefont {Yuan}}]{PhysRevResearch.3.043200}%
  \BibitemOpen
  \bibfield  {author} {\bibinfo {author} {\bibfnamefont {X.-M.}\ \bibnamefont
  {Zhang}}, \bibinfo {author} {\bibfnamefont {M.-H.}\ \bibnamefont {Yung}}, \
  and\ \bibinfo {author} {\bibfnamefont {X.}~\bibnamefont {Yuan}},\ }\href
  {\doibase 10.1103/PhysRevResearch.3.043200} {\bibfield  {journal} {\bibinfo
  {journal} {Phys. Rev. Research}\ }\textbf {\bibinfo {volume} {3}},\ \bibinfo
  {pages} {043200} (\bibinfo {year} {2021})}\BibitemShut {NoStop}%
\bibitem [{\citenamefont {Rosenthal}(2022)}]{rosenthal2022query}%
  \BibitemOpen
  \bibfield  {author} {\bibinfo {author} {\bibfnamefont {G.}~\bibnamefont
  {Rosenthal}},\ }\href@noop {} {\enquote {\bibinfo {title} {Query and depth
  upper bounds for quantum unitaries via grover search},}\ } (\bibinfo {year}
  {2022}),\ \Eprint {http://arxiv.org/abs/2111.07992} {arXiv:2111.07992
  [quant-ph]} \BibitemShut {NoStop}%
\bibitem [{\citenamefont {Araujo}\ \emph {et~al.}(2021)\citenamefont {Araujo},
  \citenamefont {Park}, \citenamefont {Petruccione},\ and\ \citenamefont
  {da~Silva}}]{Araujo2021}%
  \BibitemOpen
  \bibfield  {author} {\bibinfo {author} {\bibfnamefont {I.~F.}\ \bibnamefont
  {Araujo}}, \bibinfo {author} {\bibfnamefont {D.~K.}\ \bibnamefont {Park}},
  \bibinfo {author} {\bibfnamefont {F.}~\bibnamefont {Petruccione}}, \ and\
  \bibinfo {author} {\bibfnamefont {A.~J.}\ \bibnamefont {da~Silva}},\ }\href
  {\doibase 10.1038/s41598-021-85474-1} {\bibfield  {journal} {\bibinfo
  {journal} {Scientific Reports}\ }\textbf {\bibinfo {volume} {11}},\ \bibinfo
  {pages} {6329} (\bibinfo {year} {2021})}\BibitemShut {NoStop}%
\bibitem [{\citenamefont {Palmieri}\ \emph {et~al.}(2020)\citenamefont
  {Palmieri}, \citenamefont {Kovlakov}, \citenamefont {Bianchi}, \citenamefont
  {Yudin}, \citenamefont {Straupe}, \citenamefont {Biamonte},\ and\
  \citenamefont {Kulik}}]{Palmieri2020}%
  \BibitemOpen
  \bibfield  {author} {\bibinfo {author} {\bibfnamefont {A.~M.}\ \bibnamefont
  {Palmieri}}, \bibinfo {author} {\bibfnamefont {E.}~\bibnamefont {Kovlakov}},
  \bibinfo {author} {\bibfnamefont {F.}~\bibnamefont {Bianchi}}, \bibinfo
  {author} {\bibfnamefont {D.}~\bibnamefont {Yudin}}, \bibinfo {author}
  {\bibfnamefont {S.}~\bibnamefont {Straupe}}, \bibinfo {author} {\bibfnamefont
  {J.~D.}\ \bibnamefont {Biamonte}}, \ and\ \bibinfo {author} {\bibfnamefont
  {S.}~\bibnamefont {Kulik}},\ }\href {\doibase 10.1038/s41534-020-0248-6}
  {\bibfield  {journal} {\bibinfo  {journal} {npj Quantum Information}\
  }\textbf {\bibinfo {volume} {6}},\ \bibinfo {pages} {20} (\bibinfo {year}
  {2020})}\BibitemShut {NoStop}%
\bibitem [{\citenamefont {Cramer}\ \emph {et~al.}(2010)\citenamefont {Cramer},
  \citenamefont {Plenio}, \citenamefont {Flammia}, \citenamefont {Somma},
  \citenamefont {Gross}, \citenamefont {Bartlett}, \citenamefont
  {Landon-Cardinal}, \citenamefont {Poulin},\ and\ \citenamefont
  {Liu}}]{Cramer2010}%
  \BibitemOpen
  \bibfield  {author} {\bibinfo {author} {\bibfnamefont {M.}~\bibnamefont
  {Cramer}}, \bibinfo {author} {\bibfnamefont {M.~B.}\ \bibnamefont {Plenio}},
  \bibinfo {author} {\bibfnamefont {S.~T.}\ \bibnamefont {Flammia}}, \bibinfo
  {author} {\bibfnamefont {R.}~\bibnamefont {Somma}}, \bibinfo {author}
  {\bibfnamefont {D.}~\bibnamefont {Gross}}, \bibinfo {author} {\bibfnamefont
  {S.~D.}\ \bibnamefont {Bartlett}}, \bibinfo {author} {\bibfnamefont
  {O.}~\bibnamefont {Landon-Cardinal}}, \bibinfo {author} {\bibfnamefont
  {D.}~\bibnamefont {Poulin}}, \ and\ \bibinfo {author} {\bibfnamefont {Y.-K.}\
  \bibnamefont {Liu}},\ }\href {\doibase 10.1038/ncomms1147} {\bibfield
  {journal} {\bibinfo  {journal} {Nature Communications}\ }\textbf {\bibinfo
  {volume} {1}},\ \bibinfo {pages} {149} (\bibinfo {year} {2010})}\BibitemShut
  {NoStop}%
\bibitem [{\citenamefont {Jackson}\ and\ \citenamefont {van
  Enk}(2015)}]{PhysRevA.92.042312}%
  \BibitemOpen
  \bibfield  {author} {\bibinfo {author} {\bibfnamefont {C.}~\bibnamefont
  {Jackson}}\ and\ \bibinfo {author} {\bibfnamefont {S.~J.}\ \bibnamefont {van
  Enk}},\ }\href {\doibase 10.1103/PhysRevA.92.042312} {\bibfield  {journal}
  {\bibinfo  {journal} {Phys. Rev. A}\ }\textbf {\bibinfo {volume} {92}},\
  \bibinfo {pages} {042312} (\bibinfo {year} {2015})}\BibitemShut {NoStop}%
\bibitem [{\citenamefont {Moroder}\ \emph {et~al.}(2012)\citenamefont
  {Moroder}, \citenamefont {Hyllus}, \citenamefont {T{\'{o}}th}, \citenamefont
  {Schwemmer}, \citenamefont {Niggebaum}, \citenamefont {Gaile}, \citenamefont
  {Gühne},\ and\ \citenamefont {Weinfurter}}]{Moroder_2012}%
  \BibitemOpen
  \bibfield  {author} {\bibinfo {author} {\bibfnamefont {T.}~\bibnamefont
  {Moroder}}, \bibinfo {author} {\bibfnamefont {P.}~\bibnamefont {Hyllus}},
  \bibinfo {author} {\bibfnamefont {G.}~\bibnamefont {T{\'{o}}th}}, \bibinfo
  {author} {\bibfnamefont {C.}~\bibnamefont {Schwemmer}}, \bibinfo {author}
  {\bibfnamefont {A.}~\bibnamefont {Niggebaum}}, \bibinfo {author}
  {\bibfnamefont {S.}~\bibnamefont {Gaile}}, \bibinfo {author} {\bibfnamefont
  {O.}~\bibnamefont {Gühne}}, \ and\ \bibinfo {author} {\bibfnamefont
  {H.}~\bibnamefont {Weinfurter}},\ }\href {\doibase
  10.1088/1367-2630/14/10/105001} {\bibfield  {journal} {\bibinfo  {journal}
  {New Journal of Physics}\ }\textbf {\bibinfo {volume} {14}},\ \bibinfo
  {pages} {105001} (\bibinfo {year} {2012})}\BibitemShut {NoStop}%
\bibitem [{\citenamefont {Ahmed}\ \emph {et~al.}(2021)\citenamefont {Ahmed},
  \citenamefont {S\'anchez Mu\~noz}, \citenamefont {Nori},\ and\ \citenamefont
  {Kockum}}]{PhysRevResearch.3.033278}%
  \BibitemOpen
  \bibfield  {author} {\bibinfo {author} {\bibfnamefont {S.}~\bibnamefont
  {Ahmed}}, \bibinfo {author} {\bibfnamefont {C.}~\bibnamefont {S\'anchez
  Mu\~noz}}, \bibinfo {author} {\bibfnamefont {F.}~\bibnamefont {Nori}}, \ and\
  \bibinfo {author} {\bibfnamefont {A.~F.}\ \bibnamefont {Kockum}},\ }\href
  {\doibase 10.1103/PhysRevResearch.3.033278} {\bibfield  {journal} {\bibinfo
  {journal} {Phys. Rev. Research}\ }\textbf {\bibinfo {volume} {3}},\ \bibinfo
  {pages} {033278} (\bibinfo {year} {2021})}\BibitemShut {NoStop}%
\bibitem [{\citenamefont {T\'oth}\ \emph {et~al.}(2010)\citenamefont {T\'oth},
  \citenamefont {Wieczorek}, \citenamefont {Gross}, \citenamefont {Krischek},
  \citenamefont {Schwemmer},\ and\ \citenamefont
  {Weinfurter}}]{PhysRevLett.105.250403}%
  \BibitemOpen
  \bibfield  {author} {\bibinfo {author} {\bibfnamefont {G.}~\bibnamefont
  {T\'oth}}, \bibinfo {author} {\bibfnamefont {W.}~\bibnamefont {Wieczorek}},
  \bibinfo {author} {\bibfnamefont {D.}~\bibnamefont {Gross}}, \bibinfo
  {author} {\bibfnamefont {R.}~\bibnamefont {Krischek}}, \bibinfo {author}
  {\bibfnamefont {C.}~\bibnamefont {Schwemmer}}, \ and\ \bibinfo {author}
  {\bibfnamefont {H.}~\bibnamefont {Weinfurter}},\ }\href {\doibase
  10.1103/PhysRevLett.105.250403} {\bibfield  {journal} {\bibinfo  {journal}
  {Phys. Rev. Lett.}\ }\textbf {\bibinfo {volume} {105}},\ \bibinfo {pages}
  {250403} (\bibinfo {year} {2010})}\BibitemShut {NoStop}%
\bibitem [{\citenamefont {Torlai}\ \emph {et~al.}(2018)\citenamefont {Torlai},
  \citenamefont {Mazzola}, \citenamefont {Carrasquilla}, \citenamefont
  {Troyer}, \citenamefont {Melko},\ and\ \citenamefont
  {Carleo}}]{Torlai2018-sm}%
  \BibitemOpen
  \bibfield  {author} {\bibinfo {author} {\bibfnamefont {G.}~\bibnamefont
  {Torlai}}, \bibinfo {author} {\bibfnamefont {G.}~\bibnamefont {Mazzola}},
  \bibinfo {author} {\bibfnamefont {J.}~\bibnamefont {Carrasquilla}}, \bibinfo
  {author} {\bibfnamefont {M.}~\bibnamefont {Troyer}}, \bibinfo {author}
  {\bibfnamefont {R.}~\bibnamefont {Melko}}, \ and\ \bibinfo {author}
  {\bibfnamefont {G.}~\bibnamefont {Carleo}},\ }\href@noop {} {\bibfield
  {journal} {\bibinfo  {journal} {Nature Physics}\ }\textbf {\bibinfo {volume}
  {14}},\ \bibinfo {pages} {447} (\bibinfo {year} {2018})}\BibitemShut
  {NoStop}%
\bibitem [{\citenamefont {Blume-Kohout}(2010)}]{Blume_Kohout_2010}%
  \BibitemOpen
  \bibfield  {author} {\bibinfo {author} {\bibfnamefont {R.}~\bibnamefont
  {Blume-Kohout}},\ }\href {\doibase 10.1088/1367-2630/12/4/043034} {\bibfield
  {journal} {\bibinfo  {journal} {New Journal of Physics}\ }\textbf {\bibinfo
  {volume} {12}},\ \bibinfo {pages} {043034} (\bibinfo {year}
  {2010})}\BibitemShut {NoStop}%
\bibitem [{\citenamefont {Fiderer}\ \emph {et~al.}(2021)\citenamefont
  {Fiderer}, \citenamefont {Schuff},\ and\ \citenamefont
  {Braun}}]{PRXQuantum.2.020303}%
  \BibitemOpen
  \bibfield  {author} {\bibinfo {author} {\bibfnamefont {L.~J.}\ \bibnamefont
  {Fiderer}}, \bibinfo {author} {\bibfnamefont {J.}~\bibnamefont {Schuff}}, \
  and\ \bibinfo {author} {\bibfnamefont {D.}~\bibnamefont {Braun}},\ }\href
  {\doibase 10.1103/PRXQuantum.2.020303} {\bibfield  {journal} {\bibinfo
  {journal} {PRX Quantum}\ }\textbf {\bibinfo {volume} {2}},\ \bibinfo {pages}
  {020303} (\bibinfo {year} {2021})}\BibitemShut {NoStop}%
\bibitem [{\citenamefont {Gross}\ \emph {et~al.}(2010)\citenamefont {Gross},
  \citenamefont {Liu}, \citenamefont {Flammia}, \citenamefont {Becker},\ and\
  \citenamefont {Eisert}}]{PhysRevLett.105.150401}%
  \BibitemOpen
  \bibfield  {author} {\bibinfo {author} {\bibfnamefont {D.}~\bibnamefont
  {Gross}}, \bibinfo {author} {\bibfnamefont {Y.-K.}\ \bibnamefont {Liu}},
  \bibinfo {author} {\bibfnamefont {S.~T.}\ \bibnamefont {Flammia}}, \bibinfo
  {author} {\bibfnamefont {S.}~\bibnamefont {Becker}}, \ and\ \bibinfo {author}
  {\bibfnamefont {J.}~\bibnamefont {Eisert}},\ }\href {\doibase
  10.1103/PhysRevLett.105.150401} {\bibfield  {journal} {\bibinfo  {journal}
  {Phys. Rev. Lett.}\ }\textbf {\bibinfo {volume} {105}},\ \bibinfo {pages}
  {150401} (\bibinfo {year} {2010})}\BibitemShut {NoStop}%
\bibitem [{\citenamefont {Flammia}\ \emph {et~al.}(2012)\citenamefont
  {Flammia}, \citenamefont {Gross}, \citenamefont {Liu},\ and\ \citenamefont
  {Eisert}}]{Flammia_2012}%
  \BibitemOpen
  \bibfield  {author} {\bibinfo {author} {\bibfnamefont {S.~T.}\ \bibnamefont
  {Flammia}}, \bibinfo {author} {\bibfnamefont {D.}~\bibnamefont {Gross}},
  \bibinfo {author} {\bibfnamefont {Y.-K.}\ \bibnamefont {Liu}}, \ and\
  \bibinfo {author} {\bibfnamefont {J.}~\bibnamefont {Eisert}},\ }\href
  {\doibase 10.1088/1367-2630/14/9/095022} {\bibfield  {journal} {\bibinfo
  {journal} {New Journal of Physics}\ }\textbf {\bibinfo {volume} {14}},\
  \bibinfo {pages} {095022} (\bibinfo {year} {2012})}\BibitemShut {NoStop}%
\bibitem [{\citenamefont {Czerwinski}(2020)}]{Czerwinski2020}%
  \BibitemOpen
  \bibfield  {author} {\bibinfo {author} {\bibfnamefont {A.}~\bibnamefont
  {Czerwinski}},\ }\href {\doibase 10.1007/s10773-020-04625-8} {\bibfield
  {journal} {\bibinfo  {journal} {International Journal of Theoretical
  Physics}\ }\textbf {\bibinfo {volume} {59}},\ \bibinfo {pages} {3646}
  (\bibinfo {year} {2020})}\BibitemShut {NoStop}%
\bibitem [{\citenamefont {Flurin}\ \emph {et~al.}(2020)\citenamefont {Flurin},
  \citenamefont {Martin}, \citenamefont {Hacohen-Gourgy},\ and\ \citenamefont
  {Siddiqi}}]{PhysRevX.10.011006}%
  \BibitemOpen
  \bibfield  {author} {\bibinfo {author} {\bibfnamefont {E.}~\bibnamefont
  {Flurin}}, \bibinfo {author} {\bibfnamefont {L.~S.}\ \bibnamefont {Martin}},
  \bibinfo {author} {\bibfnamefont {S.}~\bibnamefont {Hacohen-Gourgy}}, \ and\
  \bibinfo {author} {\bibfnamefont {I.}~\bibnamefont {Siddiqi}},\ }\href
  {\doibase 10.1103/PhysRevX.10.011006} {\bibfield  {journal} {\bibinfo
  {journal} {Phys. Rev. X}\ }\textbf {\bibinfo {volume} {10}},\ \bibinfo
  {pages} {011006} (\bibinfo {year} {2020})}\BibitemShut {NoStop}%
\bibitem [{\citenamefont {M\"akinen}\ \emph {et~al.}(2019)\citenamefont
  {M\"akinen}, \citenamefont {Ikonen}, \citenamefont {Partanen},\ and\
  \citenamefont {M\"ott\"onen}}]{PhysRevA.100.042109}%
  \BibitemOpen
  \bibfield  {author} {\bibinfo {author} {\bibfnamefont {A.}~\bibnamefont
  {M\"akinen}}, \bibinfo {author} {\bibfnamefont {J.}~\bibnamefont {Ikonen}},
  \bibinfo {author} {\bibfnamefont {M.}~\bibnamefont {Partanen}}, \ and\
  \bibinfo {author} {\bibfnamefont {M.}~\bibnamefont {M\"ott\"onen}},\ }\href
  {\doibase 10.1103/PhysRevA.100.042109} {\bibfield  {journal} {\bibinfo
  {journal} {Phys. Rev. A}\ }\textbf {\bibinfo {volume} {100}},\ \bibinfo
  {pages} {042109} (\bibinfo {year} {2019})}\BibitemShut {NoStop}%
\bibitem [{\citenamefont {Liu}\ \emph {et~al.}(2020)\citenamefont {Liu},
  \citenamefont {Wang}, \citenamefont {Xue}, \citenamefont {Huang},
  \citenamefont {Fu}, \citenamefont {Qiang}, \citenamefont {Xu}, \citenamefont
  {Huang}, \citenamefont {Deng}, \citenamefont {Guo}, \citenamefont {Yang},\
  and\ \citenamefont {Wu}}]{PhysRevA.101.052316}%
  \BibitemOpen
  \bibfield  {author} {\bibinfo {author} {\bibfnamefont {Y.}~\bibnamefont
  {Liu}}, \bibinfo {author} {\bibfnamefont {D.}~\bibnamefont {Wang}}, \bibinfo
  {author} {\bibfnamefont {S.}~\bibnamefont {Xue}}, \bibinfo {author}
  {\bibfnamefont {A.}~\bibnamefont {Huang}}, \bibinfo {author} {\bibfnamefont
  {X.}~\bibnamefont {Fu}}, \bibinfo {author} {\bibfnamefont {X.}~\bibnamefont
  {Qiang}}, \bibinfo {author} {\bibfnamefont {P.}~\bibnamefont {Xu}}, \bibinfo
  {author} {\bibfnamefont {H.-L.}\ \bibnamefont {Huang}}, \bibinfo {author}
  {\bibfnamefont {M.}~\bibnamefont {Deng}}, \bibinfo {author} {\bibfnamefont
  {C.}~\bibnamefont {Guo}}, \bibinfo {author} {\bibfnamefont {X.}~\bibnamefont
  {Yang}}, \ and\ \bibinfo {author} {\bibfnamefont {J.}~\bibnamefont {Wu}},\
  }\href {\doibase 10.1103/PhysRevA.101.052316} {\bibfield  {journal} {\bibinfo
   {journal} {Phys. Rev. A}\ }\textbf {\bibinfo {volume} {101}},\ \bibinfo
  {pages} {052316} (\bibinfo {year} {2020})}\BibitemShut {NoStop}%
\bibitem [{\citenamefont {Lee}\ \emph {et~al.}(2018)\citenamefont {Lee},
  \citenamefont {Lee},\ and\ \citenamefont {Bang}}]{PhysRevA.98.052302}%
  \BibitemOpen
  \bibfield  {author} {\bibinfo {author} {\bibfnamefont {S.~M.}\ \bibnamefont
  {Lee}}, \bibinfo {author} {\bibfnamefont {J.}~\bibnamefont {Lee}}, \ and\
  \bibinfo {author} {\bibfnamefont {J.}~\bibnamefont {Bang}},\ }\href {\doibase
  10.1103/PhysRevA.98.052302} {\bibfield  {journal} {\bibinfo  {journal} {Phys.
  Rev. A}\ }\textbf {\bibinfo {volume} {98}},\ \bibinfo {pages} {052302}
  (\bibinfo {year} {2018})}\BibitemShut {NoStop}%
\bibitem [{\citenamefont {Lee}\ \emph {et~al.}(2021)\citenamefont {Lee},
  \citenamefont {Park}, \citenamefont {Lee}, \citenamefont {Kim},\ and\
  \citenamefont {Bang}}]{PhysRevLett.126.170504}%
  \BibitemOpen
  \bibfield  {author} {\bibinfo {author} {\bibfnamefont {S.~M.}\ \bibnamefont
  {Lee}}, \bibinfo {author} {\bibfnamefont {H.~S.}\ \bibnamefont {Park}},
  \bibinfo {author} {\bibfnamefont {J.}~\bibnamefont {Lee}}, \bibinfo {author}
  {\bibfnamefont {J.}~\bibnamefont {Kim}}, \ and\ \bibinfo {author}
  {\bibfnamefont {J.}~\bibnamefont {Bang}},\ }\href {\doibase
  10.1103/PhysRevLett.126.170504} {\bibfield  {journal} {\bibinfo  {journal}
  {Phys. Rev. Lett.}\ }\textbf {\bibinfo {volume} {126}},\ \bibinfo {pages}
  {170504} (\bibinfo {year} {2021})}\BibitemShut {NoStop}%
\bibitem [{\citenamefont
  {Aaronson}(2017)}]{https://doi.org/10.48550/arxiv.1711.01053}%
  \BibitemOpen
  \bibfield  {author} {\bibinfo {author} {\bibfnamefont {S.}~\bibnamefont
  {Aaronson}},\ }\href {\doibase 10.48550/ARXIV.1711.01053} {\enquote {\bibinfo
  {title} {Shadow tomography of quantum states},}\ } (\bibinfo {year}
  {2017})\BibitemShut {NoStop}%
\bibitem [{\citenamefont {Huang}\ \emph {et~al.}(2020)\citenamefont {Huang},
  \citenamefont {Kueng},\ and\ \citenamefont {Preskill}}]{Huang2020}%
  \BibitemOpen
  \bibfield  {author} {\bibinfo {author} {\bibfnamefont {H.-Y.}\ \bibnamefont
  {Huang}}, \bibinfo {author} {\bibfnamefont {R.}~\bibnamefont {Kueng}}, \ and\
  \bibinfo {author} {\bibfnamefont {J.}~\bibnamefont {Preskill}},\ }\href
  {\doibase 10.1038/s41567-020-0932-7} {\bibfield  {journal} {\bibinfo
  {journal} {Nature Physics}\ }\textbf {\bibinfo {volume} {16}},\ \bibinfo
  {pages} {1050} (\bibinfo {year} {2020})}\BibitemShut {NoStop}%
\bibitem [{\citenamefont {Diker}(2016)}]{diker2016deterministic}%
  \BibitemOpen
  \bibfield  {author} {\bibinfo {author} {\bibfnamefont {F.}~\bibnamefont
  {Diker}},\ }\href@noop {} {\enquote {\bibinfo {title} {Deterministic
  construction of arbitrary w states with quadratically increasing number of
  two-qubit gates},}\ } (\bibinfo {year} {2016}),\ \Eprint
  {http://arxiv.org/abs/1606.09290} {arXiv:1606.09290 [quant-ph]} \BibitemShut
  {NoStop}%
\bibitem [{\citenamefont {Kuzmak}(2021)}]{Kuzmak2021}%
  \BibitemOpen
  \bibfield  {author} {\bibinfo {author} {\bibfnamefont {A.~R.}\ \bibnamefont
  {Kuzmak}},\ }\href {\doibase 10.1007/s11128-021-03196-9} {\bibfield
  {journal} {\bibinfo  {journal} {Quantum Information Processing}\ }\textbf
  {\bibinfo {volume} {20}},\ \bibinfo {pages} {269} (\bibinfo {year}
  {2021})}\BibitemShut {NoStop}%
\bibitem [{\citenamefont {Paris}\ and\ \citenamefont
  {Rehacek}(2004)}]{paris2004quantum}%
  \BibitemOpen
  \bibfield  {author} {\bibinfo {author} {\bibfnamefont {M.}~\bibnamefont
  {Paris}}\ and\ \bibinfo {author} {\bibfnamefont {J.}~\bibnamefont
  {Rehacek}},\ }\href@noop {} {\emph {\bibinfo {title} {Quantum state
  estimation}}},\ Vol.\ \bibinfo {volume} {649}\ (\bibinfo  {publisher}
  {Springer Science \& Business Media},\ \bibinfo {year} {2004})\BibitemShut
  {NoStop}%
\bibitem [{\citenamefont {White}\ and\ \citenamefont
  {Wilson}(2020)}]{white2020mana}%
  \BibitemOpen
  \bibfield  {author} {\bibinfo {author} {\bibfnamefont {C.~D.}\ \bibnamefont
  {White}}\ and\ \bibinfo {author} {\bibfnamefont {J.~H.}\ \bibnamefont
  {Wilson}},\ }\href@noop {} {\bibfield  {journal} {\bibinfo  {journal} {arXiv
  preprint arXiv:2011.13937}\ } (\bibinfo {year} {2020})}\BibitemShut {NoStop}%
\bibitem [{\citenamefont {Haug}\ \emph {et~al.}(2021)\citenamefont {Haug},
  \citenamefont {Bharti},\ and\ \citenamefont {Kim}}]{PRXQuantum.2.040309}%
  \BibitemOpen
  \bibfield  {author} {\bibinfo {author} {\bibfnamefont {T.}~\bibnamefont
  {Haug}}, \bibinfo {author} {\bibfnamefont {K.}~\bibnamefont {Bharti}}, \ and\
  \bibinfo {author} {\bibfnamefont {M.}~\bibnamefont {Kim}},\ }\href {\doibase
  10.1103/PRXQuantum.2.040309} {\bibfield  {journal} {\bibinfo  {journal} {PRX
  Quantum}\ }\textbf {\bibinfo {volume} {2}},\ \bibinfo {pages} {040309}
  (\bibinfo {year} {2021})}\BibitemShut {NoStop}%
\bibitem [{\citenamefont {Schuld}\ \emph {et~al.}(2020)\citenamefont {Schuld},
  \citenamefont {Bocharov}, \citenamefont {Svore},\ and\ \citenamefont
  {Wiebe}}]{schuld2020circuit}%
  \BibitemOpen
  \bibfield  {author} {\bibinfo {author} {\bibfnamefont {M.}~\bibnamefont
  {Schuld}}, \bibinfo {author} {\bibfnamefont {A.}~\bibnamefont {Bocharov}},
  \bibinfo {author} {\bibfnamefont {K.~M.}\ \bibnamefont {Svore}}, \ and\
  \bibinfo {author} {\bibfnamefont {N.}~\bibnamefont {Wiebe}},\ }\href
  {\doibase 10.1103/physreva.101.032308} {\bibfield  {journal} {\bibinfo
  {journal} {Physical Review A}\ }\textbf {\bibinfo {volume} {101}} (\bibinfo
  {year} {2020}),\ 10.1103/physreva.101.032308}\BibitemShut {NoStop}%
\bibitem [{\citenamefont {Sim}\ \emph {et~al.}(2019)\citenamefont {Sim},
  \citenamefont {Johnson},\ and\ \citenamefont
  {Aspuru-Guzik}}]{sim2019expressibility}%
  \BibitemOpen
  \bibfield  {author} {\bibinfo {author} {\bibfnamefont {S.}~\bibnamefont
  {Sim}}, \bibinfo {author} {\bibfnamefont {P.~D.}\ \bibnamefont {Johnson}}, \
  and\ \bibinfo {author} {\bibfnamefont {A.}~\bibnamefont {Aspuru-Guzik}},\
  }\href@noop {} {\bibfield  {journal} {\bibinfo  {journal} {Advanced Quantum
  Technologies}\ }\textbf {\bibinfo {volume} {2}},\ \bibinfo {pages} {1900070}
  (\bibinfo {year} {2019})}\BibitemShut {NoStop}%
\bibitem [{\citenamefont {Mitarai}\ \emph {et~al.}(2018)\citenamefont
  {Mitarai}, \citenamefont {Negoro}, \citenamefont {Kitagawa},\ and\
  \citenamefont {Fujii}}]{mitarai2018quantum}%
  \BibitemOpen
  \bibfield  {author} {\bibinfo {author} {\bibfnamefont {K.}~\bibnamefont
  {Mitarai}}, \bibinfo {author} {\bibfnamefont {M.}~\bibnamefont {Negoro}},
  \bibinfo {author} {\bibfnamefont {M.}~\bibnamefont {Kitagawa}}, \ and\
  \bibinfo {author} {\bibfnamefont {K.}~\bibnamefont {Fujii}},\ }\href@noop {}
  {\bibfield  {journal} {\bibinfo  {journal} {Physical Review A}\ }\textbf
  {\bibinfo {volume} {98}},\ \bibinfo {pages} {032309} (\bibinfo {year}
  {2018})}\BibitemShut {NoStop}%
\bibitem [{\citenamefont {Schuld}\ \emph {et~al.}(2019)\citenamefont {Schuld},
  \citenamefont {Bergholm}, \citenamefont {Gogolin}, \citenamefont {Izaac},\
  and\ \citenamefont {Killoran}}]{schuld2019evaluating}%
  \BibitemOpen
  \bibfield  {author} {\bibinfo {author} {\bibfnamefont {M.}~\bibnamefont
  {Schuld}}, \bibinfo {author} {\bibfnamefont {V.}~\bibnamefont {Bergholm}},
  \bibinfo {author} {\bibfnamefont {C.}~\bibnamefont {Gogolin}}, \bibinfo
  {author} {\bibfnamefont {J.}~\bibnamefont {Izaac}}, \ and\ \bibinfo {author}
  {\bibfnamefont {N.}~\bibnamefont {Killoran}},\ }\href@noop {} {\bibfield
  {journal} {\bibinfo  {journal} {Physical Review A}\ }\textbf {\bibinfo
  {volume} {99}},\ \bibinfo {pages} {032331} (\bibinfo {year}
  {2019})}\BibitemShut {NoStop}%
\bibitem [{\citenamefont {Anselmetti}\ \emph {et~al.}(2021)\citenamefont
  {Anselmetti}, \citenamefont {Wierichs}, \citenamefont {Gogolin},\ and\
  \citenamefont {Parrish}}]{Anselmetti_2021}%
  \BibitemOpen
  \bibfield  {author} {\bibinfo {author} {\bibfnamefont {G.-L.~R.}\
  \bibnamefont {Anselmetti}}, \bibinfo {author} {\bibfnamefont
  {D.}~\bibnamefont {Wierichs}}, \bibinfo {author} {\bibfnamefont
  {C.}~\bibnamefont {Gogolin}}, \ and\ \bibinfo {author} {\bibfnamefont
  {R.~M.}\ \bibnamefont {Parrish}},\ }\href {\doibase 10.1088/1367-2630/ac2cb3}
  {\bibfield  {journal} {\bibinfo  {journal} {New Journal of Physics}\ }\textbf
  {\bibinfo {volume} {23}},\ \bibinfo {pages} {113010} (\bibinfo {year}
  {2021})}\BibitemShut {NoStop}%
\bibitem [{\citenamefont {Kingma}\ and\ \citenamefont
  {Ba}(2014)}]{kingma2014adam}%
  \BibitemOpen
  \bibfield  {author} {\bibinfo {author} {\bibfnamefont {D.~P.}\ \bibnamefont
  {Kingma}}\ and\ \bibinfo {author} {\bibfnamefont {J.}~\bibnamefont {Ba}},\
  }\href@noop {} {\bibfield  {journal} {\bibinfo  {journal} {arXiv preprint
  arXiv:1412.6980}\ } (\bibinfo {year} {2014})}\BibitemShut {NoStop}%
\bibitem [{\citenamefont {Stokes}\ \emph
  {et~al.}(2020{\natexlab{a}})\citenamefont {Stokes}, \citenamefont {Izaac},
  \citenamefont {Killoran},\ and\ \citenamefont {Carleo}}]{stokes2020quantum}%
  \BibitemOpen
  \bibfield  {author} {\bibinfo {author} {\bibfnamefont {J.}~\bibnamefont
  {Stokes}}, \bibinfo {author} {\bibfnamefont {J.}~\bibnamefont {Izaac}},
  \bibinfo {author} {\bibfnamefont {N.}~\bibnamefont {Killoran}}, \ and\
  \bibinfo {author} {\bibfnamefont {G.}~\bibnamefont {Carleo}},\ }\href@noop {}
  {\bibfield  {journal} {\bibinfo  {journal} {Quantum}\ }\textbf {\bibinfo
  {volume} {4}},\ \bibinfo {pages} {269} (\bibinfo {year}
  {2020}{\natexlab{a}})}\BibitemShut {NoStop}%
\bibitem [{\citenamefont {Harrow}\ and\ \citenamefont
  {Napp}(2021)}]{harrow2021low}%
  \BibitemOpen
  \bibfield  {author} {\bibinfo {author} {\bibfnamefont {A.~W.}\ \bibnamefont
  {Harrow}}\ and\ \bibinfo {author} {\bibfnamefont {J.~C.}\ \bibnamefont
  {Napp}},\ }\href@noop {} {\bibfield  {journal} {\bibinfo  {journal} {Physical
  Review Letters}\ }\textbf {\bibinfo {volume} {126}},\ \bibinfo {pages}
  {140502} (\bibinfo {year} {2021})}\BibitemShut {NoStop}%
\bibitem [{\citenamefont {Stokes}\ \emph
  {et~al.}(2020{\natexlab{b}})\citenamefont {Stokes}, \citenamefont {Izaac},
  \citenamefont {Killoran},\ and\ \citenamefont
  {Carleo}}]{Stokes2020quantumnatural}%
  \BibitemOpen
  \bibfield  {author} {\bibinfo {author} {\bibfnamefont {J.}~\bibnamefont
  {Stokes}}, \bibinfo {author} {\bibfnamefont {J.}~\bibnamefont {Izaac}},
  \bibinfo {author} {\bibfnamefont {N.}~\bibnamefont {Killoran}}, \ and\
  \bibinfo {author} {\bibfnamefont {G.}~\bibnamefont {Carleo}},\ }\href
  {\doibase 10.22331/q-2020-05-25-269} {\bibfield  {journal} {\bibinfo
  {journal} {{Quantum}}\ }\textbf {\bibinfo {volume} {4}},\ \bibinfo {pages}
  {269} (\bibinfo {year} {2020}{\natexlab{b}})}\BibitemShut {NoStop}%
\bibitem [{\citenamefont {D\"ur}\ \emph {et~al.}(2000)\citenamefont {D\"ur},
  \citenamefont {Vidal},\ and\ \citenamefont {Cirac}}]{PhysRevA.62.062314}%
  \BibitemOpen
  \bibfield  {author} {\bibinfo {author} {\bibfnamefont {W.}~\bibnamefont
  {D\"ur}}, \bibinfo {author} {\bibfnamefont {G.}~\bibnamefont {Vidal}}, \ and\
  \bibinfo {author} {\bibfnamefont {J.~I.}\ \bibnamefont {Cirac}},\ }\href
  {\doibase 10.1103/PhysRevA.62.062314} {\bibfield  {journal} {\bibinfo
  {journal} {Phys. Rev. A}\ }\textbf {\bibinfo {volume} {62}},\ \bibinfo
  {pages} {062314} (\bibinfo {year} {2000})}\BibitemShut {NoStop}%
\bibitem [{\citenamefont {Cerezo}\ \emph
  {et~al.}(2021{\natexlab{b}})\citenamefont {Cerezo}, \citenamefont {Sone},
  \citenamefont {Volkoff}, \citenamefont {Cincio},\ and\ \citenamefont
  {Coles}}]{Cerezo2021}%
  \BibitemOpen
  \bibfield  {author} {\bibinfo {author} {\bibfnamefont {M.}~\bibnamefont
  {Cerezo}}, \bibinfo {author} {\bibfnamefont {A.}~\bibnamefont {Sone}},
  \bibinfo {author} {\bibfnamefont {T.}~\bibnamefont {Volkoff}}, \bibinfo
  {author} {\bibfnamefont {L.}~\bibnamefont {Cincio}}, \ and\ \bibinfo {author}
  {\bibfnamefont {P.~J.}\ \bibnamefont {Coles}},\ }\href {\doibase
  10.1038/s41467-021-21728-w} {\bibfield  {journal} {\bibinfo  {journal}
  {Nature Communications}\ }\textbf {\bibinfo {volume} {12}},\ \bibinfo {pages}
  {1791} (\bibinfo {year} {2021}{\natexlab{b}})}\BibitemShut {NoStop}%
\bibitem [{\citenamefont {Haug}\ and\ \citenamefont
  {Kim}(2021)}]{haug2021natural}%
  \BibitemOpen
  \bibfield  {author} {\bibinfo {author} {\bibfnamefont {T.}~\bibnamefont
  {Haug}}\ and\ \bibinfo {author} {\bibfnamefont {M.}~\bibnamefont {Kim}},\
  }\href@noop {} {\bibfield  {journal} {\bibinfo  {journal} {arXiv preprint
  arXiv:2107.14063}\ } (\bibinfo {year} {2021})}\BibitemShut {NoStop}%
\bibitem [{Note1()}]{Note1}%
  \BibitemOpen
  \bibinfo {note} {For W states, if one qubit is lost, the remaining system
  still entangles, from which contrasts with GHZ states, that fully separable
  after disentangle one qubit. See also Ref.~\cite
  {PhysRevA.62.062314}.}\BibitemShut {Stop}%
\bibitem [{\citenamefont {Steinbrecher}\ \emph {et~al.}(2019)\citenamefont
  {Steinbrecher}, \citenamefont {Olson}, \citenamefont {Englund},\ and\
  \citenamefont {Carolan}}]{Steinbrecher2019}%
  \BibitemOpen
  \bibfield  {author} {\bibinfo {author} {\bibfnamefont {G.~R.}\ \bibnamefont
  {Steinbrecher}}, \bibinfo {author} {\bibfnamefont {J.~P.}\ \bibnamefont
  {Olson}}, \bibinfo {author} {\bibfnamefont {D.}~\bibnamefont {Englund}}, \
  and\ \bibinfo {author} {\bibfnamefont {J.}~\bibnamefont {Carolan}},\ }\href
  {\doibase 10.1038/s41534-019-0174-7} {\bibfield  {journal} {\bibinfo
  {journal} {npj Quantum Information}\ }\textbf {\bibinfo {volume} {5}},\
  \bibinfo {pages} {60} (\bibinfo {year} {2019})}\BibitemShut {NoStop}%
\bibitem [{\citenamefont {Torlai}\ \emph {et~al.}(2020)\citenamefont {Torlai},
  \citenamefont {Wood}, \citenamefont {Acharya}, \citenamefont {Carleo},
  \citenamefont {Carrasquilla},\ and\ \citenamefont {Aolita}}]{Giacomo2020}%
  \BibitemOpen
  \bibfield  {author} {\bibinfo {author} {\bibfnamefont {G.}~\bibnamefont
  {Torlai}}, \bibinfo {author} {\bibfnamefont {C.~J.}\ \bibnamefont {Wood}},
  \bibinfo {author} {\bibfnamefont {A.}~\bibnamefont {Acharya}}, \bibinfo
  {author} {\bibfnamefont {G.}~\bibnamefont {Carleo}}, \bibinfo {author}
  {\bibfnamefont {J.}~\bibnamefont {Carrasquilla}}, \ and\ \bibinfo {author}
  {\bibfnamefont {L.}~\bibnamefont {Aolita}},\ }\href@noop {} {\enquote
  {\bibinfo {title} {Quantum process tomography with unsupervised learning and
  tensor networks},}\ } (\bibinfo {year} {2020}),\ \Eprint
  {http://arxiv.org/abs/2006.02424} {arXiv:2006.02424 [quant-ph]} \BibitemShut
  {NoStop}%
\bibitem [{\citenamefont {Zhou}\ \emph {et~al.}(2021)\citenamefont {Zhou},
  \citenamefont {Hong},\ and\ \citenamefont {Ran}}]{PhysRevA.104.042601}%
  \BibitemOpen
  \bibfield  {author} {\bibinfo {author} {\bibfnamefont {P.-F.}\ \bibnamefont
  {Zhou}}, \bibinfo {author} {\bibfnamefont {R.}~\bibnamefont {Hong}}, \ and\
  \bibinfo {author} {\bibfnamefont {S.-J.}\ \bibnamefont {Ran}},\ }\href
  {\doibase 10.1103/PhysRevA.104.042601} {\bibfield  {journal} {\bibinfo
  {journal} {Phys. Rev. A}\ }\textbf {\bibinfo {volume} {104}},\ \bibinfo
  {pages} {042601} (\bibinfo {year} {2021})}\BibitemShut {NoStop}%
\bibitem [{\citenamefont {McClean}\ \emph {et~al.}(2018)\citenamefont
  {McClean}, \citenamefont {Boixo}, \citenamefont {Smelyanskiy}, \citenamefont
  {Babbush},\ and\ \citenamefont {Neven}}]{McClean2018}%
  \BibitemOpen
  \bibfield  {author} {\bibinfo {author} {\bibfnamefont {J.~R.}\ \bibnamefont
  {McClean}}, \bibinfo {author} {\bibfnamefont {S.}~\bibnamefont {Boixo}},
  \bibinfo {author} {\bibfnamefont {V.~N.}\ \bibnamefont {Smelyanskiy}},
  \bibinfo {author} {\bibfnamefont {R.}~\bibnamefont {Babbush}}, \ and\
  \bibinfo {author} {\bibfnamefont {H.}~\bibnamefont {Neven}},\ }\href
  {\doibase 10.1038/s41467-018-07090-4} {\bibfield  {journal} {\bibinfo
  {journal} {Nature Communications}\ }\textbf {\bibinfo {volume} {9}},\
  \bibinfo {pages} {4812} (\bibinfo {year} {2018})}\BibitemShut {NoStop}%
\bibitem [{\citenamefont {Holmes}\ \emph {et~al.}(2022)\citenamefont {Holmes},
  \citenamefont {Sharma}, \citenamefont {Cerezo},\ and\ \citenamefont
  {Coles}}]{PRXQuantum.3.010313}%
  \BibitemOpen
  \bibfield  {author} {\bibinfo {author} {\bibfnamefont {Z.}~\bibnamefont
  {Holmes}}, \bibinfo {author} {\bibfnamefont {K.}~\bibnamefont {Sharma}},
  \bibinfo {author} {\bibfnamefont {M.}~\bibnamefont {Cerezo}}, \ and\ \bibinfo
  {author} {\bibfnamefont {P.~J.}\ \bibnamefont {Coles}},\ }\href {\doibase
  10.1103/PRXQuantum.3.010313} {\bibfield  {journal} {\bibinfo  {journal} {PRX
  Quantum}\ }\textbf {\bibinfo {volume} {3}},\ \bibinfo {pages} {010313}
  (\bibinfo {year} {2022})}\BibitemShut {NoStop}%
\bibitem [{\citenamefont {Ortiz~Marrero}\ \emph {et~al.}(2021)\citenamefont
  {Ortiz~Marrero}, \citenamefont {Kieferov\'a},\ and\ \citenamefont
  {Wiebe}}]{PRXQuantum.2.040316}%
  \BibitemOpen
  \bibfield  {author} {\bibinfo {author} {\bibfnamefont {C.}~\bibnamefont
  {Ortiz~Marrero}}, \bibinfo {author} {\bibfnamefont {M.}~\bibnamefont
  {Kieferov\'a}}, \ and\ \bibinfo {author} {\bibfnamefont {N.}~\bibnamefont
  {Wiebe}},\ }\href {\doibase 10.1103/PRXQuantum.2.040316} {\bibfield
  {journal} {\bibinfo  {journal} {PRX Quantum}\ }\textbf {\bibinfo {volume}
  {2}},\ \bibinfo {pages} {040316} (\bibinfo {year} {2021})}\BibitemShut
  {NoStop}%
\bibitem [{\citenamefont {Wang}\ \emph
  {et~al.}(2021{\natexlab{a}})\citenamefont {Wang}, \citenamefont {Fontana},
  \citenamefont {Cerezo}, \citenamefont {Sharma}, \citenamefont {Sone},
  \citenamefont {Cincio},\ and\ \citenamefont {Coles}}]{Wang2021}%
  \BibitemOpen
  \bibfield  {author} {\bibinfo {author} {\bibfnamefont {S.}~\bibnamefont
  {Wang}}, \bibinfo {author} {\bibfnamefont {E.}~\bibnamefont {Fontana}},
  \bibinfo {author} {\bibfnamefont {M.}~\bibnamefont {Cerezo}}, \bibinfo
  {author} {\bibfnamefont {K.}~\bibnamefont {Sharma}}, \bibinfo {author}
  {\bibfnamefont {A.}~\bibnamefont {Sone}}, \bibinfo {author} {\bibfnamefont
  {L.}~\bibnamefont {Cincio}}, \ and\ \bibinfo {author} {\bibfnamefont {P.~J.}\
  \bibnamefont {Coles}},\ }\href {\doibase 10.1038/s41467-021-27045-6}
  {\bibfield  {journal} {\bibinfo  {journal} {Nature Communications}\ }\textbf
  {\bibinfo {volume} {12}},\ \bibinfo {pages} {6961} (\bibinfo {year}
  {2021}{\natexlab{a}})}\BibitemShut {NoStop}%
\bibitem [{\citenamefont {Uvarov}\ and\ \citenamefont
  {Biamonte}(2021)}]{Uvarov_2021}%
  \BibitemOpen
  \bibfield  {author} {\bibinfo {author} {\bibfnamefont {A.~V.}\ \bibnamefont
  {Uvarov}}\ and\ \bibinfo {author} {\bibfnamefont {J.~D.}\ \bibnamefont
  {Biamonte}},\ }\href {\doibase 10.1088/1751-8121/abfac7} {\bibfield
  {journal} {\bibinfo  {journal} {Journal of Physics A: Mathematical and
  Theoretical}\ }\textbf {\bibinfo {volume} {54}},\ \bibinfo {pages} {245301}
  (\bibinfo {year} {2021})}\BibitemShut {NoStop}%
\bibitem [{\citenamefont {Volkoff}\ and\ \citenamefont
  {Coles}(2021)}]{Volkoff_2021}%
  \BibitemOpen
  \bibfield  {author} {\bibinfo {author} {\bibfnamefont {T.}~\bibnamefont
  {Volkoff}}\ and\ \bibinfo {author} {\bibfnamefont {P.~J.}\ \bibnamefont
  {Coles}},\ }\href {\doibase 10.1088/2058-9565/abd891} {\bibfield  {journal}
  {\bibinfo  {journal} {Quantum Science and Technology}\ }\textbf {\bibinfo
  {volume} {6}},\ \bibinfo {pages} {025008} (\bibinfo {year}
  {2021})}\BibitemShut {NoStop}%
\bibitem [{\citenamefont {Verdon}\ \emph {et~al.}(2019)\citenamefont {Verdon},
  \citenamefont {Broughton}, \citenamefont {McClean}, \citenamefont {Sung},
  \citenamefont {Babbush}, \citenamefont {Jiang}, \citenamefont {Neven},\ and\
  \citenamefont {Mohseni}}]{verdon2019learning}%
  \BibitemOpen
  \bibfield  {author} {\bibinfo {author} {\bibfnamefont {G.}~\bibnamefont
  {Verdon}}, \bibinfo {author} {\bibfnamefont {M.}~\bibnamefont {Broughton}},
  \bibinfo {author} {\bibfnamefont {J.~R.}\ \bibnamefont {McClean}}, \bibinfo
  {author} {\bibfnamefont {K.~J.}\ \bibnamefont {Sung}}, \bibinfo {author}
  {\bibfnamefont {R.}~\bibnamefont {Babbush}}, \bibinfo {author} {\bibfnamefont
  {Z.}~\bibnamefont {Jiang}}, \bibinfo {author} {\bibfnamefont
  {H.}~\bibnamefont {Neven}}, \ and\ \bibinfo {author} {\bibfnamefont
  {M.}~\bibnamefont {Mohseni}},\ }\href@noop {} {\bibfield  {journal} {\bibinfo
   {journal} {arXiv preprint arXiv:1907.05415}\ } (\bibinfo {year}
  {2019})}\BibitemShut {NoStop}%
\bibitem [{\citenamefont {Grant}\ \emph {et~al.}(2019)\citenamefont {Grant},
  \citenamefont {Wossnig}, \citenamefont {Ostaszewski},\ and\ \citenamefont
  {Benedetti}}]{grant2019initialization}%
  \BibitemOpen
  \bibfield  {author} {\bibinfo {author} {\bibfnamefont {E.}~\bibnamefont
  {Grant}}, \bibinfo {author} {\bibfnamefont {L.}~\bibnamefont {Wossnig}},
  \bibinfo {author} {\bibfnamefont {M.}~\bibnamefont {Ostaszewski}}, \ and\
  \bibinfo {author} {\bibfnamefont {M.}~\bibnamefont {Benedetti}},\ }\href@noop
  {} {\bibfield  {journal} {\bibinfo  {journal} {Quantum}\ }\textbf {\bibinfo
  {volume} {3}},\ \bibinfo {pages} {214} (\bibinfo {year} {2019})}\BibitemShut
  {NoStop}%
\bibitem [{\citenamefont {Nachman}\ \emph {et~al.}(2020)\citenamefont
  {Nachman}, \citenamefont {Urbanek}, \citenamefont {de~Jong},\ and\
  \citenamefont {Bauer}}]{Nachman2020}%
  \BibitemOpen
  \bibfield  {author} {\bibinfo {author} {\bibfnamefont {B.}~\bibnamefont
  {Nachman}}, \bibinfo {author} {\bibfnamefont {M.}~\bibnamefont {Urbanek}},
  \bibinfo {author} {\bibfnamefont {W.~A.}\ \bibnamefont {de~Jong}}, \ and\
  \bibinfo {author} {\bibfnamefont {C.~W.}\ \bibnamefont {Bauer}},\ }\href
  {\doibase 10.1038/s41534-020-00309-7} {\bibfield  {journal} {\bibinfo
  {journal} {npj Quantum Information}\ }\textbf {\bibinfo {volume} {6}},\
  \bibinfo {pages} {84} (\bibinfo {year} {2020})}\BibitemShut {NoStop}%
\bibitem [{\citenamefont {Endo}\ \emph {et~al.}(2021)\citenamefont {Endo},
  \citenamefont {Cai}, \citenamefont {Benjamin},\ and\ \citenamefont
  {Yuan}}]{doi:10.7566/JPSJ.90.032001}%
  \BibitemOpen
  \bibfield  {author} {\bibinfo {author} {\bibfnamefont {S.}~\bibnamefont
  {Endo}}, \bibinfo {author} {\bibfnamefont {Z.}~\bibnamefont {Cai}}, \bibinfo
  {author} {\bibfnamefont {S.~C.}\ \bibnamefont {Benjamin}}, \ and\ \bibinfo
  {author} {\bibfnamefont {X.}~\bibnamefont {Yuan}},\ }\href {\doibase
  10.7566/JPSJ.90.032001} {\bibfield  {journal} {\bibinfo  {journal} {Journal
  of the Physical Society of Japan}\ }\textbf {\bibinfo {volume} {90}},\
  \bibinfo {pages} {032001} (\bibinfo {year} {2021})},\ \Eprint
  {http://arxiv.org/abs/https://doi.org/10.7566/JPSJ.90.032001}
  {https://doi.org/10.7566/JPSJ.90.032001} \BibitemShut {NoStop}%
\bibitem [{\citenamefont {Czarnik}\ \emph {et~al.}(2021)\citenamefont
  {Czarnik}, \citenamefont {Arrasmith}, \citenamefont {Coles},\ and\
  \citenamefont {Cincio}}]{Czarnik2021errormitigation}%
  \BibitemOpen
  \bibfield  {author} {\bibinfo {author} {\bibfnamefont {P.}~\bibnamefont
  {Czarnik}}, \bibinfo {author} {\bibfnamefont {A.}~\bibnamefont {Arrasmith}},
  \bibinfo {author} {\bibfnamefont {P.~J.}\ \bibnamefont {Coles}}, \ and\
  \bibinfo {author} {\bibfnamefont {L.}~\bibnamefont {Cincio}},\ }\href
  {\doibase 10.22331/q-2021-11-26-592} {\bibfield  {journal} {\bibinfo
  {journal} {{Quantum}}\ }\textbf {\bibinfo {volume} {5}},\ \bibinfo {pages}
  {592} (\bibinfo {year} {2021})}\BibitemShut {NoStop}%
\bibitem [{\citenamefont {Nation}\ \emph {et~al.}(2021)\citenamefont {Nation},
  \citenamefont {Kang}, \citenamefont {Sundaresan},\ and\ \citenamefont
  {Gambetta}}]{PRXQuantum.2.040326}%
  \BibitemOpen
  \bibfield  {author} {\bibinfo {author} {\bibfnamefont {P.~D.}\ \bibnamefont
  {Nation}}, \bibinfo {author} {\bibfnamefont {H.}~\bibnamefont {Kang}},
  \bibinfo {author} {\bibfnamefont {N.}~\bibnamefont {Sundaresan}}, \ and\
  \bibinfo {author} {\bibfnamefont {J.~M.}\ \bibnamefont {Gambetta}},\ }\href
  {\doibase 10.1103/PRXQuantum.2.040326} {\bibfield  {journal} {\bibinfo
  {journal} {PRX Quantum}\ }\textbf {\bibinfo {volume} {2}},\ \bibinfo {pages}
  {040326} (\bibinfo {year} {2021})}\BibitemShut {NoStop}%
\bibitem [{\citenamefont {Maciejewski}\ \emph {et~al.}(2020)\citenamefont
  {Maciejewski}, \citenamefont {Zimbor{\'{a}}s},\ and\ \citenamefont
  {Oszmaniec}}]{Maciejewski2020mitigationofreadout}%
  \BibitemOpen
  \bibfield  {author} {\bibinfo {author} {\bibfnamefont {F.~B.}\ \bibnamefont
  {Maciejewski}}, \bibinfo {author} {\bibfnamefont {Z.}~\bibnamefont
  {Zimbor{\'{a}}s}}, \ and\ \bibinfo {author} {\bibfnamefont {M.}~\bibnamefont
  {Oszmaniec}},\ }\href {\doibase 10.22331/q-2020-04-24-257} {\bibfield
  {journal} {\bibinfo  {journal} {{Quantum}}\ }\textbf {\bibinfo {volume}
  {4}},\ \bibinfo {pages} {257} (\bibinfo {year} {2020})}\BibitemShut {NoStop}%
\bibitem [{\citenamefont {Geller}(2020)}]{Geller_2020}%
  \BibitemOpen
  \bibfield  {author} {\bibinfo {author} {\bibfnamefont {M.~R.}\ \bibnamefont
  {Geller}},\ }\href {\doibase 10.1088/2058-9565/ab9591} {\bibfield  {journal}
  {\bibinfo  {journal} {Quantum Science and Technology}\ }\textbf {\bibinfo
  {volume} {5}},\ \bibinfo {pages} {03LT01} (\bibinfo {year}
  {2020})}\BibitemShut {NoStop}%
\bibitem [{\citenamefont {Wang}\ \emph
  {et~al.}(2021{\natexlab{b}})\citenamefont {Wang}, \citenamefont {Chen},\ and\
  \citenamefont {Wang}}]{wang2021measurement}%
  \BibitemOpen
  \bibfield  {author} {\bibinfo {author} {\bibfnamefont {K.}~\bibnamefont
  {Wang}}, \bibinfo {author} {\bibfnamefont {Y.-A.}\ \bibnamefont {Chen}}, \
  and\ \bibinfo {author} {\bibfnamefont {X.}~\bibnamefont {Wang}},\ }\href@noop
  {} {\bibfield  {journal} {\bibinfo  {journal} {arXiv preprint
  arXiv:2103.13856}\ } (\bibinfo {year} {2021}{\natexlab{b}})}\BibitemShut
  {NoStop}%
\end{thebibliography}%
\end{document}